\documentclass[fleqn,usenatbib]{mnras}
\usepackage{fix-cm}
\usepackage{newtxtext,newtxmath}

\usepackage[T1]{fontenc}
\usepackage[percent]{overpic}

\DeclareRobustCommand{\VAN}[3]{#2}
\let\VANthebibliography\thebibliography
\def\thebibliography{\DeclareRobustCommand{\VAN}[3]{##3}\VANthebibliography}

\usepackage{graphicx}	
\usepackage{amsmath}	
\usepackage{hyperref}
\usepackage{upgreek}

\title[30 Massive Galaxy Clusters]{The MeerKAT Massive Distant Clusters Survey: a search for diffuse radio emission in 30 massive SZ-selected clusters at \texorpdfstring{$z > 1$}{z > 1}}

\author[D. G. Phuravhathu et al.]{Dakalo G. Phuravhathu$^{1}$\thanks{E-mail: dakalophuravha2@gmail.com},
M. Hilton$^{1,2}$,
S. P. Sikhosana$^{2,3}$,
D. Y. Klutse$^{2,3}$,
K. Knowles$^{4,5,3}$,
J. van Marrewĳk$^{6}$,
\newauthor
K. Moodley$^{2,3}$,
T. Mroczkowski$^{7}$,
N. Oozeer$^{5,4}$,
B. Partridge$^{8}$,
Y. C. Perrott$^{9}$,
C. Sifón$^{10}$,
U. Sureshkumar$^{11,1}$
\\
$^{1}$Wits Centre for Astrophysics, School of Physics, University of the Witwatersrand, Private Bag 3, 2050, Johannesburg, South Africa\\
$^{2}$School of Agriculture and Science University of KwaZulu-Natal, Westville Campus, Durban 4041, South Africa\\
$^{3}$Astrophysics Research Centre, University of KwaZulu-Natal, Durban, 3696, South Africa\\
$^{4}$Centre for Radio Astronomy Techniques and Technologies, Department of Physics and Electronics, Rhodes University, P.O. Box 94, Makhanda 6140, South Africa\\
$^{5}$South African Radio Astronomy Observatory, 2 Fir Street, Black River Park, Observatory 7925, South Africa\\
$^{6}$Leiden Observatory, Leiden University P.O. Box 9513, 2300 RA Leiden The Netherlands\\
$^{7}$Institute of Space Sciences (ICE, CSIC), Carrer de Can Magrans, s/n, 08193 Cerdanyola del Vallès, Barcelona, Spain\\
$^{8}$Department of Astronomy, Haverford College, Haverford, PA 19041, USA\\
$^{9}$School of Chemical and Physical Sciences, Victoria University of Wellington, PO Box 600, Wellington 6140, New Zealand\\
$^{10}$Instituto de Física, Pontificia Universidad Católica de Valparaíso, Casilla 4059, Valparaíso, Chile\\
$^{11}$National Centre for Nuclear Research, ul. Pasteura 7, 02-093 Warsaw, Poland
}

\date{Accepted XXX. Received YYY; in original form ZZZ}

\pubyear{\the\year{}}

\begin{document}
\label{firstpage}
\pagerange{\pageref{firstpage}--\pageref{lastpage}}
\maketitle

\begin{abstract}
We present the results of a search for diffuse radio emission in a uniformly selected sample of 30 of the most massive Sunyaev--Zel'dovich selected galaxy clusters at $z > 1$, providing the first statistical constraints on the evolution of cluster-scale diffuse emission beyond this redshift. We also analyse the scaling relations between radio power ($P_{1.4\,\mathrm{GHz}}$) and cluster mass ($M_{\rm 500c}$) in this high-redshift sample. It is well established that radio halos are primarily found in the most massive clusters, where turbulent energy from major mergers can re-accelerate relativistic electrons and amplify magnetic fields on megaparsec scales. Deep MeerKAT 1.28\,GHz observations reveal diffuse radio halos in eight clusters (27\%), while the remaining 21 (70\%) show no emission; one additional cluster (3\%) was excluded from the radio analysis due to poor data quality. The halo detection rate in this high-redshift sample is lower than at intermediate redshift, but remains higher than the $\lesssim 10\%$ occurrence generally predicted by theoretical models at $z \gtrsim 1$. The detected radio halos scatter around the best-fitting $P_{1.4\,\mathrm{GHz}}$--$M_{\rm 500c}^{\rm {Unc}}$ relation derived for the MMDCS sample, whereas non-detections populate the lower envelope of the radio power–mass plane, similar to trends seen at lower redshift. No cluster-scale radio relics or mini-halos are identified. Our findings highlight MeerKAT's ability to probe non-thermal processes in the most distant clusters and the need for deeper, lower-frequency surveys to uncover faint diffuse emission and test the persistence of the $P_{1.4\,\mathrm{GHz}}$--$M_{\rm 500}$ relation across cosmic time.
\end{abstract}





\section{Introduction}

Galaxy clusters are the largest gravitationally bound structures in the universe, assembling at the intersections of cosmic filaments over billions of years through hierarchical merging processes \citep{2007PhR...443....1M, 2014IJMPD..2330007B, 2019SSRv..215...16V}. Major cluster mergers rank among the most energetic phenomena since the Big Bang, each releasing up to $10^{64}\,\mathrm{erg}$, sufficient to heat the intracluster medium (ICM) to $10^7$--$10^8$\,K, stir turbulence, drive shocks, and amplify magnetic fields on megaparsec scales \citep{2006MNRAS.369.1577C, 2008A&A...484..327V, 2013ApJ...777..141C}. These non-thermal processes produce diffuse synchrotron radio emission, observed as radio halos and relics, which serve as direct tracers of turbulence, cosmic ray populations, and magnetic field amplification in the ICM \citep{2012A&ARv..20...54F, 2019SSRv..215...16V, 2023A&A...680A..30C}.
\newline

Radio halos are extended ($\sim$Mpc-scale), centrally located sources with steep spectra ($-1.0 \gtrsim \alpha \gtrsim -1.4$\footnote[1]{$S_{\nu} \propto \nu^{\alpha}$, where $S_{\nu}$ is the flux density at frequency $\nu$ and $\alpha$ is the spectral index}), mainly found in dynamically disturbed clusters undergoing major mergers \citep{2001MNRAS.320..365B, 2013ApJ...777..141C, 2014IJMPD..2330007B}. Current evidence strongly supports turbulent (re)acceleration as the dominant origin: turbulence injected during mergers re-energizes relativistic electrons and amplifies cluster-wide magnetic fields \citep{2004JKAS...37..493B, 2011MNRAS.410..127B, 2013ApJ...777..141C}. Hadronic models, in which relativistic protons produce secondary electrons via collisions, likely play only a subdominant role, constrained by deep gamma-ray and radio observations \citep{1980ApJ...239L..93D, 2000A&A...362..151D}. In contrast, radio relics are elongated, polarized structures found in cluster outskirts, tracing shock fronts driven by mergers \citep{1998A&A...332..395E, 2019SSRv..215...16V, 2023A&A...672A..43C}. Their spectral and polarization properties are explained by diffusive shock acceleration, but observed low Mach numbers imply that re-acceleration of fossil electrons or pre-existing cosmic-ray populations is critical \citep{2010Sci...330..347V, 2014ApJ...785....1B, 2020A&A...634A..64B}. Mini-halos are smaller ($\sim$100--500 kpc) diffuse sources centred on brightest cluster galaxies in cool-core systems, likely supported by turbulence from core sloshing \citep{2008SSRv..134...93F, 2019ApJ...880...70G, 2002A&A...386..456G}.
\newline

These diffuse phenomena are fundamental to understanding the link between large-scale structure formation and the microphysics of cosmic plasmas. They reveal the interplay between turbulence, shocks, and magnetic fields and provide direct evidence of relativistic particles permeating the ICM. Their detection remains observationally challenging because their emission is extremely faint, typically a few $\upmu$Jy\,arcsec$^{-2}$ at GHz frequencies, and further reduced by cosmological surface-brightness dimming proportional to $(1+z)^{4}$, which becomes increasingly severe toward high-redshift \citep{2012A&ARv..20...54F, 2015aska.confE.105G, 2014ApJ...785....1B, 2021NatAs...5..268D}. Consequently, statistical samples of radio halos and relics have mostly been limited to bright and nearby clusters \citep{2006MNRAS.369.1577C, 2008A&A...484..327V, 2013ApJ...777..141C}. Recent sensitive observations with facilities such as MeerKAT \citep{2016mks..confE...1J,2018ApJ...856..180C} and LOFAR \citep{2013A&A...556A...2V} are changing this view by revealing faint extended sources, halos in less massive systems, and diffuse structures even in apparently relaxed clusters \citep{2021MNRAS.504.1749K, 2022A&A...657A..56K, 2023A&A...680A..30C}.
\newline

Statistical analyses of large, well-selected samples have become key to understanding the connection between cluster mass, dynamical state, and radio emission. Early works revealed that the probability of detecting giant halos increases sharply with X-ray luminosity and cluster mass \citep{2006MNRAS.369.1577C, 2013ApJ...777..141C}. The correlation between radio power and mass supports the scenario that the non-thermal energy content scales with the depth of the cluster's gravitational potential well. Furthermore, analyses have revealed a striking bimodality in the radio power versus mass distribution: clusters either host bright halos that follow this scaling relation or they show only upper-limit constraints, implying a possible physical boundary or sensitivity bias \citep{2013ApJ...777..141C, 2015A&A...579A..92K, 2023A&A...680A..30C}. Extending bimodality studies across a broad redshift range is essential for testing whether it reflects true differences in cluster states or primarily observational limitations, although at high-redshift this becomes increasingly challenging because of surface‑brightness dimming and the need for much deeper integrations.
\newline

Comprehensive surveys that include both detections and non-detections, as emphasized by \citet{2008A&A...484..327V, 2013ApJ...777..141C, 2015A&A...579A..92K, 2023A&A...680A..30C}, are required to constrain theoretical models of turbulent re-acceleration, magnetic field amplification, and the evolution of non-thermal components in clusters. The MeerKAT Massive Distant Clusters Survey (MMDCS) extends such statistical analyses to $z > 1$, targeting the 30 most massive Sunyaev--Zel’dovich (SZ) selected clusters from the Atacama Cosmology Telescope (ACT) DR5 catalogue \citep{2021ApJS..253....3H} that also lie within the Dark Energy Camera Legacy Survey (DECaLS) DR9 footprint \citep{2019AJ....157..168D}. A key strength of MMDCS is that the cluster sample is uniformly selected from the ACT DR5 SZ catalogue using well-defined mass and redshift cuts, within a region of sky with homogeneous optical/infrared coverage, enabling a clean statistical assessment of the radio halo occurrence at $z>1$. Early results from a pilot study of six of the most massive clusters ($M_{\rm 500c} \approx 6.7$--$8.5 \times 10^{14} \, M_{\odot}$) reveal diffuse radio emission in four clusters (confidently detected) and tentative signals in two others \citep[P25 hereafter]{2025MNRAS.542.1544P}. The detected halos span $0.47$--$1.08 \, \mathrm{Mpc}$ and exhibit $1.4 \, \mathrm{GHz}$ radio powers from $(0.30 \pm 0.08)$ to $(3.55 \pm 1.06) \times 10^{25} \, \mathrm{W\,Hz^{-1}}$. These systems broadly follow the established radio power–mass scaling seen at lower redshifts, supporting models of turbulent re-acceleration in high $z$ mergers \citep{2021NatAs...5..268D}. The present work expands this analysis to the full sample of 30 clusters, thereby providing the first statistical constraints on the evolution of cluster-scale diffuse emission beyond $z>1$.
\newline

This paper is organized as follows: Section~\ref{sec:Radio observations and data reduction} describes the cluster sample, observations, and data processing; Section~\ref{sec:Results} presents the main results; Section~\ref{sec:Discussion} discusses physical implications and statistical trends; and Section~\ref{sec:Conclusions} summarizes our conclusions. Throughout, $R_{\rm 500c}$ denotes the radius within which the mean cluster density equals 500 times the critical density of the universe at redshift $z$, while $M_{\rm 500c}$ refers to the enclosed mass. A $\Lambda$CDM cosmology is assumed with $H_{0} = 70\,{\rm km\,s^{-1}\,Mpc^{-1}}$, $\Omega_{\mathrm{m}} = 0.3$, and $\Omega_{\Lambda} = 0.7$.

\section{Radio observations and data reduction}
\label{sec:Radio observations and data reduction}
\subsection{The cluster sample}
The MeerKAT Massive Distant Cluster Survey (MMDCS) is a focused MeerKAT survey at 1.28~GHz targeting a comprehensive sample of thirty of the most massive galaxy clusters at $z > 1$. The clusters are selected from the ACT DR5 catalogue \citep{2021ApJS..253....3H}, restricted to the portion of the sky overlapping with the DECaLS Data Release 9 optical and infrared footprint \citep{2019AJ....157..168D}. This region covers approximately $10\,861~\mathrm{deg^{2}}$, equivalent to about one quarter of the sky. Restricting the selection to this area ensures uniform optical coverage and access to multi-wavelength information, which is necessary for photometric redshifts and source classification, including identification of active galaxies through mid-infrared colours.
\newline

\begin{figure}
 \includegraphics[width=\columnwidth]{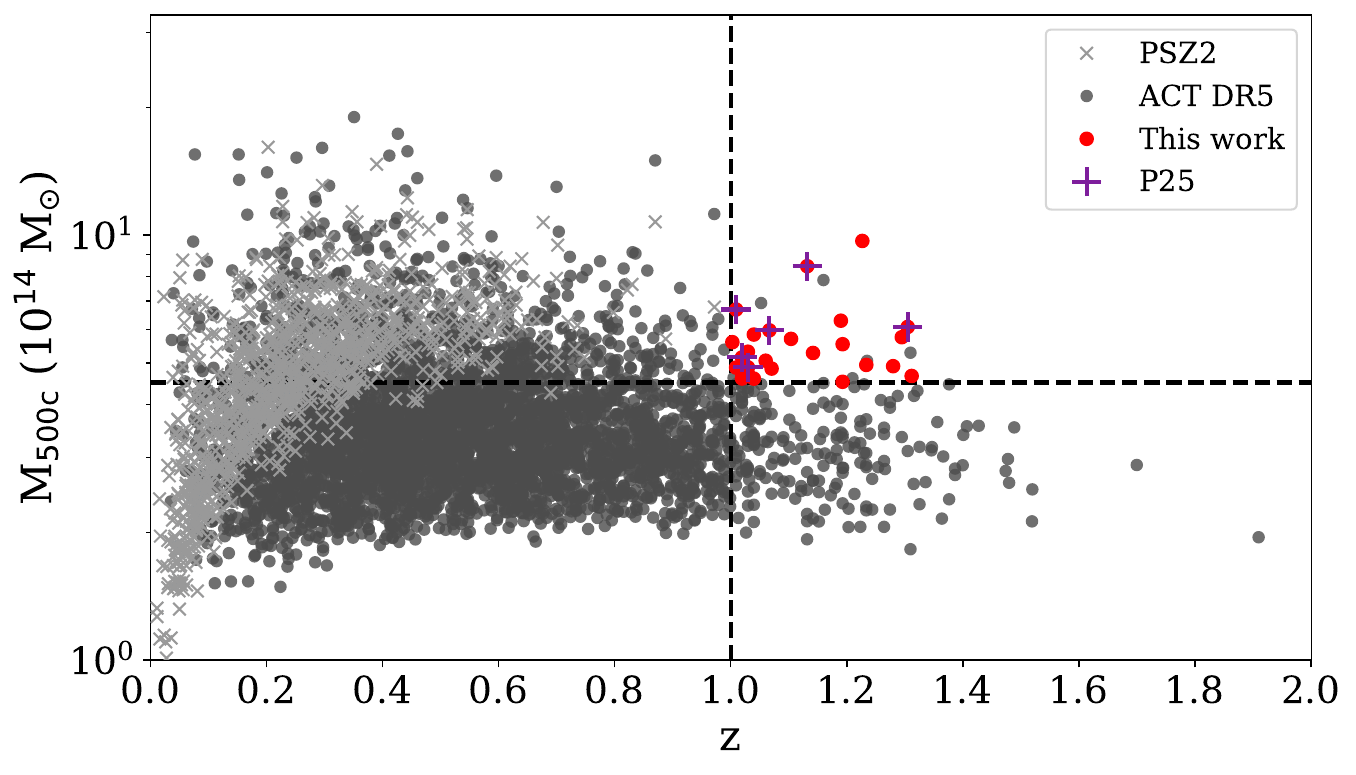}
 \caption{The distribution of ACT SZ-selected clusters in mass and redshift (grey points), highlighting in red the 30 clusters observed with MeerKAT that meet the criteria $z > 1$, $M_{\rm 500c} > 4.5 \times 10^{14}\,{\rm M_{\odot}}$, and overlap with the DECaLS DR9 footprint. Dashed lines indicate the adopted selection boundaries. Six additional ACT clusters fall within the same mass and redshift range but lie outside the DECaLS footprint and were therefore excluded from the survey to maintain uniform optical coverage.}

\label{fig:1}
\end{figure}

Full details of the survey design, selection criteria, and the pilot analysis of the six most massive clusters are presented in P25. The full MMDCS sample comprises 30 massive clusters at $z > 1$, but one target, ACT-CL~J2245.2$-$0433, was excluded from the present radio analysis due to poor data quality. In this work, we expand to the sample of twenty-nine clusters, spanning a mass\footnote[2]{Cluster masses $M_{\rm 500c}=M_{\rm 500c}^{\rm Cal}$ used throughout this work are the richness-based, weak-lensing-calibrated values reported in \cite{2021ApJS..253....3H}, corresponding to the \texttt{M\rm 500c Cal} column in the ACT DR5 catalogue. $M_{\rm 500c}^{\rm Unc}$ corresponds to the \texttt{M\rm 500c Unc} column (uncorrected for Eddington bias), which is on the same scale as masses reported in the \textit{Planck} PSZ2 catalogue and $\sim20$--$30\%$ lower than $M_{\rm 500c}^{\rm Cal}$.} range of $4.5 < M_{\rm 500c}/10^{14}~M_{\odot} < 9.7$ and a redshift interval of $1.00 < z < 1.31$. Figure~\ref{fig:1} shows the distribution of mass and redshift for the entire sample. The properties of all clusters in the sample are summarised in Table \ref{table:1}.

\subsection{Observations}

All clusters were observed with the MeerKAT $L$-band receiver (900–1670 MHz; PI: M.~Hilton; Proposal IDs: SCI-20220822-MH-01 and SCI-20230907-MH-02), following the observational setup, calibration strategy, and integration time described in P25. The central frequency of these observations was 1284~MHz. For reference, each target was observed for approximately 3.5 hours, and calibration was performed using primary and secondary calibrators as in the pilot study. Full technical and observational details are provided in P25; a summary of the observations for the twenty three (23) clusters reported here is listed in Table~\ref{table:2}.

\subsection{Data Reduction and Imaging}
Data reduction followed the procedures outlined in P25, utilizing the \texttt{OXKAT} v0.4\footnote[3]{\url{https://github.com/IanHeywood/oxkat}}  pipeline \citep{2020ascl.soft09003H}. The \texttt{OXKAT} workflow incorporates major radio astronomy software packages, including \texttt{CASA} v5.6\footnote[4]
{\url{https://casa.nrao.edu}} 
\citep{2007ASPC..376..127M,2022PASP..134k4501C} and \texttt{WSCLEAN} v3\footnote[5]
{\url{https://sourceforge.net/p/wsclean/wiki/Home/}} \citep{2014MNRAS.444..606O}. Direction-dependent calibration tools used in the analysis include \texttt{DDFACET}\footnote[6]{\url{https://github.com/saopicc/DDFacet}}  \citep{2018A&A...611A..87T}, \texttt{KILLMS}\footnote[7]{\url{https://github.com/saopicc/killMS}}  \citep{2014A&A...566A.127T}, and \texttt{CUBICAL}\footnote[8]{\url{https://github.com/ratt-ru/CubiCal}} \citep{2018MNRAS.478.2399K}. 
\newline

Standard flagging and primary or secondary calibration were performed first. Self-calibration and imaging followed to increase data fidelity. Correction for direction-dependent effects (DDEs), such as primary beam rotation, pointing errors, and spatially varying gains, was a critical step in the workflow.  Direction-dependent calibration \citep[often referred to as third-generation calibration, or 3GC;][]{2010A&A...524A..61N,2015MNRAS.449.2668S} corrects for these effects by solving for gain solutions that vary across the field of view. Both peeling and facet-based techniques were employed for 3GC. Peeling removes the effects of bright, problematic sources from the visibilities by iteratively modelling and subtracting them in specific directions. The facet-based calibration approach utilized \texttt{KILLMS} to derive directional gain corrections, and \texttt{DDFACET} applied these corrections during wide-field imaging. Peeling was performed prior to facet-based calibration when both approaches were necessary for optimal data quality. 
\newline

Compact sources within the cluster region were modeled and subtracted in the $uv$-plane using \texttt{CRYSTALBALL} \footnote[9]{\url{https://github.com/caracal-pipeline/crystalball}} and \texttt{MSUTILS}\footnote[10]
{\url{https://github.com/SpheMakh/msutils}} \citep{caracal2025crystalball,makhathini2021msutils}. This step ensured clean flux measurements of diffuse emission. Outer $uv$ tapering was applied to generate low-resolution maps and increase sensitivity to extended structures. Flux densities for diffuse sources were measured within polygons that trace the $3\sigma$ ($\sigma$ denotes the local root-mean-square (rms) noise of the image, measured in a source-free region) contours and uncertainties on the measured flux densities were computed following the method and error propagation formula given in Equation 1 of P25. Radio powers were calculated at 1.4~GHz, assuming a Gaussian distribution of spectral index ($\alpha = -1.3 \pm 0.4$), in line with the approach established in P25. A summary of data reduction parameters and observation details for the twenty-three cluster sample can be found in Table~\ref{table:2}. Full pipeline details are available in P25.

\subsection{Upper limit estimation}
\label{sec:Upper limit estimation}

To assess the sensitivity of our observations to diffuse emission in clusters with no clear detections, we derived upper limits using the Mock UV-data Injector Tool (\texttt{MUVIT}\footnote[11]{\url{https://github.com/lucabruno2501/MUVIT}}) package \citep{2023A&A...672A..41B}. \texttt{MUVIT} is an open-source \texttt{PYTHON} code that injects simulated visibilities for model radio halos. It supports 2D-exponential, 2D-Gaussian, and 2D-cylindrical surface-brightness profiles, which can be applied to real interferometric datasets. In this work, the mock halo visibilities are added directly to the calibrated MeerKAT data for each cluster, so that the simulations inherit the actual noise properties, residual calibration artefacts, and $uv$-coverage of the observations, rather than relying on idealised noise realisations. The package was developed for LOFAR \citep{2013A&A...556A...2V} datasets but is fully compatible with MeerKAT data and other interferometers. \texttt{MUVIT} leverages \texttt{WSCLEAN} for imaging and Fourier transforms, with core functionality making use of scientific \texttt{PYTHON} packages such as \texttt{NUMPY}\footnote[12]{\url{https://numpy.org}}, \texttt{ASTROPY}\footnote[13]{\url{https://github.com/astropy/}}, and \texttt{CASACORE}\footnote[14]{\url{https://github.com/casacore/casacore}}.
\newline

For each cluster, initial imaging is performed in \texttt{WSCLEAN} without $uv$-tapering, following the workflow recommended in the \texttt{MUVIT} documentation. Simulated mock halos, with parameters (flux, and profile) tailored to each target, are iteratively injected into the visibility data. In our simulations we assume that undetected radio halos, if present, would have a characteristic physical extent comparable to known radio halos in massive clusters, and we model their surface-brightness distribution with an exponential profile with a fixed scale radius $r_{e} = 200\,\mathrm{kpc}$ for all systems. This choice sets the concentration of the model halos and therefore their effective size in the images. We do not impose an explicit mass dependence of the halo radius in the injection; instead, we adopt a single representative physical scale and explore how bright such a halo would need to be to be detected in each cluster. In the commonly used exponential description, the halo size and scale radius are related through $R_H \sim 3\,r_e$ \citep{2007MNRAS.378.1565C}, so our assumption corresponds to injecting halos with a characteristic radius of order $\sim 600$ kpc at all redshifts. This is a pragmatic choice that simplifies the injection procedure but should not be regarded as a tightly calibrated size–mass relation. For each cluster, we centred the mock halo on the ACT SZ peak. In a few systems with strong central artefacts we shifted the injection centre slightly to a nearby clean region (see Table~\ref{tab:upperlimits}), and used these shifted positions consistently when deriving the upper limits.
\newline

After the mock visibilities are added, the datasets are re-imaged using the same low-resolution, tapered settings adopted for the diffuse-emission search, so that the response to the injected halo is measured in the tapered images (see Fig.~\ref{fig:injection_process} for an illustration of the procedure). For each trial, the flux of the injected model halo was increased stepwise, and the resulting tapered images were used to evaluate detection thresholds relative to the measured local image noise. The procedure was repeated until the flux density corresponding to a $3\sigma$ detection level was reached, where $\sigma$ is the local rms noise of the image, providing a constraint on the maximum non-detected diffuse emission for each system for a halo of this assumed size and profile. In practice, larger halos of the same total power would be more strongly surface-brightness diluted and thus harder to detect, whereas more compact halos would be easier to detect; our upper limits should therefore be interpreted as explicitly model-dependent, tied to the assumed exponential profile with $r_e = 200\,\mathrm{kpc}$, rather than as fully generic constraints for arbitrary halo morphologies. The full injection methodology, including visibility manipulation and significance criteria, follows \citet{2023A&A...672A..41B} in detail, and references therein.
\newline
\newline
The injected central surface-brightness is calculated as
\begin{equation}
I_{0,\mathrm{inj}} = S_{\mathrm{inj,tot}}/{2\pi r_{e,\mathrm{inj}}^2},
\end{equation}
where $S_{\mathrm{inj,tot}}$ is the total flux density injected and $r_{e,\mathrm{inj}}$ is the exponential scale radius. The upper limit flux density at $1.4\,\mathrm{GHz}$ is estimated as
\begin{equation}
S_{1.4\,\mathrm{GHz,UL}} = 2\pi f(3r_e)\, I_{0,\mathrm{inj}}\, r_{e,\mathrm{inj}}^2,
\end{equation}
where $f(3r_e) = 0.8$ is the fraction of the total exponential halo flux contained within three effective radii. Radio power for upper limits is calculated as
\begin{equation}
P_{1.4\,\mathrm{GHz}} = 4\pi D_L^2\, S_{1.4\,\mathrm{GHz}}\,/(1+z)^{(\alpha+1)},
\label{eq:P14}
\end{equation}
where $D_L$ is the luminosity distance at redshift $z$ and $S_{1.4\,\mathrm{GHz}}$ is the flux density scaled to $1.4\,\mathrm{GHz}$ using spectral index $\alpha = -1.3$.

\begin{table*}
\caption{Properties of the twenty four (24) MMDCS clusters analysed in this work. Clusters are listed in order of decreasing $M_{\mathrm{500c}}$ value. The first six clusters in the table (initially presented in P25) are also included here for completeness and are marked with an \textit{*}. The last column gives the adopted diffuse radio-emission classification for each cluster.}
\label{table:1}

\begin{tabular}{lccccccccc}
\hline
Cluster Name                  & RA (J2000)                                        & Dec (J2000)                                                           & redshift        & redshift type       & \textit{$M_{\rm 500c}$} & $M_{\rm 500c}^{\rm{Unc}}$ & $R_{\rm 500c}$ & SNR  & Class \\
(ACT-CL)                      & $(^{\mathrm{h}}{:}^{\mathrm{m}}{:}^{\mathrm{s}})$ & $(^{\mathrm{\circ}}{:}^{\mathrm{\prime}}{:}^{\mathrm{\prime\prime}})$ & $z$             &                     & \multicolumn{2}{c}{$10^{14} M_\odot$}                     & (kpc)          &      &       \\ \hline
J0329.2$-$2330$^{\textit{*}}$ & 03:29:17.7                                        & $-$23:30:09.7                                                         & 1.23            & spec$^{\mathrm{a}}$ & $9.7^{+1.7}_{-1.6}$            & $8.0^{+1.3}_{-1.1}$      & 946            & 22.1 & RH    \\
J2106.0$-$5844$^{\textit{*}}$ & 21:06:04.3                                        & $-$58:44:35.0                                                         & 1.13            & spec$^{\mathrm{a}}$ & $8.5^{+1.5}_{-1.4}$            & $6.8^{+1.1}_{-0.9}$      & 939            & 18.7 & RH    \\
J1137.8+0728$^{\textit{*}}$   & 11:37:50.8                                        & +07:28:36.0                                                           & $1.01 \pm 0.03$ & phot$^{\mathrm{b}}$ & $6.7^{+1.2}_{-1.1}$            & $5.3^{+0.9}_{-0.7}$      & 909            & 16.9 & RH    \\
J1142.7+1527$^{\textit{*}}$   & 11:42:46.0                                        & +15:27:22.0                                                           & 1.19            & spec$^{\mathrm{c}}$ & $6.3^{+1.1}_{-1.0}$            & $5.0^{+0.8}_{-0.7}$      & 831            & 19.6 & RH    \\
J0003.9+1642$^{\textit{*}}$   & 00:03:51.8                                        & +16:42:06.5                                                           & $1.31 \pm 0.03$ & phot$^{\mathrm{b}}$ & $6.1^{+1.1}_{-1.0}$            & $4.9^{+0.8}_{-0.7}$      & 787            & 14.7 & RH    \\
J0546.6$-$5345$^{\textit{*}}$ & 05:46:37.6                                        & $-$53:45:33.2                                                         & 1.07            & spec$^{\mathrm{d}}$ & $6.0^{+1.1}_{-1.0}$            & $4.7^{+0.8}_{-0.7}$      & 856            & 12.8 & U     \\
J0851.9+1500                  & 08:51:55.2                                        & +15:00:42.0                                                           & $1.04 \pm 0.03$ & phot$^{\mathrm{b}}$ & $5.8^{+1.1}_{-1.0}$            & $4.6^{+0.8}_{-0.7}$      & 858            & 12.8 & NDE   \\
J0930.2+0615                  & 09:30:15.6                                        & +06:15:32.2                                                           & $1.30 \pm 0.03$ & phot$^{\mathrm{b}}$ & $5.8^{+1.1}_{-1.0}$            & $4.6^{+0.7}_{-0.6}$      & 720            & 13.4 & NDE   \\
J0947.9-0120                  & 09:47:58.3                                        & $-$01:20:02.8                                                         & $1.10 \pm 0.04$ & phot$^{\mathrm{e}}$ & $5.7^{+1.1}_{-1.0}$            & $4.5^{+0.7}_{-0.6}$      & 832            & 12.6 & NDE   \\
J2341.2-5119                  & 23:41:12.1                                        & $-$51:19:40.2                                                         & 1.00            & spec$^{\mathrm{a}}$ & $5.6^{+1.0}_{-0.9}$            & $4.4^{+0.7}_{-0.6}$      & 859            & 12.8 & NDE   \\
J0244.1-0923                  & 02:44:08.0                                        & $-$09:23:13.8                                                         & $1.19 \pm 0.03$ & phot$^{\mathrm{b}}$ & $5.5^{+1.0}_{-0.9}$            & $4.4^{+0.7}_{-0.6}$      & 796            & 10.7 & NDE   \\
J0044.4+0113                  & 00:44:25.7                                        & +01:13:01.3                                                           & $1.03 \pm 0.03$ & phot$^{\mathrm{b}}$ & $5.3^{+1.0}_{-0.9}$            & $4.2^{+0.6}_{-0.6}$      & 835            & 15.0 & NDE   \\
J0314.0+0203                  & 03:14:01.0                                        & +02:03:39.8                                                           & $1.14 \pm 0.03$ & phot$^{\mathrm{b}}$ & $5.3^{+1.0}_{-0.9}$            & $4.2^{+0.7}_{-0.6}$      & 799            & 9.5  & NDE   \\
J0241.2-3916                  & 02:41:13.5                                        & $-$39:16:05.7                                                         & $1.02 \pm 0.03$ & phot$^{\mathrm{b}}$ & $5.2^{+1.0}_{-0.9}$            & $4.1^{+0.7}_{-0.6}$      & 830            & 11.2 & RH    \\
J0125.3-0802                  & 01:25:18.0                                        & $-$08:02:12.5                                                         & $1.06 \pm 0.03$ & phot$^{\mathrm{b}}$ & $5.1^{+0.9}_{-0.9}$            & $4.0^{+0.6}_{-0.6}$      & 812            & 11.5 & NDE   \\
J1049.1+1106                  & 10:49:06.9                                        & +11:06:14.3                                                           & $1.23 \pm 0.03$ & phot$^{\mathrm{b}}$ & $5.0^{+0.9}_{-0.8}$            & $3.9^{+0.6}_{-0.6}$      & 756            & 10.6 & NDE   \\
J0154.3-4824                  & 01:54:20.5                                        & $-$48:24:42.8                                                         & $1.28 \pm 010$  & phot$^{\mathrm{a}}$ & $4.9^{+0.9}_{-0.8}$            & $3.8^{+0.6}_{-0.5}$      & 740            & 12.8 & NDE   \\
J0113.9+1451                  & 01:13:54.5                                        & +14:51:10.4                                                           & $1.03 \pm 0.03$ & phot$^{\mathrm{b}}$ & $4.9^{+1.0}_{-0.9}$            & $3.9^{+0.7}_{-0.6}$      & 812            & 7.8  & U     \\
J0204.3-1918                  & 02:04:22.2                                        & $-$19:18:42.0                                                         & $1.01 \pm 0.03$ & phot$^{\mathrm{b}}$ & $4.9^{+1.0}_{-0.9}$            & $3.9^{+0.7}_{-0.6}$      & 817            & 6.9  & NDE   \\
J2133.0+1805                  & 21:33:01.6                                        & +18:05:59.5                                                           & $1.07 \pm 0.03$ & phot$^{\mathrm{b}}$ & $4.9^{+0.9}_{-0.8}$            & $3.8^{+0.6}_{-0.6}$      & 797            & 8.6  & NDE   \\
J1527.2+1600                  & 15:27:17.7                                        & +16:00:10.6                                                           & $1.02 \pm 0.03$ & phot$^{\mathrm{b}}$ & $4.8^{+0.9}_{-0.8}$            & $3.7^{+0.6}_{-0.5}$      & 810            & 13.9 & NDE   \\
J1105.3+0636                  & 11:05:21.1                                        & +06:36:12.1                                                           & $1.03 \pm 0.03$ & phot$^{\mathrm{b}}$ & $4.7^{+0.9}_{-0.8}$            & $3.7^{+0.7}_{-0.6}$      & 802            & 8.8  & NDE   \\
J1521.1+0451                  & 15:21:07.1                                        & +04:51:48.1                                                           & 1.31            & spec$^{\mathrm{c}}$ & $4.7^{+0.8}_{-0.8}$            & $3.7^{+0.6}_{-0.5}$      & 719            & 15.5 & NDE   \\
J1525.8+1540                  & 15:25:50.9                                        & +15:40:54.9                                                           & $1.02 \pm 0.05$ & phot$^{\mathrm{c}}$ & $4.6^{+0.9}_{-0.8}$            & $3.6^{+0.6}_{-0.5}$      & 800            & 13.1 & NDE   \\
J0200.7-3106                  & 02:00:46.8                                        & $-$31:06:26.5                                                         & $1.02 \pm 0.03$ & phot$^{\mathrm{b}}$ & $4.6^{+0.9}_{-0.8}$            & $3.6^{+0.6}_{-0.5}$      & 798            & 8.6  & NDE   \\
J0344.3-5453                  & 03:44:21.9                                        & $-$54:53:00.4                                                         & $1.04 \pm 0.03$ & phot$^{\mathrm{b}}$ & $4.6^{+0.9}_{-0.8}$            & $3.6^{+0.6}_{-0.5}$      & 792            & 8.1  & NDE   \\
J0543.0-2941                  & 05:43:00.0                                        & $-$29:41:37.4                                                         & $1.19 \pm 0.15$ & phot$^{\mathrm{a}}$ & $4.5^{+1.0}_{-0.8}$            & $3.6^{+0.7}_{-0.6}$      & 744            & 6.8  & NDE   \\
J2146.6-0321                  & 21:46:38.6                                        & $-$03:21:08.1                                                         & $1.16 \pm 0.05$ & phot$^{\mathrm{c}}$ & $4.5^{+1.0}_{-0.8}$            & $3.7^{+0.7}_{-0.6}$      & 751            & 5.8  & NDE   \\
J2245.2-0433                  & 22:45:14.5                                        & $-$04:33:39.3                                                         & $1.02 \pm 0.03$ & phot$^{\mathrm{b}}$ & $4.5^{+1.0}_{-0.9}$            & $3.7^{+0.7}_{-0.6}$      & 791            & 5.6  & EX    \\
J1139.3+0154                  & 11:39:19.5                                        & +01:54:16.5                                                           & $1.05 \pm 0.02$ & phot$^{\mathrm{e}}$ & $4.5^{+0.9}_{-0.8}$            & $3.5^{+0.6}_{-0.5}$      & 782            & 8.1  & NDE   \\ \hline
\end{tabular}

\begin{minipage}{\textwidth}
\footnotesize
\textbf{Notes:} All SZ quantities are sourced from the ACT DR5 catalogue \citep{2021ApJS..253....3H}. Table columns correspond to: (1) ACT DR5 cluster name; (2) Right Ascension of the SZ centre (J2000); (3) Declination of the SZ centre (J2000); (4) cluster redshift; (5) source/type of redshift information, with values drawn from $^{\mathrm{a}}$South Pole Telescope (SPT) \citep{2019ApJ...878...55B,2020ApJS..247...25B}, $^{\mathrm{b}}$zCluster \citep{2021ApJS..253....3H}, $^{\mathrm{c}}$Massive and Distant Clusters of WISE Survey (MaDCoWS), $^{\mathrm{d}}$ACT \citep{2013ApJ...765...67M,2016MNRAS.461..248S,2018ApJS..235...20H}, or $^{\mathrm{e}}$CAMIRA \citep{2018PASJ...70S..20O}. Columns (6) and (7) list the ACT SZ-based mass calibrated via weak lensing, and the uncorrected ACT SZ mass, respectively. The latter quantity is not corrected for Eddington bias, enabling direct comparison with \textit{Planck} PSZ2 masses. Column (8) and (9) reports the $R_{\rm 500c}$ in kpc and the ACT signal-to-noise ratio (SNR), receptively. Column (10) gives the adopted diffuse radio-emission classification: RH = radio halo, U = uncertain/candidate, NDE = no diffuse emission detected, and EX = excluded from the radio analysis because of poor data quality.
\end{minipage}
\end{table*}

\begin{table*}
\caption{Overview of the twenty three (23) MeerKAT 1.28 GHz observations used for the radio analysis in this work, summarising the main observational parameters and the full-resolution imaging properties. ACT-CL J2245.2$-$0433, although part of the full MMDCS sample, is excluded from this table because it was not included in the radio analysis due to poor data quality. Clusters are listed in order of decreasing $M_{\mathrm{500c }}$ value.}
\label{table:2}
\begin{tabular}{ccccccc}
\hline
Cluster Name                                    & Observing date & Bandpass calibrator     & Phase calibrator        & Flagged & $\mathrm{\theta_{synth, FR}}$         & $\mathrm{\sigma_{rms, FR}}$ \\
(ACT-CL)                                        & (Y-M-D)        &    &   & (per cent)  & ($\mathrm{arcsec \times arcsec}$, $^{\circ}$) & $\mathrm{\upmu Jy/beam}$      \\ \hline
\multicolumn{1}{l}{J0851.9+1500$^{\textit{a}}$} & 2024-05-17     & J0408$-$6545 & J0842+1835   & 49.0    & 6.3 $\times$ 6.3, 0.0                 & 7.3                         \\
J0930.2+0615$^{\textit{a}}$                     & 2024-08-05     & J0408$-$6545 & J1008+0730   & 72.4    & 6.1 $\times$ 6.1, 0.0                 & 14.9                        \\
J0947.9-0120$^{\textit{a}}$                     & 2024-08-17     & J0408$-$6545 & J1008+0730   & 66.4    & 6.0 $\times$ 6.0, 0.0                 & 7.7                         \\
J2341.2-5119                                    & 2024-03-11     & J1939$-$6342 & J0010$-$4153 & 51.5    & 5.0 $\times$ 5.0, 0.0                 & 5.9                         \\
J0244.1-0923                                    & 2024-03-09     & J0408$-$6545 & J0240$-$2309 & 45.1    & 5.7 $\times$ 5.7, 0.0                 & 6.5                         \\
J0044.4+0113                                    & 2024-03-16     & J0408$-$6545 & J0022+0014   & 53.4    & 6.0 $\times$ 6.0, 0.0                 & 7.6                         \\
J0314.0+0203$^{\textit{a}}$                     & 2024-08-23     & J0408$-$6545 & J0323+0534   & 49.4    & 6.2 $\times$ 6.2, 0.0                 & 6.7                         \\
J0241.2-3916$^{\textit{a}}$                     & 2024-03-07     & J0408$-$6545 & J0203$-$4349 & 49.4    & 5.4 $\times$ 5.4, 0.0                 & 4.7                         \\
J0125.3-0802$^{\textit{a}}$                     & 2025-02-23     & J1939$-$6342 & J0108+0134   & 51.0    & 6.0 $\times$ 6.0, 0.0                 & 5.7                         \\
J1049.1+1106$^{\textit{a}}$                     & 2025-03-06     & J0408$-$6545 & J1058+0133   & 46.4    & 9.8 $\times$ 5.6, 159.3               & 6.4                         \\
J0154.3-4824                                    & 2025-03-08     & J0408$-$6545 & J0155$-$4048 & 51.5    & 5.5 $\times$ 5.5, 0.0                 & 5.0                         \\
J0113.9+1451$^{\textit{a}}$                     & 2025-03-07     & J0408$-$6545 & J0108+0134   & 55.3    & 6.3 $\times$ 6.3, 0.0                 & 7.5                         \\
J0204.3-1918$^{\textit{a}}$                     & 2025-03-21     & J0408$-$6545 & J0240$-$2309 & 51.1    & 6.8 $\times$ 5.4, 152.7               & 7.1                         \\
J2133.0+1805$^{\textit{a}}$                     & 2025-02-28     & J1939$-$6342 & J2148+0657   & 51.2    & 6.1 $\times$ 61, 0.0                  & 5.5                         \\
J1527.2+1600                                    & 2025-09-04     & J1939$-$6342 & J1550+0527   & 50.1    & 6.1 $\times$ 61, 0.0                  & 7.4                         \\
J1105.3+0636                                    & 2025-03-24     & J0408$-$6545 & J1058+0133   & 47.5    & 6.2 $\times$ 6.2, 0.0                 & 5.7                         \\
J1521.1+0451$^{\textit{a}}$                     & 2025-11-22     & J1939$-$6342 & J1550+0527   & 50.0    & 6.0 $\times$ 6.0, 0.0                 &  7.9                           \\ 
J1525.8+1540$^{\textit{a}}$                     & 20025-11-13    & J1939$-$6342 & J1550+0527   & 56.1    & 9.4 $\times$ 6.2, 180.0               & 8.7                            \\
J0200.7-3106$^{\textit{a}}$                     & 2025-02-21     & J0408$-$6545 & J0155$-$4048 & 56.1    & 6.2 $\times$ 5.5, 176.9               & 13.9                        \\
J0344.3-5453                                    & 2025-02-24     & J0408$-$6545 & J0408$-$6545 & 55.9    & 5.4 $\times$ 5.4, 0.0                 & 6.0                         \\
J0543.0-2941                                    & 2025-02-23     & J0408$-$6545 & J0538$-$4405 & 49.9    & 5.6 $\times$ 5.6, 0.0                 & 5.6                         \\
J2146.6-0321$^{\textit{a}}$                     & 2025-02-21     & J1939$-$6342 & J2131$-$1207 & 57.8    & 6.1 $\times$ 4.7, 165.7               & 7.9                         \\
J1139.3+0154$^{\textit{a}}$                     & 2025-03-12     & J0408$-$6545 & J1150$-$0023 & 45.5    & 6.2 $\times$ 6.2, 0.0                 & 5.4                         \\ \hline
\end{tabular}
\begin{minipage}{\textwidth}
\footnotesize
\textbf{Notes:} Table columns are as follows: (1) ACT DR5 cluster name; (2) date of the MeerKAT observation; (3–4) bandpass and phase calibrators used; (5) The percentage of MeerKAT data that is flagged during reduction, which includes frequency ranges that are known to be affected by satellite interference; (6–7) synthesized beam characteristics (major/minor axis dimensions, position angle) and central rms noise values for the full-resolution images. $^{\textit{a}}$ Direction-dependent corrections were applied for this cluster field. Significant noise variations in the image are due to residual artefacts left by a bright contaminating source.
\end{minipage}

\end{table*}

\section{Results}
\label{sec:Results}
The initial results from P25 presented MeerKAT observations of the six most massive clusters in the MMDCS sample. Diffuse radio emission was detected in these clusters, with four showing prominent halos and two displaying faint detections. In addition to our detections, recent studies have identified the possible diffuse cluster-scale emission in high-redshift clusters ($z > 1$), as reported by \citet{2025A&A...695A.215D} and \citet{2025ApJ...987L..40H}. These findings collectively strengthen the case that extended radio emission is present in the cluster population at early cosmic times.
\newline

Here, we extend the analysis to the full MMDCS sample of 30 clusters, incorporating the remaining 23 systems. One cluster, ACT-CL~J2245.2$-$0433, was excluded from further analysis due to poor data quality. Among these additional targets, two clusters show tentative candidates of faint diffuse emission, while the other 21 yield no significant detections above our sensitivity limits. To quantitatively constrain these non-detections, we employed mock halo injection simulations (see Section \ref{sec:Upper limit estimation}). This approach enables us to probe the survey sensitivity and place meaningful constraints on radio halo and relic detection at high-redshift. 
\newline

\begin{figure*}
\centering
\begin{tabular}{ccc}
    \begin{overpic}[width=0.32\textwidth]{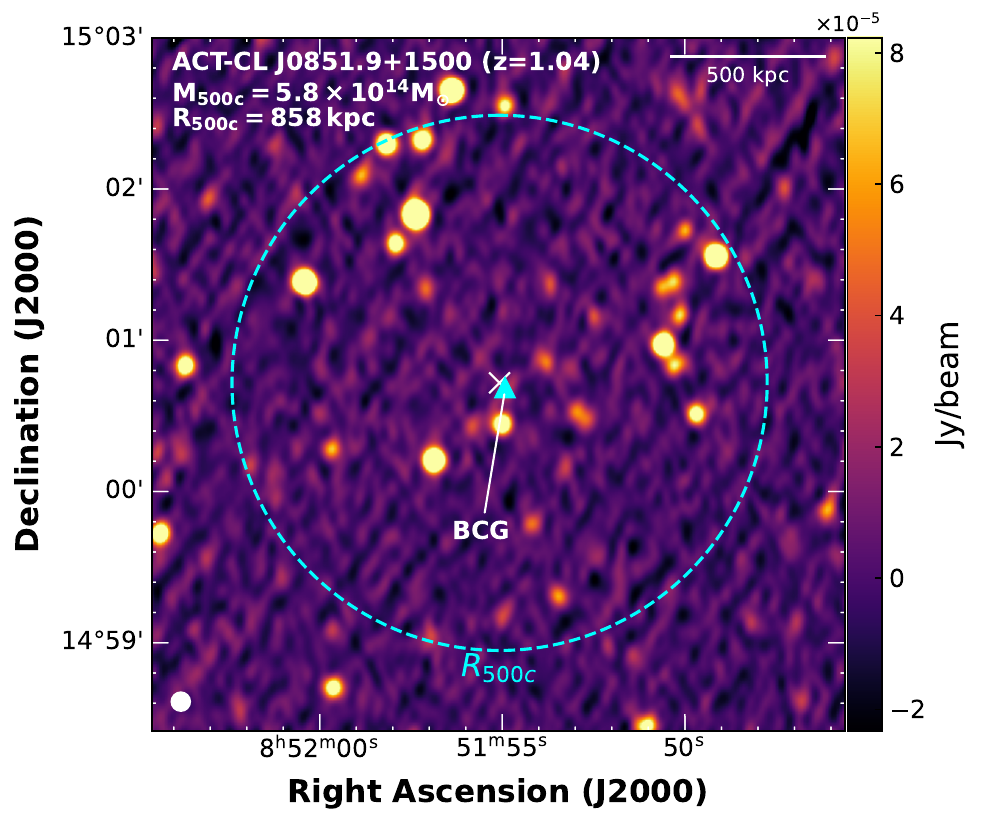}\put(5,85){\textbf{(a)}}\end{overpic} &
    \begin{overpic}[width=0.315\textwidth]{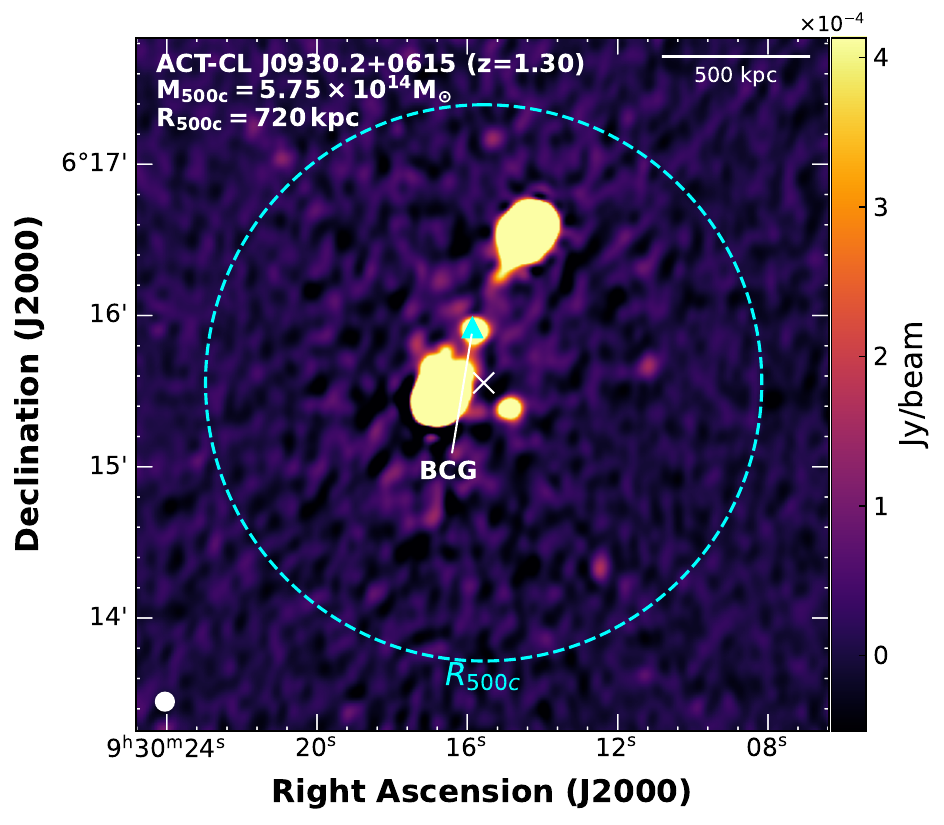}\put(5,85){\textbf{(b)}}\end{overpic} &
    \begin{overpic}[width=0.32\textwidth]{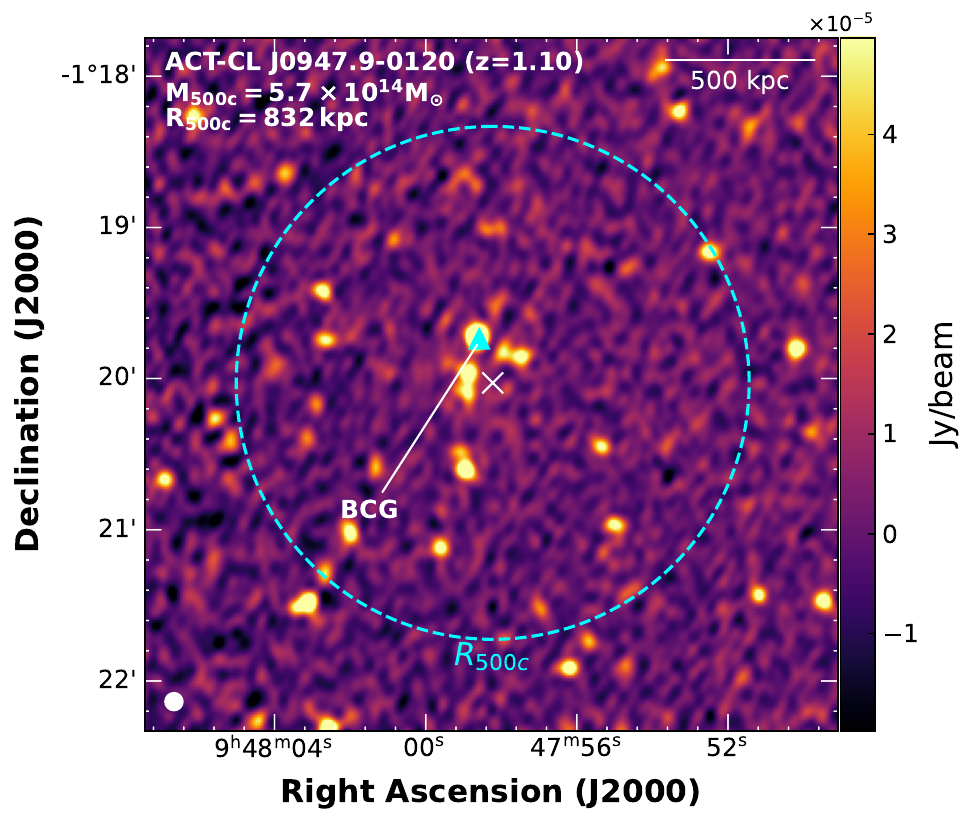}\put(5,85){\textbf{(c)}}\end{overpic} \\

    \begin{overpic}[width=0.32\textwidth]{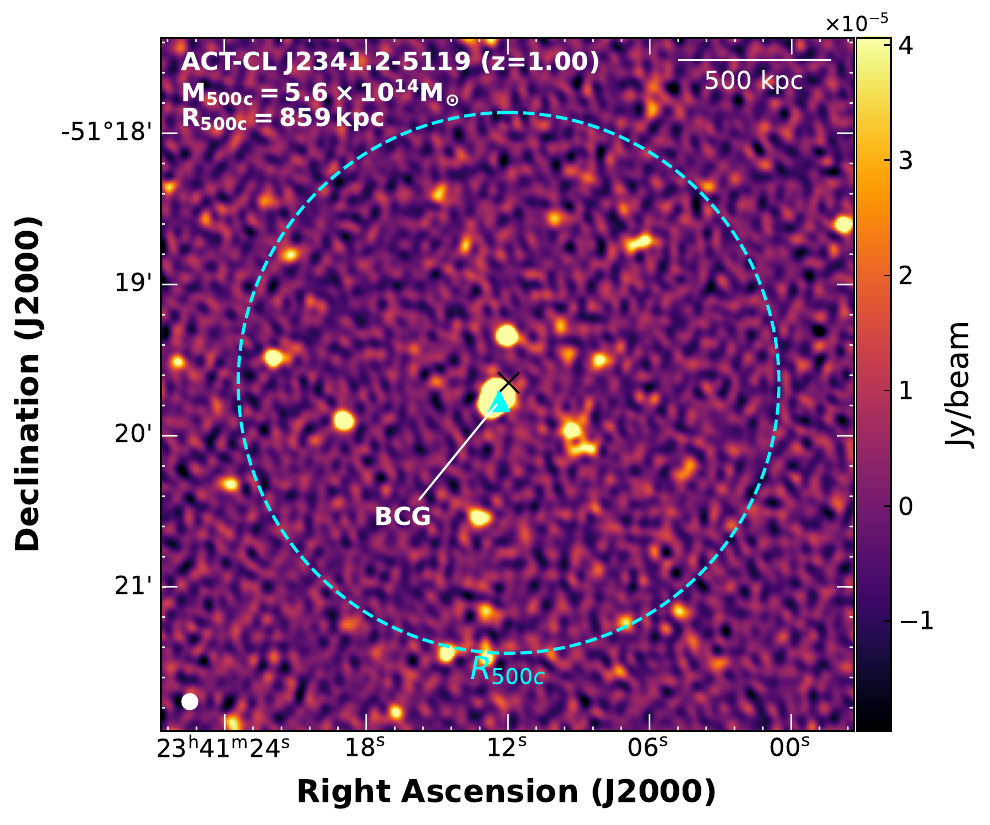}\put(5,85){\textbf{(d)}}\end{overpic} &
    \begin{overpic}[width=0.32\textwidth]{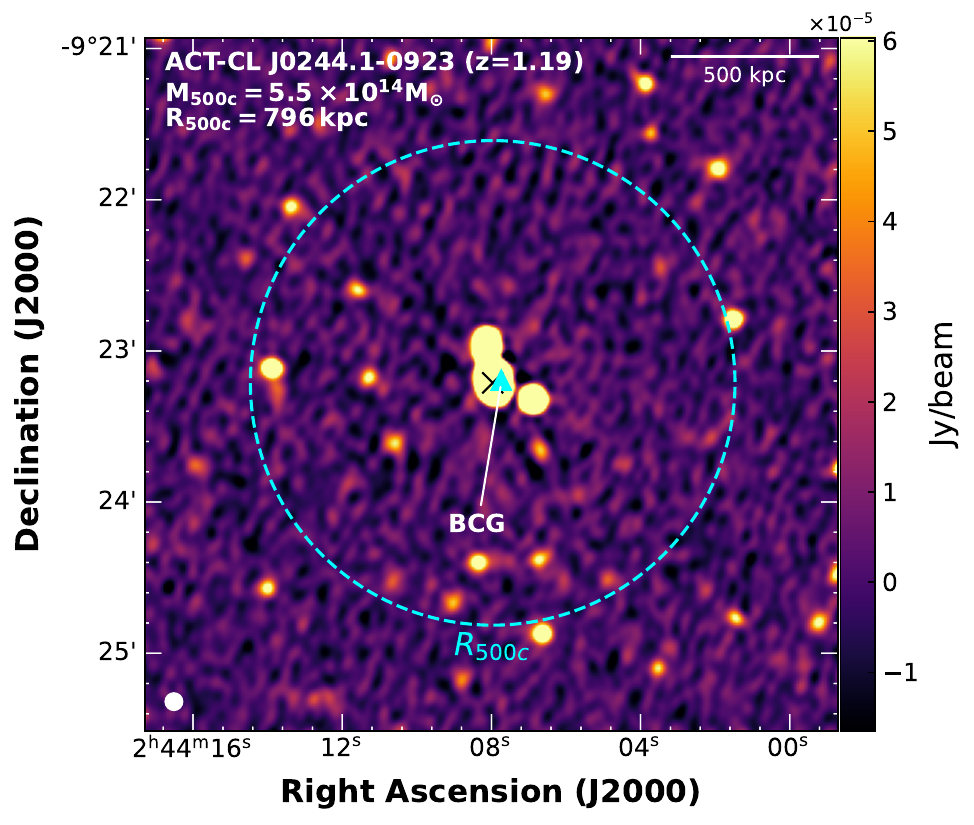}\put(5,85){\textbf{(e)}}\end{overpic} &
    \begin{overpic}[width=0.32\textwidth]{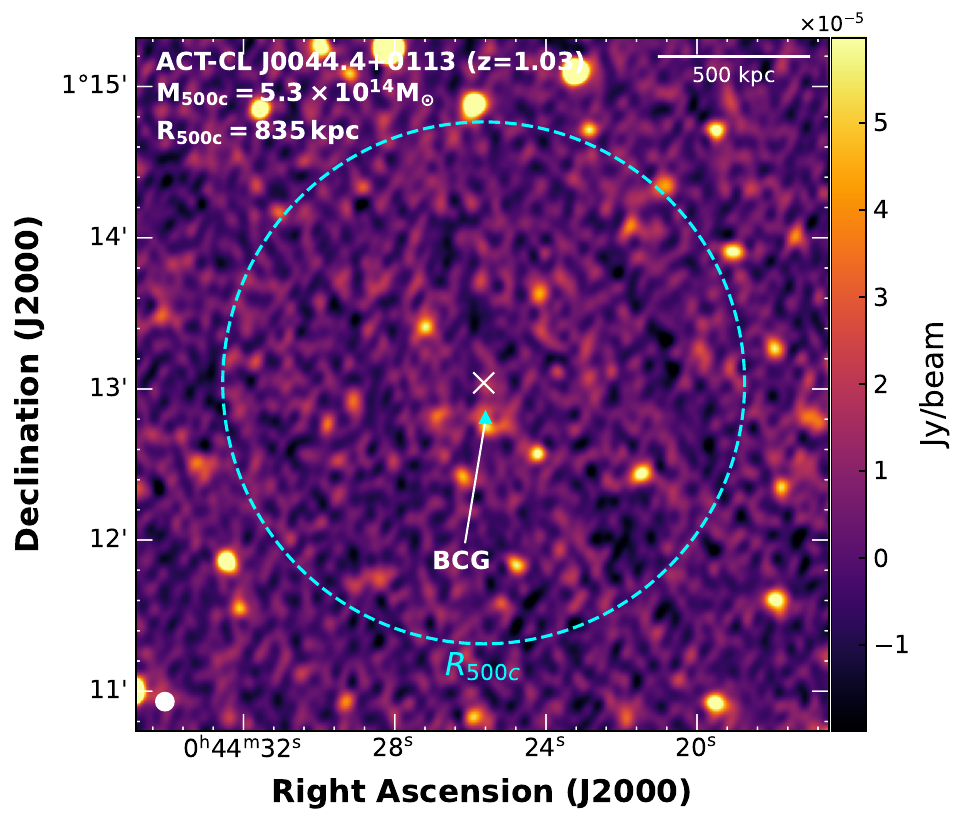}\put(5,85){\textbf{(f)}}\end{overpic} \\

    \begin{overpic}[width=0.32\textwidth]{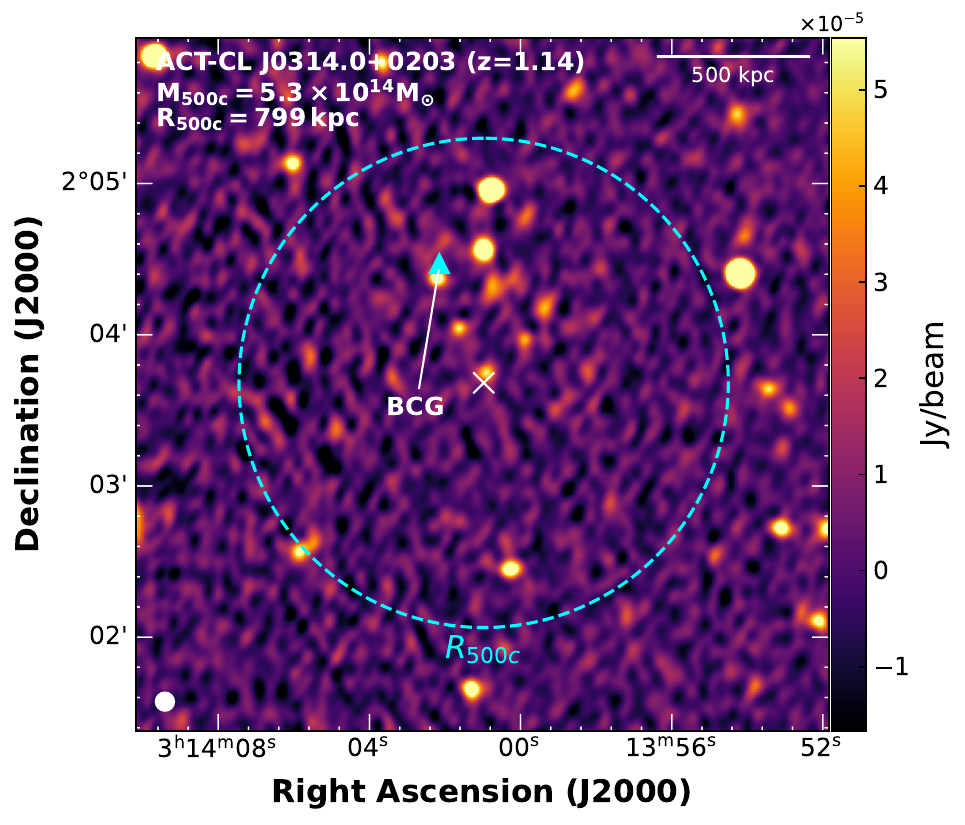}\put(5,85){\textbf{(g)}}\end{overpic} &
    \begin{overpic}[width=0.32\textwidth]{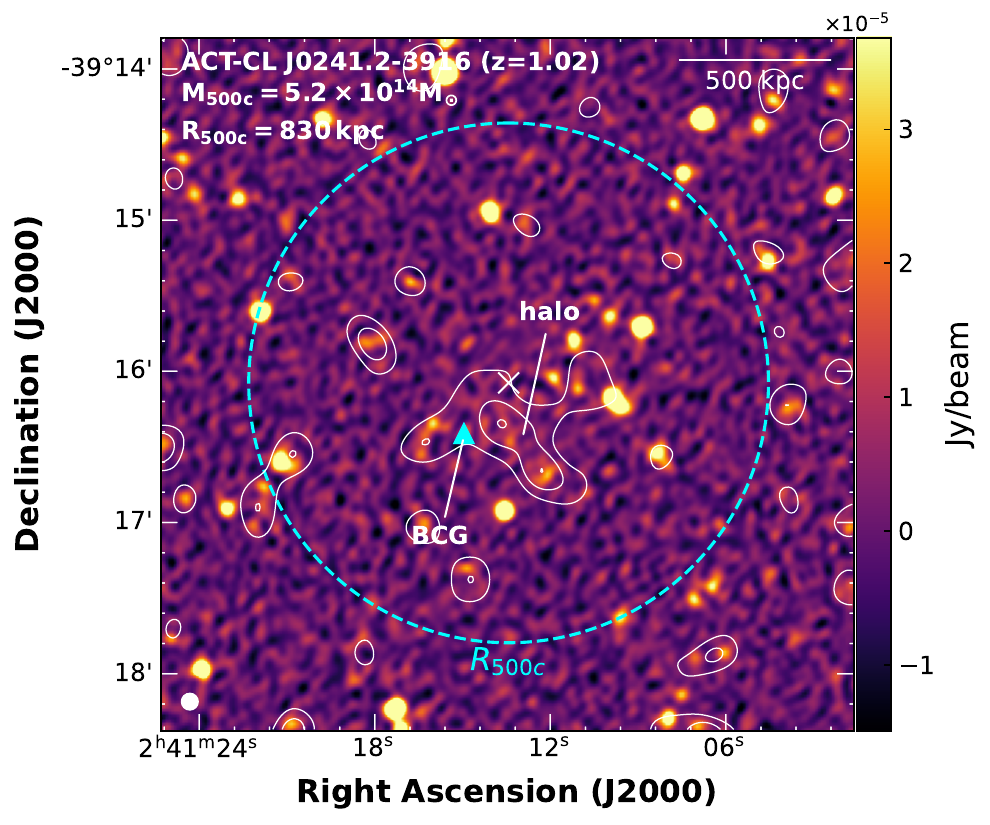}\put(5,85){\textbf{(h)}}\end{overpic} &
    \begin{overpic}[width=0.32\textwidth]{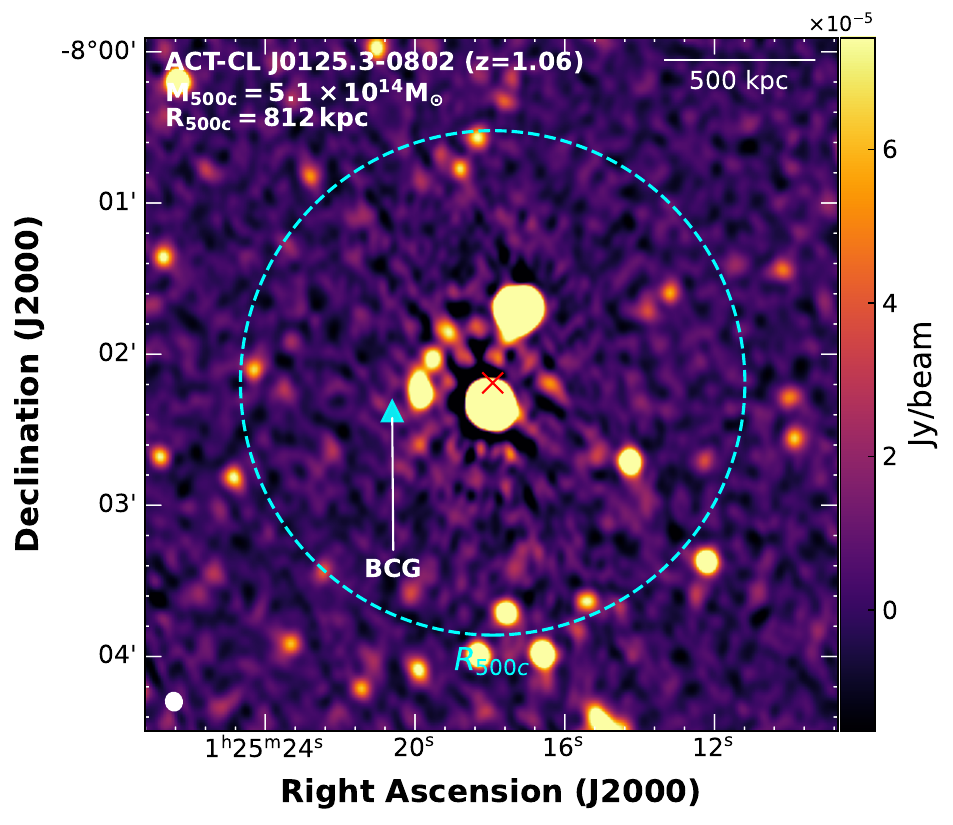}\put(5,85){\textbf{(i)}}\end{overpic} \\

    \begin{overpic}[width=0.32\textwidth]{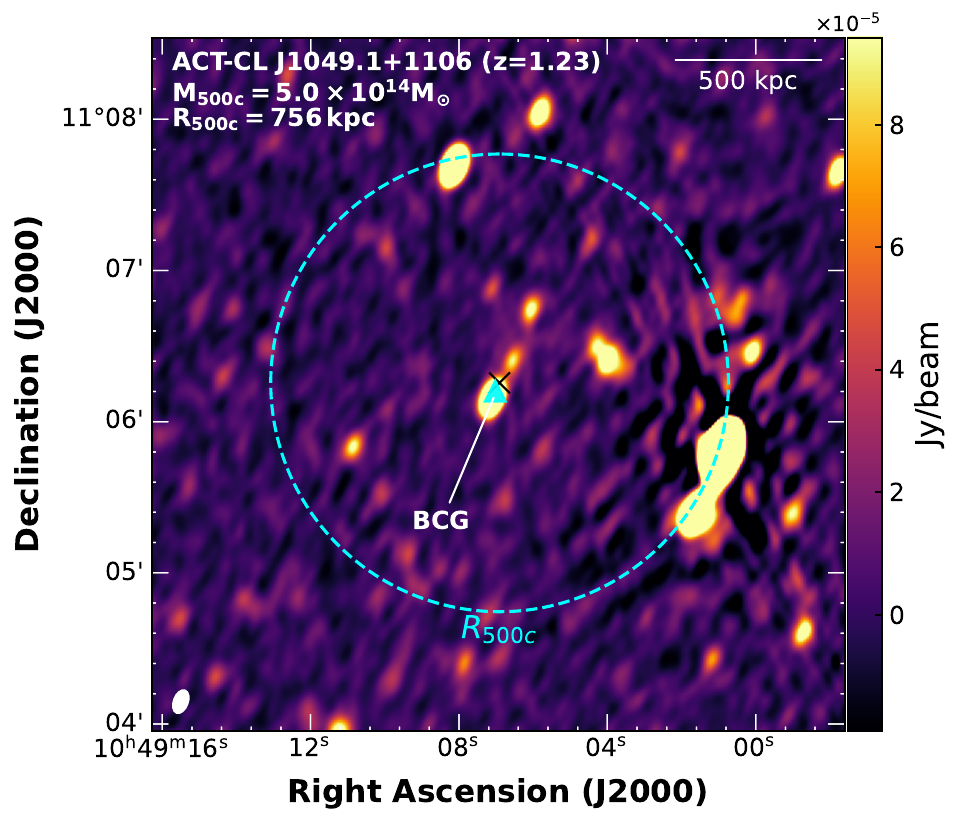}\put(5,85){\textbf{(j)}}\end{overpic} &
    \begin{overpic}[width=0.32\textwidth]{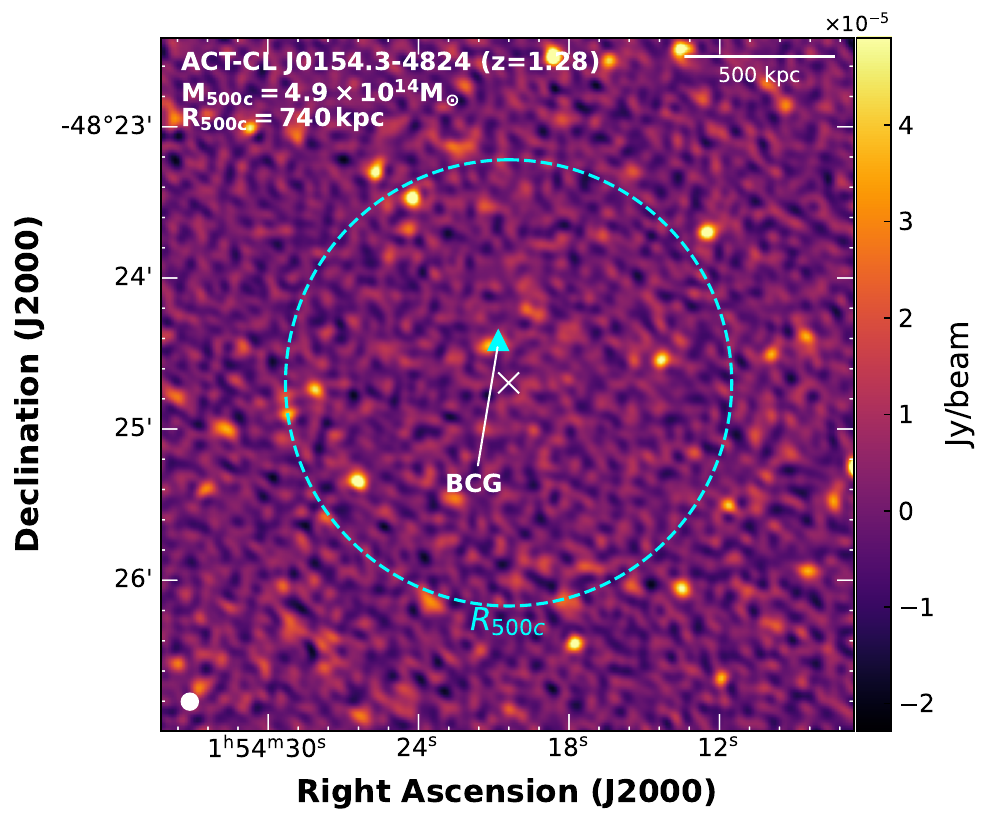}\put(5,85){\textbf{(k)}}\end{overpic} &
    \begin{overpic}[width=0.32\textwidth]{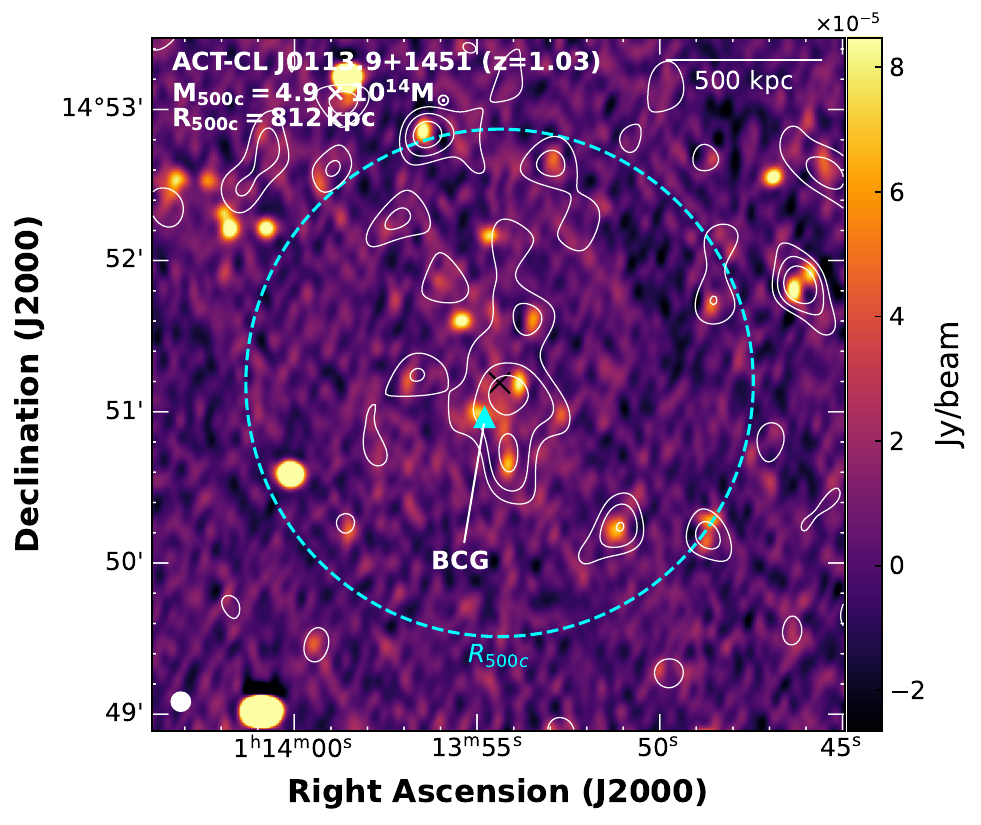}\put(5,85){\textbf{(l)}}\end{overpic} \\
\end{tabular}

\caption{MeerKAT 1.28~GHz full-resolution images for each cluster. The projected physical scale at each cluster's redshift is indicated in the top right of each panel. The synthesized beam is rendered as a filled white ellipse or circle at the lower left of each panel. The ACT SZ peak is labelled with a black or white cross, and the location of the BCG, obtained from ACT DR5 \citep{2021ApJS..253....3H}, is marked with a cyan triangle. The 3GC was applied to the following clusters to mitigate artefacts from bright sources and improve image fidelity: ACT-CL J0851, J0930, J0314, J0241, J0125, J1049, and J0113. The superimposed contours are from the low-resolution (LR) source-subtracted images, with contour levels set at $\sigma_{\mathrm{LR}} \times [3, 6, 10]$. The LR images used for the contours have local rms values of $1\sigma = 5.5\,\upmu\mathrm{Jy\,beam^{-1}}$ and $9.4\,\upmu\mathrm{Jy\,beam^{-1}}$ for ACT‑CL J0241.2–3916 and ACT‑CL J0113.9+1451, respectively. Contours are shown only for these two fields where faint diffuse emission is detected.}

\label{fig:allclusters}
\end{figure*}

The imaging results of the MMDCS sample are presented in Figures~\ref{fig:allclusters} and \ref{fig:allclusters2}, which together display full-resolution MeerKAT 1.28~GHz images for all clusters analysed in this study. The cluster images are split roughly evenly between these two figures for clarity and convenient presentation of the large sample. Each panel is labelled to refer to specific clusters and their radio features, allowing easy cross-reference with the text below.

\subsection{Individual galaxy clusters}

\subsubsection{Clusters with detected diffuse emission}
\label{sec:Clusters with detected diffuse emission}

In addition to the six radio halo systems previously reported in P25, hosted by ACT-CL J0329.2-2330, J2106.0-5844, J1137.8+0728, J1142.7+1527, J0003.9+1642, and J0546.6-5345, the two clusters discussed below bring the total number of MMDCS clusters with detected or candidate halos to eight. Table~\ref{table:3} summarises the main observational properties of these newly detected diffuse emissions, including their sizes, flux densities, radio powers, and classifications.
\newline

Both ACT-CL~J0241.2$-$3916 and ACT-CL~J0113.9+1451 display faint, low-resolution (LR) diffuse radio emission at or above the $3\sigma$ level after point source subtraction and image tapering, as shown in panel h and l of Figure~\ref{fig:allclusters}, respectively. In both systems, residual low level features are also visible around some compact sources in the LR contours. These arise where bright or slightly extended radio galaxies are not perfectly described by the point-source models used in the subtraction, so that a fraction of their flux remains in the tapered images. As a result, the LR maps contain a mixture of genuinely diffuse emission and residuals from imperfectly subtracted compact sources.
\newline

To assess the impact of residual compact emission more quantitatively, we inspected compact-only, full-resolution point-source--subtracted, and tapered low-resolution images for both targets (see Appendix~\ref{app:subtraction_quality}). In ACT-CL~J0241.2$-$3916 and ACT-CL~J0113.9+1451, the most prominent residuals closely trace the positions of the brightest compact objects, as expected when slightly extended or complex radio galaxies are imperfectly modelled, but the candidate halo emission extends beyond these residuals and fills a substantial fraction of the $R_{500\mathrm{c}}$ region. This supports our interpretation that the LR emission includes a genuine cluster-scale component, while we still regard both detections as tentative given their low surface-brightness and the complex central radio environment.
\newline

In ACT-CL~J0241.2$-$3916, a massive galaxy cluster at $z = 1.02$ with $M_{\rm 500c} = 5.2^{+1.0}_{-0.9} \times 10^{14}\, M_\odot$, the detected emission is not apparent in the original full-resolution image, but emerges upon smoothing and is spatially extended over a projected physical scale of $0.69 \times 0.53$\,Mpc, with an integrated flux density of $S_{1.28\,\mathrm{GHz}} = 0.270 \pm 0.045$\,mJy and a corresponding 1.4\,GHz radio power of $(0.17 \pm 0.06) \times 10^{25}$\,W\,Hz$^{-1}$. We therefore classify this extended emission as a radio halo. There is a positional offset between the SZ peak (marked by a cross) and the BCG (cyan triangle). Such offsets are commonly indicative of a dynamically disturbed or merging system and may affect the properties and morphology of any faint diffuse emission. As shown in Figure~\ref{fig:subtraction_0241}, the tapered emission does not break up into isolated patches at the locations of the brightest compact sources, but instead forms a single, approximately cluster‑centred structure on $\sim 0.7$\,Mpc scales, consistent with a radio halo.
\newline

In ACT-CL~J0113.9+1451, a massive galaxy cluster at $z = 1.03$ with $M_{\rm 500c} = 4.9^{+1.0}_{-0.9} \times 10^{14}\, M_\odot$, the smoothed, post-subtraction map also reveals residual diffuse emission at the $3\sigma$ level. Here, the emission has an integrated flux density of $S_{1.28\,\mathrm{GHz}} = 0.736 \pm 0.100$\,mJy and a corresponding power at 1.4\,GHz of $(0.47 \pm 0.13) \times 10^{25}$\,W\,Hz$^{-1}$. The detected emission spans an angular scale of $1.92' \times 1.42'$, corresponding to a projected physical scale of $0.93 \times 0.69$\,Mpc at $z = 1.03$. The emission is centred close to the SZ peak and overlaps with a faint compact radio source associated with the BCG (cyan triangle), but exhibits an elongation towards the north. Multiple compact radio sources are present near the cluster centre, and several leave residual structure in the LR contours after subtraction (see Figure~\ref{fig:subtraction_0113}), making the interpretation of the faint emission more challenging. We therefore classify this as a possible radio halo, but the detection remains uncertain. At present, it remains unclear whether the detected signal traces a merger-driven radio halo or predominantly remnant/residual emission related to AGN activity in cluster galaxies; deeper or lower frequency follow-up data (e.g. with the MeerKAT $UHF$ band and increased on-source time) will be required to distinguish between these scenarios.
\newline

No archival \textit{Chandra} or \textit{XMM--Newton} imaging is available for ACT-CL~J0241.2$-$3916 and ACT-CL~J0113.9+1451, so a detailed comparison with the X-ray ICM is not yet possible. Instead, we use \textit{Planck}+ACT DR6 Compton-$y$ maps \citep{2024PhRvD.109f3530C} as tracers of the projected hot gas distribution and compare these to the radio morphology. In both clusters, the diffuse radio emission is broadly centred on, and comparable in extent to, the SZ signal, as illustrated in Appendix~\ref{app:SZmaps}, consistent with expectations for cluster-scale radio halos that trace the bulk of the ICM.

\begin{figure*}
\centering
\begin{tabular}{ccc}
    \begin{overpic}[width=0.32\textwidth]{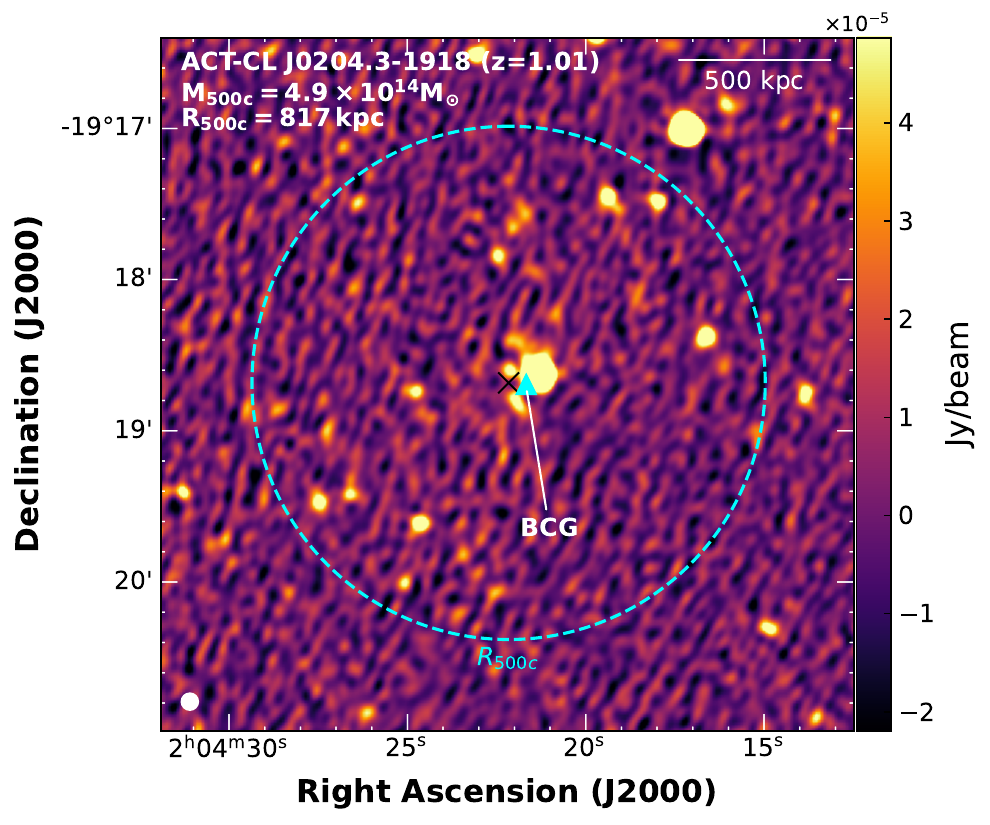}\put(5,85){\textbf{(a)}}\end{overpic} &
    \begin{overpic}[width=0.32\textwidth]{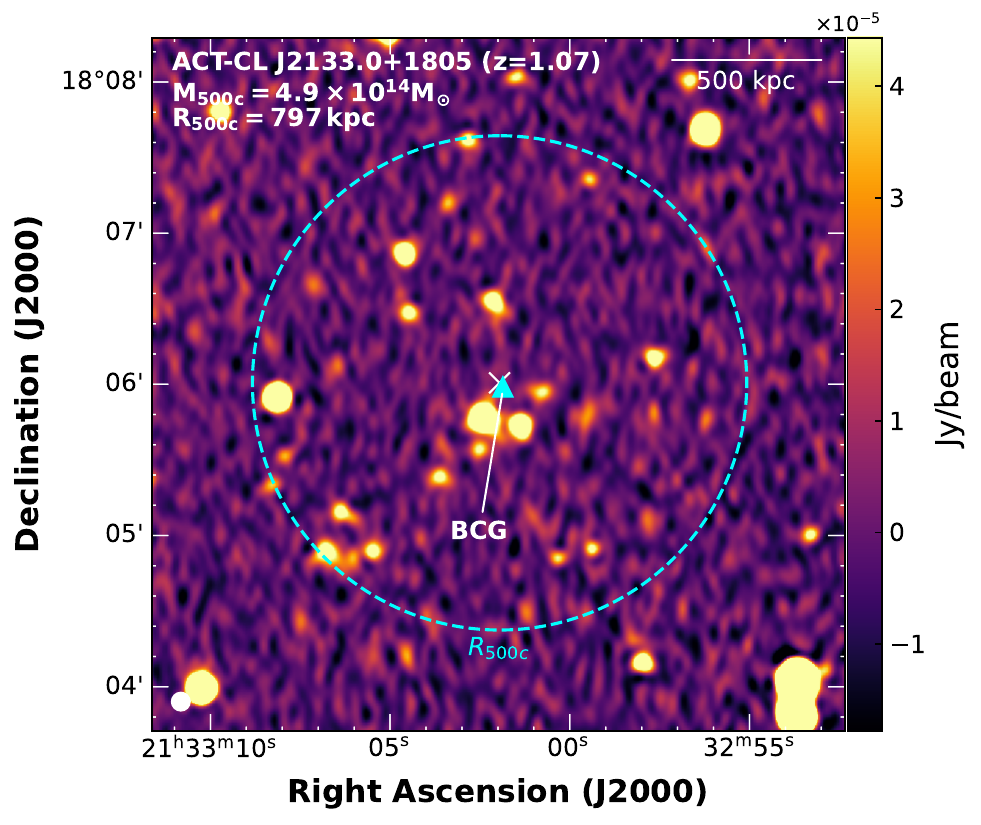}\put(5,85){\textbf{(b)}}\end{overpic} &
    \begin{overpic}[width=0.32\textwidth]{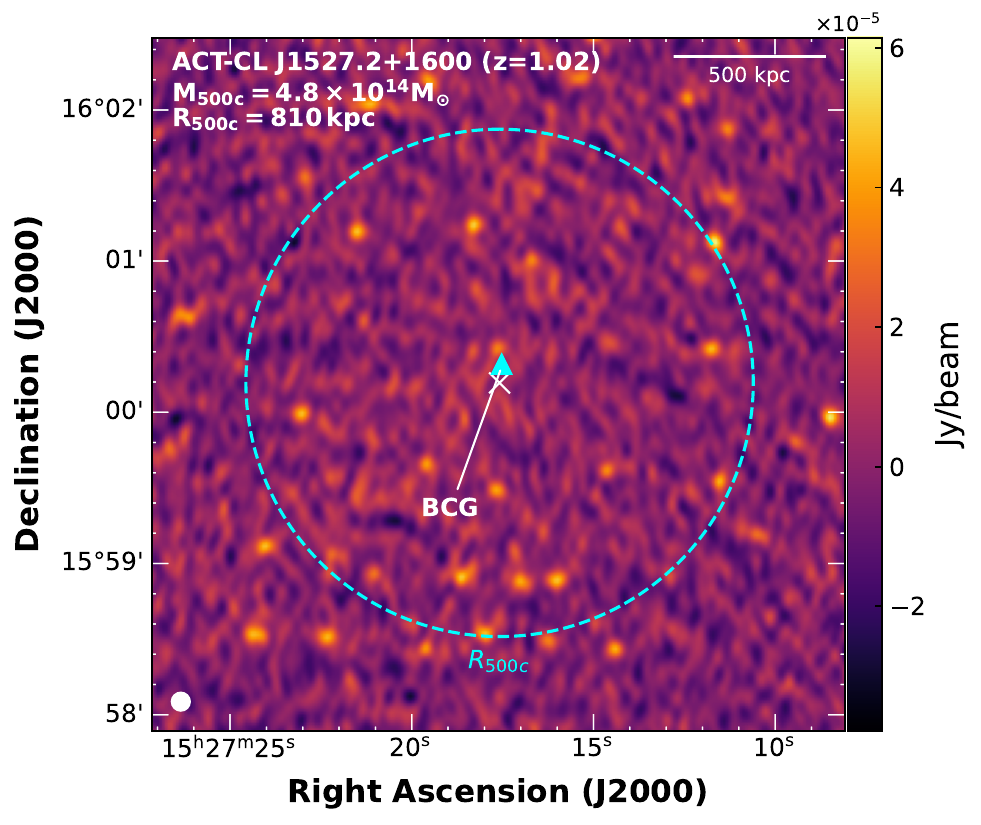}\put(5,85){\textbf{(c)}}\end{overpic} \\

    \begin{overpic}[width=0.32\textwidth]{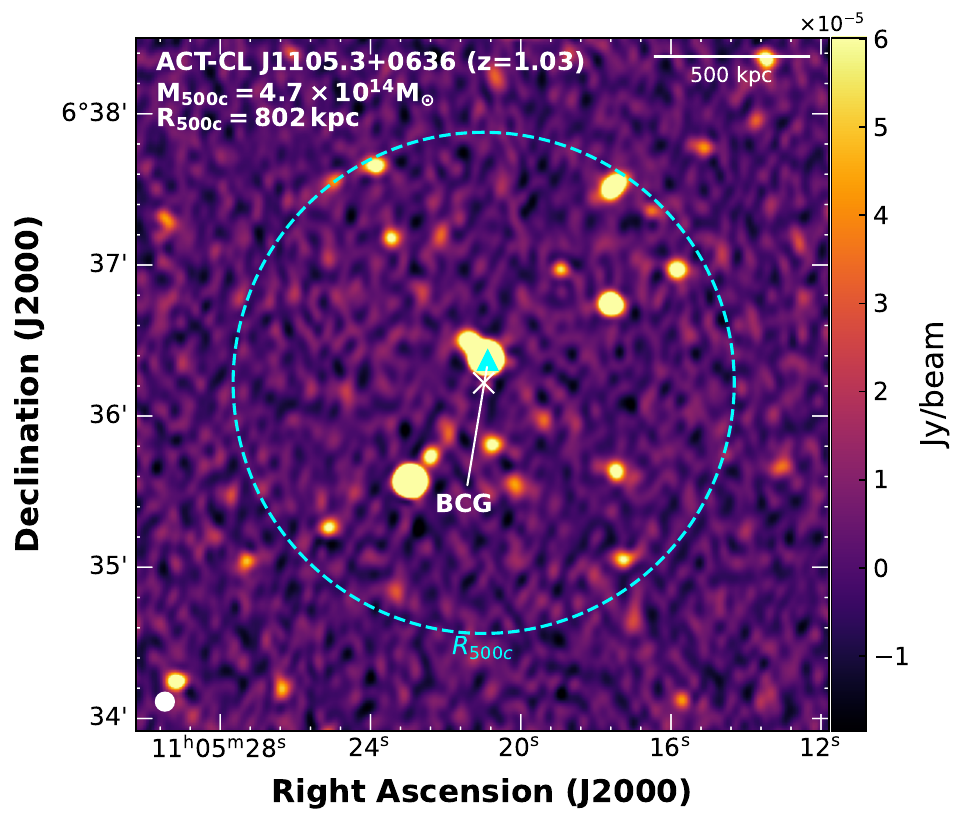}\put(5,86){\textbf{(d)}}\end{overpic} &
    \begin{overpic}[width=0.32\textwidth]{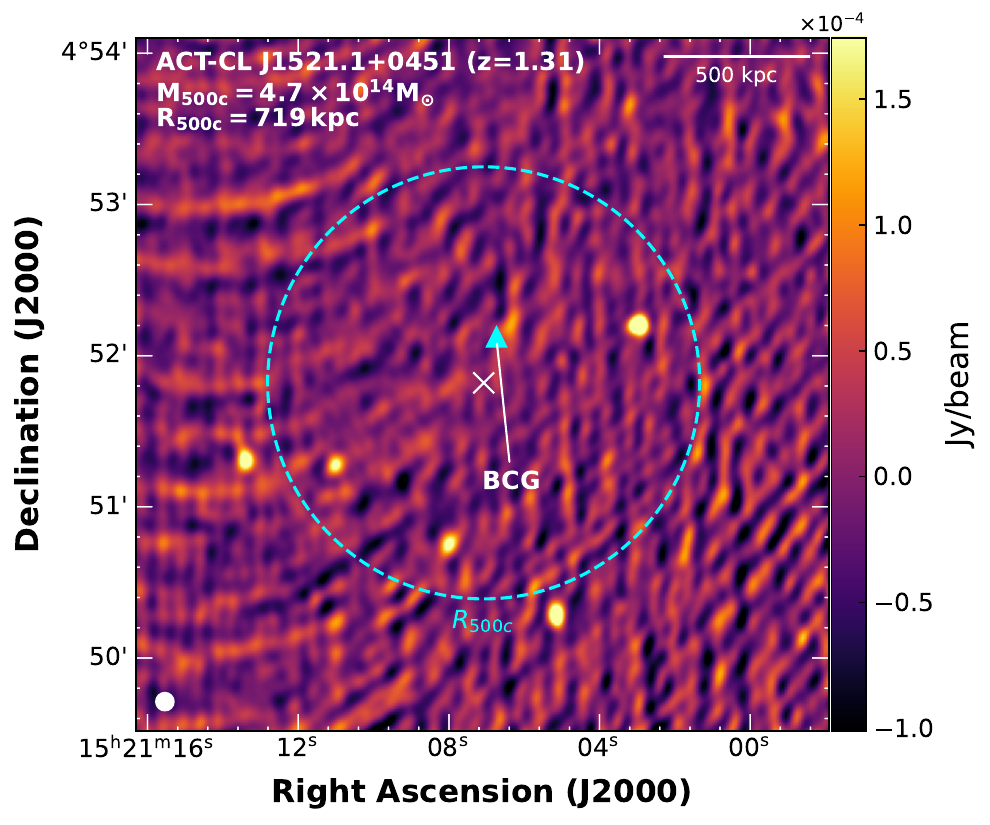}\put(5,85){\textbf{(e)}}\end{overpic} &
    \begin{overpic}[width=0.32\textwidth]{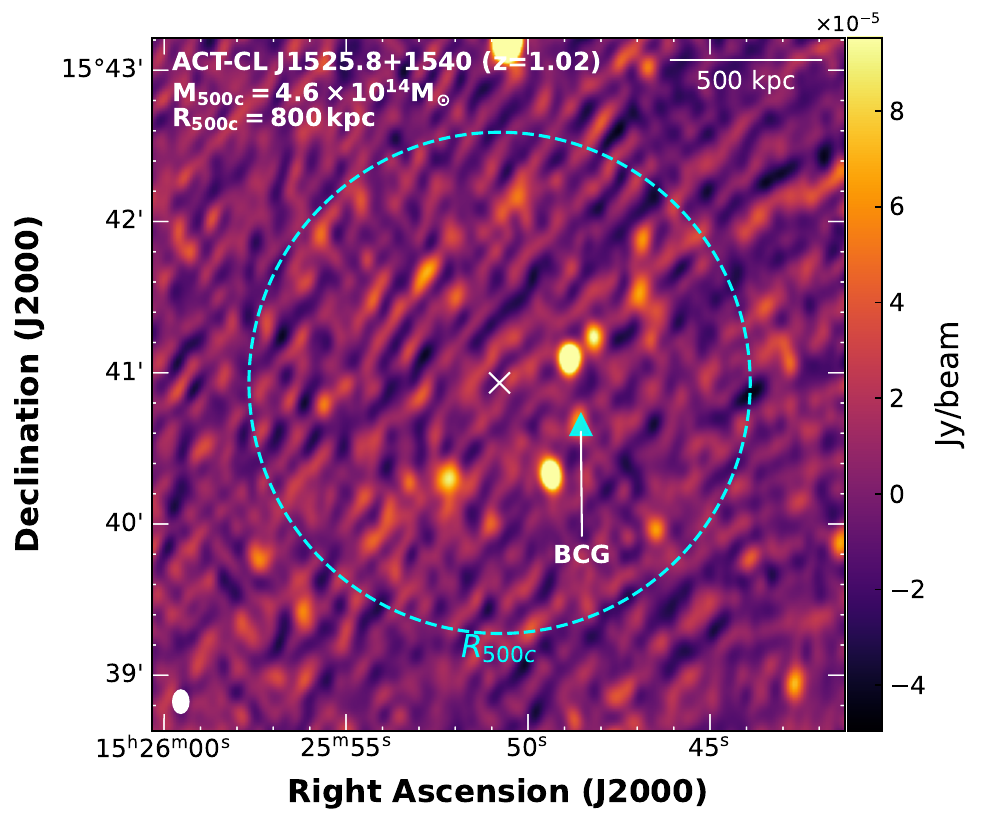}\put(5,85){\textbf{(f)}}\end{overpic} \\

    \begin{overpic}[width=0.32\textwidth]{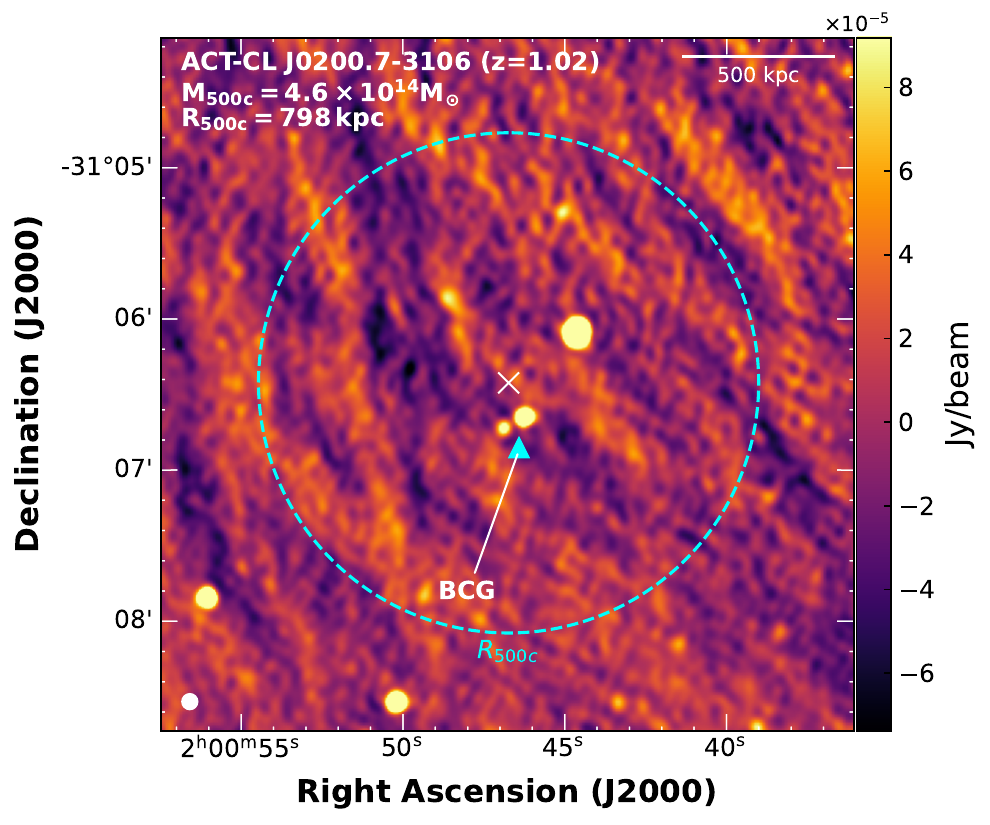}\put(5,85){\textbf{(g)}}\end{overpic} &
    \begin{overpic}[width=0.32\textwidth]{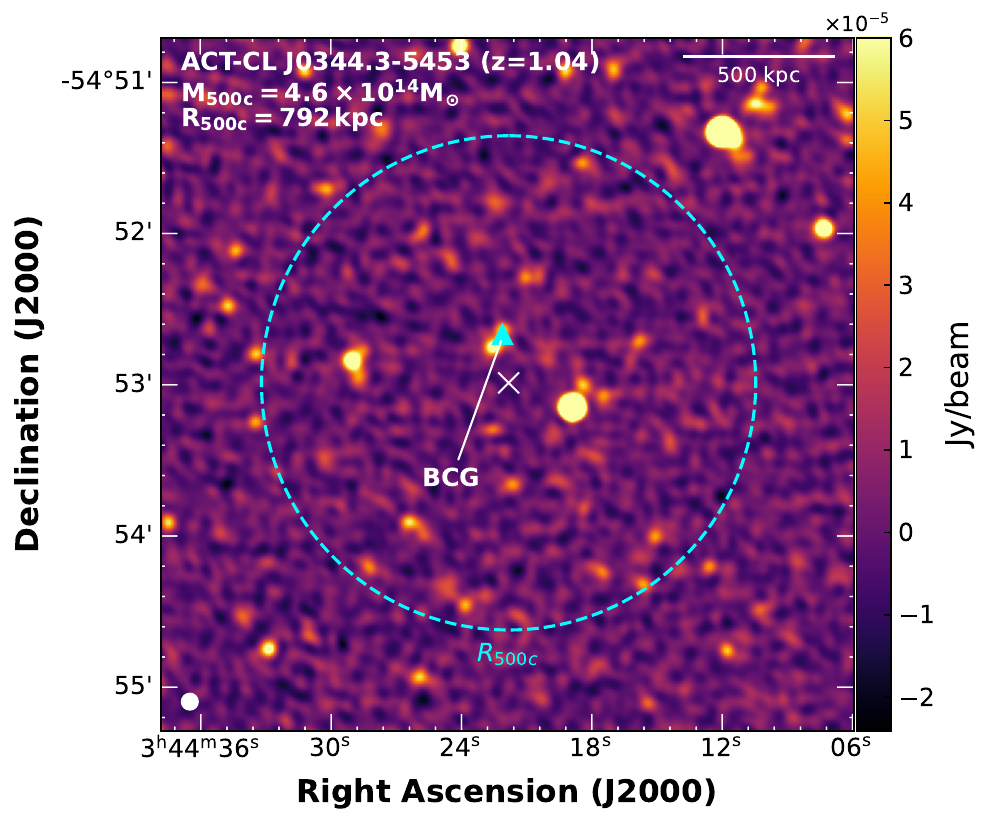}\put(5,85){\textbf{(h)}}\end{overpic} &
    \begin{overpic}[width=0.32\textwidth]{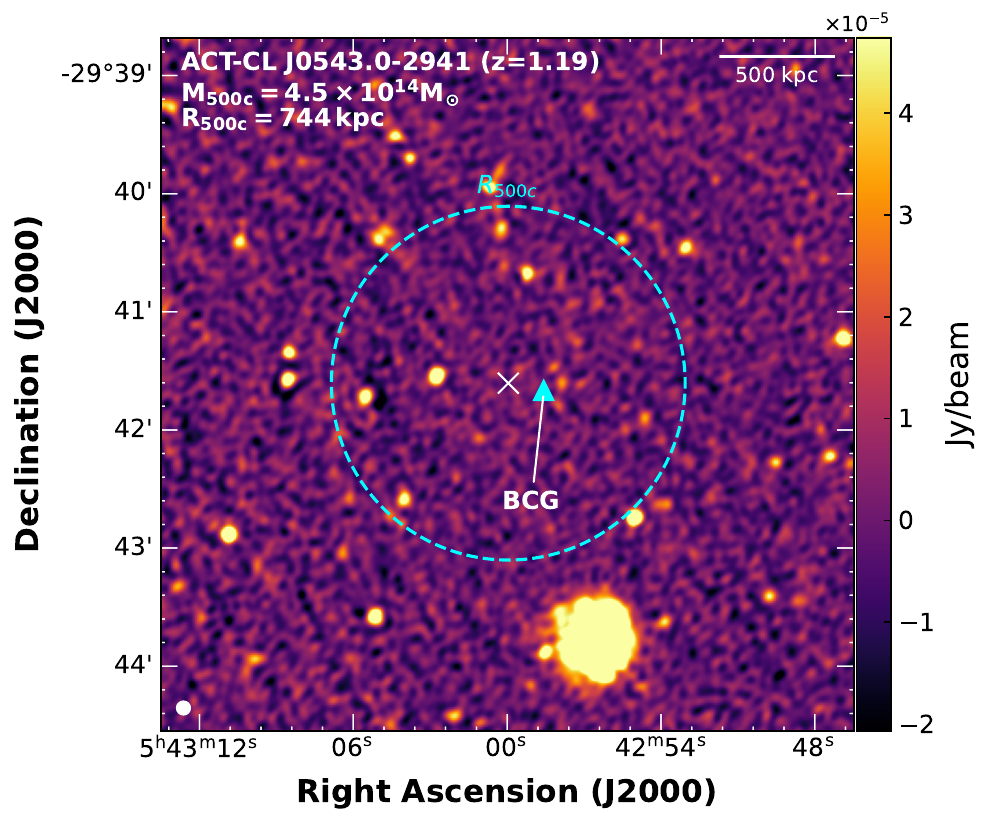}\put(5,85){\textbf{(i)}}\end{overpic} \\

   
    \begin{overpic}[width=0.32\textwidth]{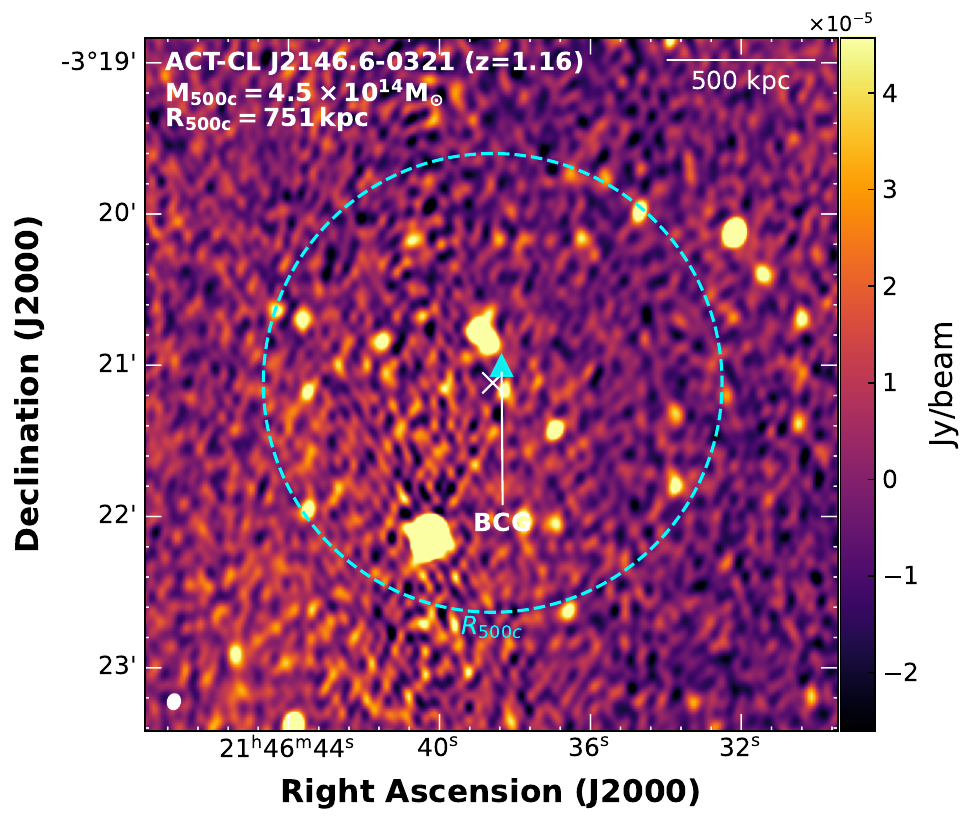}\put(5,85){\textbf{(j)}}\end{overpic} &
    \begin{overpic}[width=0.32\textwidth]{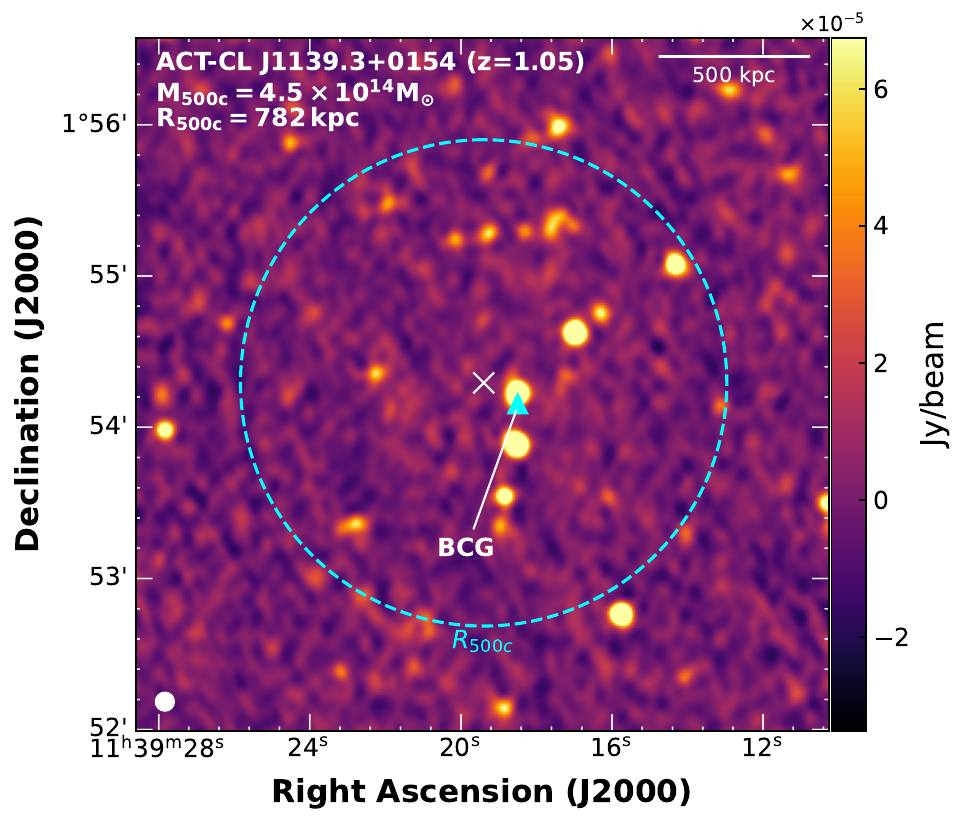}\put(5,85){\textbf{(k)}}\end{overpic} \\
\end{tabular}
\caption{MeerKAT 1.28~GHz full-resolution images for each cluster. The projected physical scale at each cluster's redshift is indicated in the top right of each panel. The synthesized beam is rendered as a filled white ellipse or circle at the lower left of each panel. The ACT SZ peak is labelled with a black or white cross, and the location of the BCG, obtained from ACT DR5 \citep{2021ApJS..253....3H}, is marked with a cyan triangle. The 3GC was applied to the following clusters to mitigate artefacts from bright sources and improve image fidelity: ACT-CL J0204, J2133, J1521, J1525, J0200, J2146, and J1139.}
\label{fig:allclusters2}
\end{figure*}

\begin{table*}
\caption{Measured properties of the detected diffuse emissions. Clusters are listed in order of decreasing $M_{\mathrm{500c}}$ value. The first six clusters in the table (initially presented in P25) are also included here for completeness and are marked with an \textit{*}.}
\label{table:3}

\begin{tabular}{cccccc}
\hline
Cluster Name   & LAS                & LLS                & Classification & S$_{\nu}$             & $P_{1.4,\mathrm{GHz}}$                    \\
(ACT-CL)       & ($\prime$)         & (Mpc)              &                & (mJy)                 & ($\times 10^{25}$ W Hz$^{-1}$)            \\ \hline
J0329.2$-$2330{\textit{*}} & $2.17 \times 1.73$ & $1.08 \times 0.86$ & RH             & $3.45\, \pm\, 0.21$   & $3.55\, \pm\, 1.06$                       \\
J2106.0$-$5844{\textit{*}} & $1.99 \times 1.88$ & $0.98 \times 0.93$ & RH             & $3.23\, \pm\, 0.19$   & $2.64\, \pm\, 0.73$                       \\
J1137.8+0728{\textit{*}}   & $1.79 \times 1.38$ & $0.86 \times 0.66$ & RH             & $0.87\, \pm\, 0.08$   & $0.53\, \pm\, 0.14$                       \\
J1142.7+1527{\textit{*}}   & $1.96 \times 1.59$ & $0.97 \times 0.79$ & RH             & $1.47\, \pm\, 0.11$   & $1.38\, \pm\, 0.41$                       \\
J0003.9+1642{\textit{*}}   & $1.96 \times 1.51$ & $0.99 \times 0.76$ & RH             & $1.43\, \pm\, 0.12 $  & $1.74\, \pm\, 0.55$                       \\
J0546.6$-$5345{\textit{*}} & $0.97 \times 0.90$ & $0.47 \times 0.44$ & U              & $0.42\, \pm\, 0.04$   & $0.30\, \pm\, 0.08$                       \\
J0241.2$-$3916 & $1.43 \times 1.10$ & $0.69 \times 0.53$ & RH             & $0.27\, \pm\, 0.05$ & $0.17\, \pm\, 0.06$ \\
J0113.9+1451   & $1.94 \times 1.42$ & $0.93 \times 0.69$ & U             & $0.74\, \pm\, 0.10$ & $0.47\, \pm\, 0.13$ \\ \hline
\end{tabular}

\begin{minipage}{\textwidth}
\footnotesize
\textbf{Notes:} Columns: (1) ACT DR5 cluster name; (2) largest angular size (LAS) of diffuse emission in arc-minutes; (3) largest physical extent of diffuse emission at cluster redshift in Mpc; (4) classification type: radio halo (RH) and uncertain/candidate (U) ; (5) total flux density in mJy; (6) $k$-corrected radio power at 1.4 GHz, assuming $\alpha = -1.3 \pm 0.4$.
\end{minipage}
\end{table*}

\subsubsection{Clusters with central radio AGN or radio lobe systems}

ACT-CL~J0930.2+0615 exhibits a prominent central radio galaxy morphology in the MeerKAT FR image (see panel b of Figure~\ref{fig:allclusters}), with extended lobes and a radio core coincident with the BCG (cyan triangle), suggesting a physical association. The integrated flux density of the central radio source is \( S_{1.28\,\mathrm{GHz}} = 1.98 \pm 0.10 \) mJy. The extended radio emission signifies AGN activity, but no diffuse cluster-scale emission is detected within $R_{\rm 500c}$. Photometric redshifts from DESI Legacy Imaging Surveys DR10 indicate a red galaxy at $z = 0.945 \pm 0.277$ near the radio core, plausibly consistent with the cluster redshift and core membership, though spectroscopic follow-up is needed for confirmation \citep{2019AJ....157..168D,2022MNRAS.512.3662D}. The morphology is complex, and the rich radio environment is a prime example of AGN–ICM interaction at high-redshift. The absence of diffuse ICM emission points to a lack of sufficient turbulence or recent major merging.
\newline

In ACT-CL~J0947.9$-$0120 (see panel c of Figure~\ref{fig:allclusters}), the BCG (cyan triangle) coincides with a compact radio source slightly offset from the SZ peak (cross), and has an integrated flux density of $\mathrm{S_{1.28\,GHz}} = 0.29 \pm 0.01$\,mJy. No diffuse or halo-scale radio emission is present at the current sensitivity. Analysis from the HSC optical study of \citet{2025MNRAS.536..572D} finds this cluster is well centred with respect to both SZ and optical positions. Nonetheless, comparison of optical and SZ morphologies suggests ongoing merger activity or mild dynamical disturbance.
\newline

For ACT-CL~J2341.2$-$5119 (see panel d of Figure~\ref{fig:allclusters}), both the SZ peak and the BCG (cyan triangle) are essentially coincident with a bright compact radio source at the cluster centre ($\mathrm{S_{1.28\,GHz}} = 4.16 \pm 0.21$\,mJy), a configuration typical of radio-loud AGN residing in the BCG \citep[e.g.][]{2007MNRAS.379..894B,2007ApJ...667L..13C,2015MNRAS.453.1201H}. There is no evidence for additional diffuse emission in the central region.
\newline

Finally, ACT-CL~J0244.1$-$0923 (see panel e of Figure~\ref{fig:allclusters}) displays a strong central source, with BCG and SZ peaks co-located with a compact radio AGN of integrated flux density $\mathrm{S_{1.28\,GHz}} = 14.69 \pm 0.73$\,mJy, but there is no hint of extended, non-thermal emission above the sensitivity limits. Overall, while these clusters host central AGN and, in one case, prominent radio lobes, none show clear cluster-scale diffuse radio halos or relics.

\subsubsection{Clusters with only compact sources and no diffuse emission}

This category includes clusters that lack detectable Mpc-scale diffuse emission but show mixed dynamical indicators, such as SZ–BCG offsets, bright central AGN, or complex radio environments.
\newline

ACT-CL~J0851.9+1500 (see panel a of Figure~\ref{fig:allclusters}) shows several compact radio sources in the field, likely associated with cluster galaxies, but no diffuse emission within $R_{\rm 500c}$ and no radio counterpart detected at the BCG (triangle) or SZ peak (cross). The lack of central radio emission and the field’s radio properties are suggestive of a relatively relaxed dynamical state, or the absence of merger-driven turbulence sufficient to power a radio halo.
\newline

ACT-CL~J0044.4+0113 (see panel f of Figure~\ref{fig:allclusters}) is among the hottest and most X-ray luminous clusters beyond $z = 1$, with XMM-Newton observations indicating a high ICM temperature of $kT_{\rm X} = 7.9 \pm 1.0$~keV \citep{2013ApJ...765...67M}. Despite its high mass and temperature, no diffuse radio emission is identified in the cluster core, and several faint sources are found in the field, none of which are spatially coincident with the cluster centre.
\newline

In ACT-CL~J0314.0+0203 (see panel g of Figure~\ref{fig:allclusters}), the BCG (cyan triangle) is offset from the SZ peak (cross) but coincides with a faint compact radio source detected in the MeerKAT image, with an integrated flux density of $S_{1.28\,\text{GHz}} = 0.06 \pm 0.01$\,mJy. No extended radio emission is detected within $R_{\rm 500c}$, and the spatial offset may indicate a disturbed system or projection effect.
\newline

ACT-CL~J0125.3$-$0802 (see panel i of Figure~\ref{fig:allclusters}) shows no evidence for diffuse radio emission in the cluster core, but several bright compact radio sources are present in the vicinity. Notably, the SZ peak is nearly coincident with one of the brightest sources, which has a flux density of $24.2 \pm 1.21$ mJy. Another bright source lies north of the cluster centre. The BCG, marked by a cyan triangle, remains offset from the SZ centroid (marked by a cross), which may indicate a dynamically disturbed system or projection effect, as commonly seen in high-redshift clusters.
\newline

ACT-CL~J1521.1+0451 (see panel e of Figure~\ref{fig:allclusters2}) shows no evidence for diffuse radio emission in the cluster core, with only a few compact sources detected in the field and none displaying obvious extended structure within $R_{\rm 500c}$. The cluster exhibits an offset between the SZ peak (cross) and the BCG (cyan triangle).
\newline

ACT-CL~J1525.8+1540 (see panel i of Figure~\ref{fig:allclusters2}) likewise lacks any detectable diffuse radio emission in the central region, and only faint compact sources are present within $R_{\rm 500c}$. The SZ peak (cross) and BCG (cyan triangle) are offset, while the BCG position coincides with a faint compact radio source.
\newline

ACT-CL~J0543.0$-$2941 (see panel i of Figure~\ref{fig:allclusters2}) contains several compact radio sources within $R_{\rm 500c}$, an offset between the BCG and SZ peak, and a bright spiral galaxy just outside $R_{\rm 500c}$, but no evidence for diffuse central radio emission. 
\newline

For ACT-CL~J1139.3+0154 (see panel k of Figure~\ref{fig:allclusters2}), the SZ peak and BCG positions are offset, but the BCG is aligned with the compact radio source in the image ($S_{1.28\,\text{GHz}} = 0.28 \pm 0.01$\,mJy) and no extended emission is apparent.
\newline

For all these systems, MeerKAT imaging confirms the presence of small-scale, point-like radio sources, sometimes coincident with BCGs, but none of these clusters show Mpc-scale synchrotron or radio halo emission at the achieved sensitivity.

\subsubsection{Relaxed clusters with no extended diffuse emission}

This category comprises clusters that appear dynamically relaxed, with BCG–SZ coincidence (or small offsets) and only compact radio emission in the core, and no evidence for halo- or relic-like emission within $R_{\rm 500c}$.
\newline

In ACT-CL~J1049.1+1106 (see panel j of Figure~\ref{fig:allclusters}), the BCG (cyan triangle) and the SZ peak (cross) are spatially coincident and associated with a compact radio source detected in the MeerKAT image, with an integrated flux density of $S_{1.28\,\mathrm{GHz}} = 0.62 \pm 0.03$\,mJy. No diffuse radio halo or mini-halo is detected within $R_{\rm 500c} = 756$\,kpc, and the absence of extended emission is consistent with a radio-quiet ICM at the achieved sensitivity.
\newline

For ACT-CL~J0154.3$-$4824 (see panel k of Figure~\ref{fig:allclusters}), several compact radio sources are detected within $R_{\rm 500c} = 740$\,kpc, but no extended emission is evident in the cluster core. The BCG position (cyan triangle) coincides with a faint compact radio source ($S_{1.28\,\mathrm{GHz}} = 0.06 \pm 0.01$\,mJy) and there is an offset between the BCG and the SZ peak. Despite this spatial alignment, the field shows neither a radio halo nor any compelling evidence for large-scale diffuse synchrotron emission.
\newline

ACT-CL~J0204.3$-$1918 (see panel a of Figure~\ref{fig:allclusters2}) reveals no evidence for diffuse radio emission within the cluster centre. The BCG, marked by a cyan triangle, coincides with a bright compact radio source, with an integrated flux density of $\mathrm{S\,_{1.28\,GHz} = 3.56 \pm 0.18\, mJy}$, while the BCG and SZ peak (cross) are slightly offset from each other.
\newline

ACT-CL~J2133.0+1805 (see panel b of Figure~\ref{fig:allclusters2}) displays no diffuse or halo-like emission at the cluster centre.  The SZ peak (cross) and the BCG (cyan triangle) are co-located. Additionally, two bright radio sources are detected to the south of the cluster centre, though these are not directly associated with the BCG/SZ peak position.
\newline

ACT-CL~J1527.2+1600 (see panel c of Figure~\ref{fig:allclusters2}) is characterized by several compact radio sources distributed within $R_{\rm 500c}$. While the BCG and cluster centre are closely-aligned, no corresponding halo, relic, or other extended radio feature is observed above current surface-brightness limits. 
\newline

ACT-CL~J1105.3+0636 (see panel d of Figure~\ref{fig:allclusters2}) shows no diffuse or extended radio emission detected within the cluster core. The BCG (marked by cyan triangle) coincides with the most bright compact radio source in the field, with an integrated flux density of $\mathrm{S\,_{1.28\,GHz} = 2.40 \pm 0.12\, mJy}$, while the SZ peak (cross) is slightly offset from the BCG position.
\newline

ACT-CL~J0344.3$-$5453 (see panel h of Figure~\ref{fig:allclusters2}) contains several faint radio point sources, with no evidence for diffuse or halo-like emission at the cluster centre. Notably, there is a positional offset between the SZ peak (white cross) and the brightest cluster galaxy (BCG, cyan triangle). The BCG coincides with a faint compact radio source, with an integrated flux density of $\mathrm{S\,_{1.28\,GHz} = 0.06 \pm 0.01\, mJy}$.
\newline

Taken together, the clusters in this category show BCG–SZ coincidence (or small offsets) accompanied by only compact radio emission and no detectable cluster-scale diffuse component. This outcome may indicate intrinsically radio-quiet intracluster media in these systems, or it may simply reflect that, at these redshifts, our sensitivity (typically $\sigma \sim 6$--$8\,\upmu\mathrm{Jy}\,\mathrm{beam}^{-1}$ at 1.28\,GHz) is insufficient to reveal faint, extended emission.

\subsubsection{Clusters with artefacts}

ACT-CL~J0200.7$-$3106, presents substantial imaging  challenges in the MeerKAT FR image (see panel g of Figure~\ref{fig:allclusters2}). The field includes three prominent radio point sources within $R_{\rm 500c}$, but no evidence is found for diffuse radio emission in the cluster centre. The SZ peak (cross) and BCG (cyan triangle) are offset, showing no direct alignment between the brightest galaxy and SZ centroid. Notably, a bright radio source north of the cluster centre produced severe direction-dependent artefacts and persistently elevated noise across the field, even after applying advanced calibration (3GC) and source subtraction. These residuals strongly limit the sensitivity to faint diffuse emission, highlighting the inherent challenges of probing diffuse phenomena in fields crowded with bright sources at high-redshift.
\newline

Similarly, ACT-CL~J2146.6$-$0321, shows no diffuse emission in the cluster centre in the MeerKAT FR image (see panel j of Figure~\ref{fig:allclusters2}). Compact radio sources are detected within $R_{\rm 500c} = 751$\,kpc, the brightest of which is toward the southern outskirts. The SZ peak (white cross) and BCG (cyan triangle) are slightly offset, and the southern radio source near the centre left pronounced artefacts across the map after multiple subtraction attempts. The increased noise and structured residuals across the cluster core significantly reduce sensitivity to extended emission and contribute to high uncertainties in this region.
\newline

ACT-CL~J2245.2$-$0433 had poor data quality as a result of observational problems and was therefore excluded from the sample.

\subsection{Non-detections and upper limits}
\label{sec:non-detections}

We report $3\sigma$ upper limits on the radio halo power for all clusters lacking detected diffuse emission (see Table~\ref{tab:upperlimits}). Following the procedures of \citet{2021A&A...647A..50C} and \citet{2023A&A...672A..41B}, we apply an injection strategy in which model radio halos with exponential profiles and physically motivated sizes are introduced into the visibility data at increasing intensities until the $3\sigma$ detection threshold is reached in the central region. The flux densities are scaled to 1.4\,GHz assuming a mean spectral index of $\alpha = -1.3$; this assumption is based on the typical range of spectral indices observed in classical radio halos, which often cluster around $\alpha \sim -1.1$ to $-1.5$ in statistical samples \citep{2013A&A...551A..24V,2013ApJ...777..141C,2021MNRAS.504.1749K}. However, recent work indicates that a substantial fraction of high-redshift halos may have steeper spectra \citep[$\alpha \leq -1.5$;][]{2019ApJ...881L..18C,2021A&A...654A.166D}. The standard $k$-correction is applied to account for redshift effects when computing the rest-frame radio power. Most of our upper limits at
$z>1$ reach sensitivities that are comparable, once cosmological
dimming is taken into account, to the deepest radio-halo limits reported
in previous work \citep{2021A&A...647A..50C,2021A&A...647A..51C,2023A&A...672A..41B}.
\newline

During analysis, several clusters could not be included in our upper limit sample due to the inability to perform reliable mock halo injections. For ACT-CL~J0930.2+0615, a bright radio galaxy (core and lobes), not associated with any diffuse cluster emission, was subtracted from the data, but residual structures remain and dominate the cluster centre. ACT-CL~J0200.7$-$3106 was affected by a bright radio source north of the cluster, which generated strong direction-dependent artefacts and high noise; despite applying direction-dependent (third-generation) calibration and three rounds of point-source subtraction, these artefacts were not cleared, making a reliable limit on diffuse emission impossible. Similarly, in ACT-CL~J2146.6$-$0321, a southern radio source near the cluster centre left persistent artefacts and elevated noise even after extensive subtraction, resulting in residual structure throughout the field. In addition, ACT-CL~J2245.2$-$0433 was excluded from the radio analysis altogether because the data quality did not meet our minimum imaging requirements. Consequently, these four clusters were excluded from the statistical upper-limit analysis because strong residual point-source artefacts or poor data quality affect the expected halo region to such an extent that injected model halos are severely distorted or blended with residuals, and the recovered flux depends sensitively on the exact injection position and cannot be converted into a meaningful upper limit.
\newline

These issues prevented confident mock halo injection and upper-limit calculation in these four clusters, highlighting the impact of compact source environment, calibration artefacts, and data quality on upper-limit analyses with MeerKAT. Mock radio halo injection was performed in the remaining 18 clusters where imaging and subtraction procedures were successful; see Appendix Figures~\ref{fig:mockhalo}--\ref{fig:mockhalo4} for examples of the injected halo upper limits and corresponding images.

\begin{table*}
\caption{Summary of the host cluster properties and upper limit parameters for the eighteen (18) considered targets lacking detected diffuse emission. Clusters are listed in order of decreasing $M_{\mathrm{500c}}$ value.}
\label{tab:upperlimits}
\begin{tabular}{lcccccc}
\hline
Cluster Name & RA$_{\rm inj}$ (J2000) & Dec$_{\rm inj}$ (J2000) & $r_{\rm e,inj}$ & $I_{\rm 0,inj}$                  & S$_{1.4\,\mathrm{GHz, UL}}$ & $P_{1.4\,\mathrm{GHz, UL}}$ \\
(ACT-CL)     & $(^{\mathrm{h}}{:}^{\mathrm{m}}{:}^{\mathrm{s}})$ & $(^{\mathrm{\circ}}{:}^{\mathrm{\prime}}{:}^{\mathrm{\prime\prime}})$ & (arcsec)    & $(\mathrm{m Jy\, arcsec^{-2}})$ & (mJy)                      & ($10^{24}$ W Hz$^{-1}$)    \\ \hline
J0851.9+1500 & 08:51:55.2             & +15:00:42.0             & 24.75       & 0.09                             & 0.27                       & 1.98                     \\
J0947.9-0120 & 09:47:58.3             & $-$01:20:02.8           & 23.87       & 0.09                             & 0.26                       & 2.65                     \\
J2341.2-5119 & 23:41:12.1             & $-$51:19:40.2           & 24.97       & 0.06                             & 0.20                       & 1.32                     \\
J0244.1-0923 & 02:44:02.6             & $-$09:22:53.9           & 24.15       & 0.12                             & 0.36                       & 3.71                     \\
J0044.4+0113 & 00:44:25.7             & +01:13:01.3             & 24.81       & 0.05                             & 0.16                       & 1.11                     \\
J0314.0+0203 & 03:14:01.0             & +02:03:39.8             & 24.32       & 0.09                             & 0.28                       & 2.50                     \\
J0125.3-0802 & 01:25:13.7             & $-$08:01:38.4           & 24.65       & 0.12                             & 0.37                       & 2.80                     \\
J1049.1+1106 & 10:49:06.9             & +11:06:14.3             & 24.03       & 0.07                             & 0.19                       & 2.17                     \\
J0154.3-4824 & 01:54:20.5             & $-$48:24:42.8           & 23.92       & 0.11                             & 0.32                       & 3.95                     \\
J0204.3-1918 & 02:04:22.2             & $-$19:18:42.0           & 24.92       & 0.09                             & 0.28                       & 1.82                     \\
J2133.0+1805 & 21:33:01.6             & +18:05:59.5             & 24.61       & 0.10                             & 0.31                       & 2.33                     \\
J1527.2+1600 & 15:27:17.7             & +16:00:10.6             & 24.86       & 0.07                             & 0.21                       & 1.39                     \\
J1105.3+0636 & 11:05:21.1             & +06:36:12.1             & 24.81       & 0.10                             & 0.30                       & 2.10                     \\
J1521.1+0451 & 15:21:07.1             & +04:51:48.1             & 23.86       & 0.13                             & 0.36                       & 4.78                     \\
J1525.8+1540 & 15:25:50.9             & +15:40:54.9             & 24.86       & 0.11                             & 0.34                       & 2.29                     \\
J0344.3-5453 & 03:44:21.9             & $-$54:53:00.4           & 24.75       & 0.11                             & 0.35                       & 2.48                     \\
J0543.0-2941 & 05:43:00.0             & $-$29:41:37.4           & 24.15       & 0.10                             & 0.31                       & 3.08                     \\
J1139.3+0154 & 11:39:19.5             & +01:54:16.5             & 24.70       & 0.09                             & 0.28                       & 2.08                     \\ \hline
\end{tabular}
\begin{minipage}{\textwidth}
\footnotesize
\textbf{Notes:} Columns 1--3 list the ACT DR5 cluster name and the J2000 Right Ascension and Declination of the SZ injection centre. Column 4 gives the exponential scale radius (\( r_{e,\mathrm{inj}} \)). Columns 5--7 provide: the injected central surface-brightness, the upper limit flux density at $1.4\,\mathrm{GHz}$ (scaled with spectral index $\alpha = - 1.3$), and the corresponding radio power upper limit, respectively. All upper limits are derived with the exponential profile injection method following \cite{2023A&A...672A..41B}.
\end{minipage}
\end{table*}

\subsection{Statistical studies}

\subsubsection{Scaling relations}

\begin{figure*}
    \centering
    \includegraphics[width=0.48\textwidth]{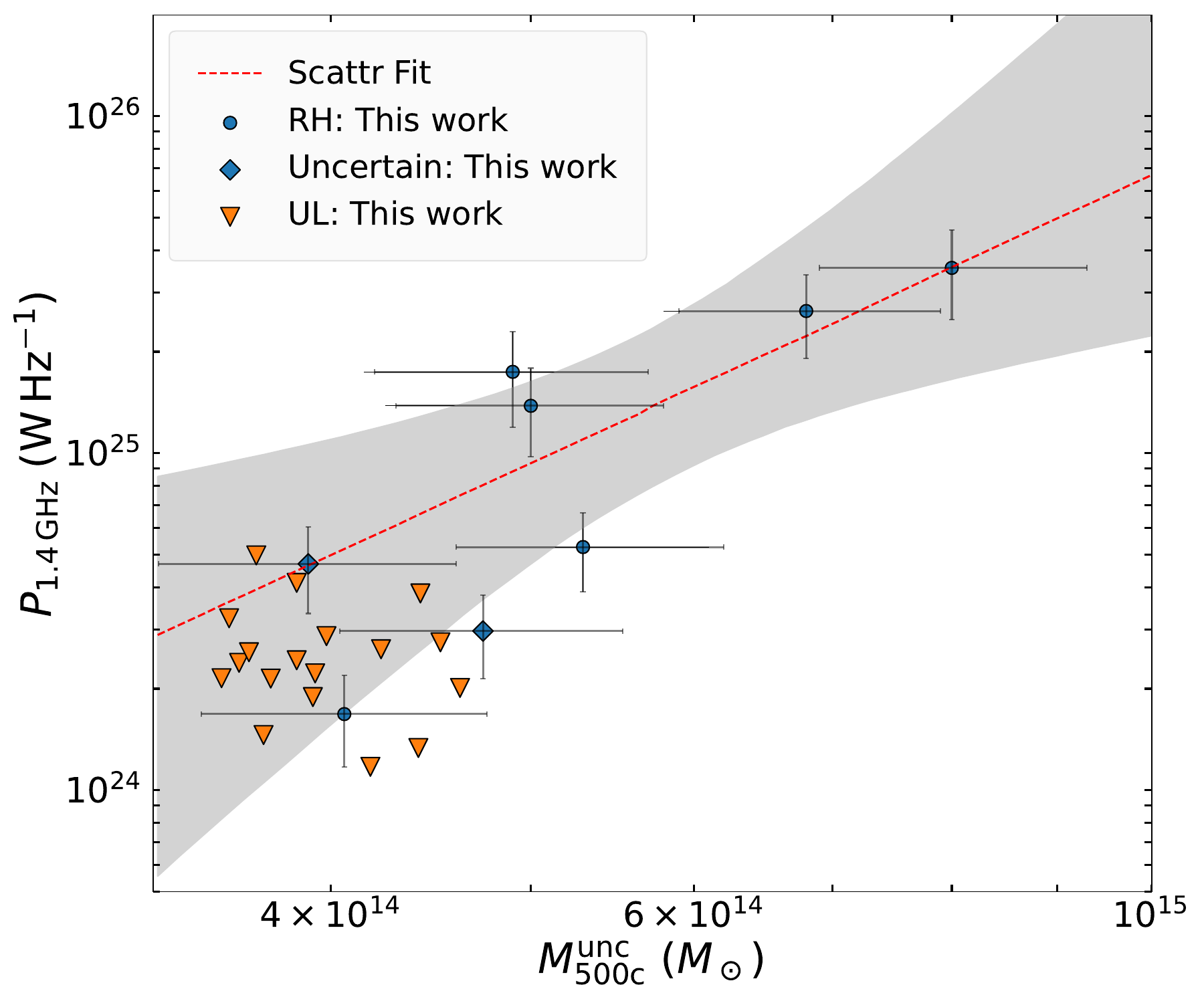}
    \includegraphics[width=0.48\textwidth]{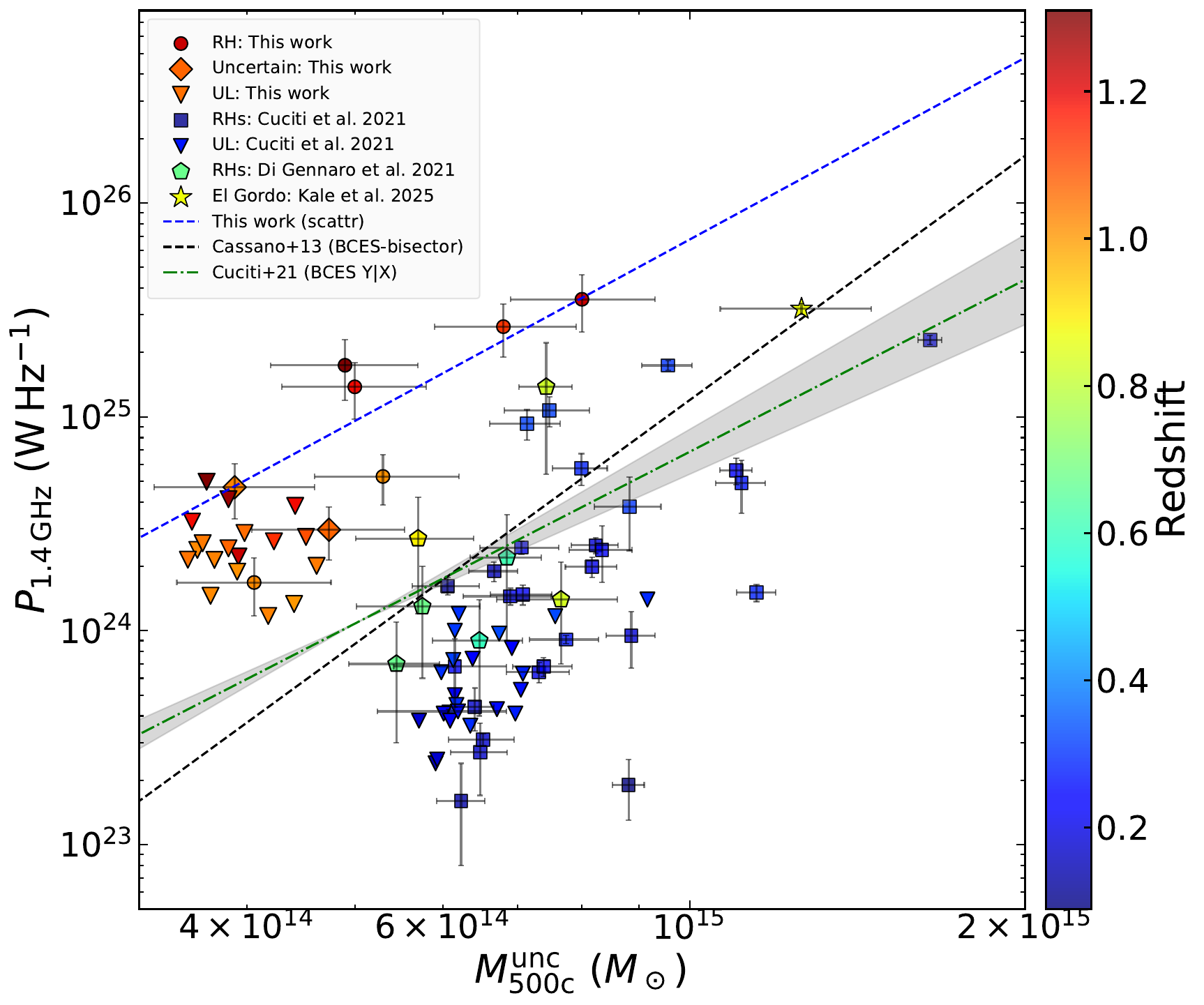}
    \caption{%
    Left: $P_{1.4\,\mathrm{GHz}}$–$M_{\rm 500c}^{\mathrm{unc}}$ relation for the MMDCS sample. Circles show radio halos and diamonds indicate uncertain diffuse emission from this work, while inverted triangles mark $3\sigma$ upper limits derived from the halo-injection analysis. The red dashed line shows the best-fitting scaling relation obtained with the \texttt{scattr} Bayesian regression, and the grey shaded region represents the 95 per cent confidence interval. Right: Comparison of the MMDCS sample with clusters from \citet{2021A&A...647A..51C} at $0.089 < z < 0.322$, the \textit{Planck} sample of \citet{2021NatAs...5..268D} at $0.6 < z < 0.9$, and the El~Gordo cluster at $z = 0.87$ \citep{2025A&A...698A.271K}. Blue squares denote radio halos and blue inverted triangles indicate upper limits from \citet{2021A&A...647A..51C}. Green diamonds show radio halos in the \citet{2021NatAs...5..268D} sample, and the yellow star marks El~Gordo. Red circles represent radio halos from this work, orange diamonds indicate uncertain detections, and orange inverted triangles show upper limits. The black dashed line shows the scaling relation from \citet{2013ApJ...777..141C}, while the green dot-dashed line denotes the relation from \citet{2021A&A...647A..51C} with its 95 per cent confidence region shown by the grey band, and the blue dotted line shows the best-fitting $P_{1.4,\mathrm{GHz}}$–$M_{\rm 500c}^{\mathrm{unc}}$ relation derived for the MMDCS radio halos.}
    \label{fig:scaling_combined}
\end{figure*}

The correlation between radio-halo power and cluster mass
($P_{1.4\,\mathrm{GHz}}$–$M_{\rm 500c}^{\rm unc}$) provides a useful statistical test of non-thermal processes in the ICM. The left panel of Figure~\ref{fig:scaling_combined} shows this relation for the MMDCS sample, including detected and uncertain radio halos from this work, together with the $3\sigma$ upper limits from our injection analysis. The red dashed line and grey band mark the power-law relation obtained with the \texttt{scattr}\footnote[15] {\url{https://github.com/lucadimascolo/scattr}}
package \citep{DiMascolo2024scattr}, which performs a Bayesian linear regression with intrinsic scatter and uncertainties in both variables, using only the radio halos (detected and uncertain). The detected radio halos are scattered around this relation, while most upper limits populate the region below it. The trend we observe in the distribution of detections and upper limits resembles what is found in low- and intermediate-redshift studies that combine radio halos with non-detections \citep{2012MNRAS.421L.112B,2013ApJ...777..141C,2016MNRAS.459.4240K,2021A&A...647A..51C,2023A&A...672A..43C}. In those studies, the apparent bimodality in the
$P_{1.4\,\mathrm{GHz}}$–$M_{\rm 500}$ plane is interpreted as evidence for a
population of faint halos with powers moderately below the scaling relation traced by bright systems. For a direct comparison with lower-redshift samples, the right panel of Figure~\ref{fig:scaling_combined} shows the same relation including clusters from \citet{2021A&A...647A..51C}, placing the MMDCS systems in the broader context of previously established $P_{1.4\,\mathrm{GHz}}$–$M_{\rm 500}$ scaling relations.
\newline

Our results therefore extend the observed separation between detections and upper limits to clusters at $z>1$, though some systems are near the detection threshold and cannot be conclusively identified as radio quiet. While radio halos are linked to merger‑driven turbulence and particle re‑acceleration, the limited ancillary data for our sample prevent a definitive classification of all non‑detections as relaxed systems. In particular, for clusters whose current upper limits remain within the scatter of the $P_{1.4\,\mathrm{GHz}}$--$M_{\rm 500c}$ relation, the typical MeerKAT rms levels of $\sigma \sim 6$--$8\,\upmu\mathrm{Jy\,beam^{-1}}$ imply that halos consistent with the correlation cannot yet be excluded. To rule these out would require reducing the image noise to $\sigma \sim 2$--$3\,\upmu\mathrm{Jy\,beam^{-1}}$, so that the corresponding power limits fall several $\sigma$ below the expected relation at each mass. Deeper and lower‑frequency surveys will be essential to improve these upper limits and fully test the persistence and origins of the $P_{1.4 \,\mathrm{GHz}}-M_{\rm 500}$ relation across redshift.
\newline

We quantified the position of the MMDCS halos in the $P_{1.4}$--$M_{\rm 500c}$ plane by fitting a power-law relation in log--log space to the confirmed radio-halo detections in P25 and this work using the \texttt{scattr} package. The best-fitting relation, $\log_{10} P_{1.4\,\mathrm{GHz}} = B\,\log_{10} M_{\rm 500c}^{\rm{unc}} + A$, yields $B = 2.82 \pm 2.27$ and $A = -0.20 \pm 0.35$. At fixed mass, this best-fit relation lies systematically above the low-redshift $P_{1.4\,\mathrm{GHz}} - M_{500}$ relations measured at $z \lesssim 0.6$ in \citet{2013ApJ...777..141C} and \citet{2021A&A...647A..51C}. However, the uncertainties on the slope and normalisation are large, owing to the small number of high-redshift halos, and the confidence region overlaps with the scatter seen in the lower-redshift samples. Therefore, we do not regard this as a statistically significant detection of evolution in the underlying scaling relation.
\newline

These high powers are comparable to those of El~Gordo and the most luminous halo in the \citet{2021NatAs...5..268D}  \textit{Planck} sample at $0.6 < z < 0.9$, and are hosted by some of the most massive clusters at $z > 1$ in a $\sim 10^4\,\mathrm{deg}^2$ survey area. This suggests that, in these extreme systems, merger-driven turbulence and amplified intracluster magnetic fields can partly compensate for severe surface-brightness dimming and enhanced inverse-Compton losses, allowing bright radio halos to remain observable even at high redshift.

\subsubsection{Detection fractions}
Diffuse radio emission is detected in 8 of the 30 SZ‑selected clusters at $z>1$, corresponding to an overall halo detection fraction of $27\%$. Of these, 4 clusters ($13\%$) host prominent halos and 4 ($13\%$) exhibit faint diffuse emission. The remaining 22 clusters ($73\%$) show no detectable central emission at the current sensitivity limits. These trends are consistent with previous studies that find low halo fractions at high-redshift, where IC losses and cosmological surface‑brightness dimming suppress diffuse synchrotron emission \citep{2013ApJ...777..141C,2025A&A...695A.215D}. 
\newline

In order to test whether the halo occurrence depends on cluster mass within our sample, we divided the MMDCS clusters into two bins using the uncorrected ACT SZ masses from Table~\ref{table:1}. Splitting at $M_{\rm 500c}^{\mathrm{Unc}} = 4.9\times10^{14}\,M_\odot$ yields a low-mass bin with 11 clusters and a high-mass bin with 19 clusters. In the low-mass bin we find one radio halo, corresponding to a halo fraction of $f_{\mathrm{RH}} = 0.09 \pm 0.09$ (1/11), whereas in the high-mass bin we detect seven halos, giving $f_{\mathrm{RH}} = 0.37 \pm 0.11$ (7/19). Although the uncertainties are large because of the small sample size, the halo fraction is higher for massive clusters, as expected if the radio-halo occurrence increases with $M_{500}$ in turbulent re-acceleration models.

\section{Discussion}
\label{sec:Discussion}

\subsection{Absence of radio relics at \texorpdfstring{$z > 1$}{z > 1}}

A noteworthy result of this work is that no radio relics are detected above the sensitivity limits of our survey in any of the 30 massive clusters at $z > 1$. Extended searches for large-scale peripheral radio emission in all fields yielded no relic candidates, in contrast to lower-redshift samples where radio relics are observed in approximately 20–30\% of massive, merging clusters and frequently accompany radio halos in major mergers \citep{2019SSRv..215...16V}. This non-detection does not necessarily imply that relics are intrinsically absent at high-redshift. Although relics are typically brighter than halos at a given redshift, at $z > 1$ their detectability is strongly affected by cosmological surface-brightness dimming, increased inverse-Compton (IC) losses, and geometric factors (such as shock orientation relative to the line of sight and location within the primary beam), which can reduce their observed surface-brightness below our current limits. Thus, in systems with lower merger energy or an unfavourable viewing angle, even strong relics may have their apparent surface-brightness reduced below our detection threshold at $z > 1$, and therefore remain undetected in our data.
\newline

The increased IC losses at higher redshift, due to the denser cosmic microwave background, strongly suppress the observed synchrotron brightness of relics. These features typically have steep spectra and are located in large-scale shock fronts in the outer parts of galaxy clusters, so their observed surface-brightness and detectability depend on observing frequency, shock strength, and projection effects. Theoretical models and recent cosmological simulations require strong, large-scale shocks from massive mergers to produce bright relics. Results from the TNG-Cluster simulation \citep{2024A&A...686A..55L} show that extremely large and luminous relics ($>$2~Mpc) are predominantly formed in the most massive systems ($M_{\rm 500c} > 8\times10^{14}\,\mathrm{M}_\odot$) and become increasingly rare at $z  > 1$. The simulated relic morphologies, luminosities, and detection rates agree well with current observations, indicating that the scarcity of relics in the MMDCS sample is consistent with expectations driven by a combination of shock strength, cluster mass, and cosmological evolution. Projection effects may also play a role, as relics viewed along the line of sight can appear projected near the cluster centre in imaging, making it challenging to distinguish and interpret these features \citep{2024A&A...686A..55L}.
\newline

In ACT‑CL~J0329.2–2330 \citep[P25]{2025A&A...698L..17S}, a nearby elongated source connects to the eastern side of the detected halo emission. Detailed imaging and morphological analysis in \cite{2025A&A...698L..17S} suggest that this feature is more likely associated with a head–tail radio galaxy than with a peripheral relic, although further studies (including magnetic field diagnostics) are required to definitively rule out a relic origin.

\subsection{Absence of radio mini-halos at \texorpdfstring{$z > 1$}{z > 1}}

No radio mini-halos are detected in any of the MMDCS clusters, despite the presence of several systems with bright central galaxies and significant SZ signal. This is in line with recent results indicating that strong cool cores, which are typically associated with mini-halos, become rarer towards high-redshift and may be disrupted by frequent mergers \citep{2025ApJ...987L..40H}. The lack of detected mini-halos at $z>1$ therefore likely reflects a combination of the evolving cool-core population and the limited surface-brightness sensitivity of current observations, rather than providing definitive evidence that mini-halos are intrinsically absent at early cosmic times.

\subsection{Evolution of surface-brightness with redshift}

Detecting diffuse cluster emission becomes increasingly challenging toward higher redshift because cosmological surface-brightness dimming scales as $(1+z)^4$. For the redshift interval probed here ($1.00 < z < 1.31$), this effect introduces more than an order-of-magnitude reduction in observed brightness relative to $z \sim 0.2$. Coupled with elevated IC losses, which also scale as $(1+z)^4$ for electron cooling and may steepen the observed radio spectrum, these effects explain the scarcity of luminous halos and relics at $z > 1$.
\newline

To quantify how this affects our survey sensitivity, we compute an analytic detection threshold in radio power as a function of redshift for a MeerKAT 1.283\,GHz observation, following the surface-brightness approach of \citet{2023A&A...672A..43C} and \citet{2025A&A...695A.215D}. We assume an exponential radio-halo brightness profile with e-folding radius $r_{\mathrm{e}} = 200$\,kpc and adopt a low-resolution noise level of $\sigma_{\mathrm{rms}} \simeq 13\,\upmu\mathrm{Jy\,beam^{-1}}$ and beam size $\theta_{\mathrm{beam}} \simeq 15\,\mathrm{arcsec}$ for the tapered MeerKAT 1.28\,GHz images. Following \citet{2023A&A...672A..43C}, the minimum integrated flux density detectable within a radius $3r_{\mathrm{e}}$ at redshift $z$ can be written as
\begin{equation}
    S_{\nu,\mathrm{lim}}^{3 r_{\mathrm{e}}}(z)
    = A\,\xi\,\sigma_{\mathrm{rms}}\,
      \frac{\theta_{\mathrm{e}}(z)}{\theta_{\mathrm{beam}}}
    \quad \mathrm{[mJy]} ,
\end{equation}
where $\theta_{\mathrm{e}}(z)$ is the angular size corresponding to $r_{\mathrm{e}}$, $\theta_{\mathrm{beam}}$ is the full width at half maximum of the low-resolution beam, $\xi$ is the adopted surface-brightness significance threshold, and $A$ is a numerical factor that encapsulates the assumed halo profile and the fraction of the total flux enclosed within $3r_{\mathrm{e}}$. We adopt $A = 4.44 \times 10^{-3}$ and $\xi = 3$, with the latter representing a conservative surface-brightness threshold for the MeerKAT 1.28\,GHz images. The corresponding limiting radio power at rest-frame 1.4\,GHz is obtained using Equation~\ref{eq:P14} and assuming spectral indices in the range
$\alpha = -1.0$ to $-1.8$ for the diffuse emission.
\newline

\begin{figure}
    \centering
    \includegraphics[width=0.48\textwidth]{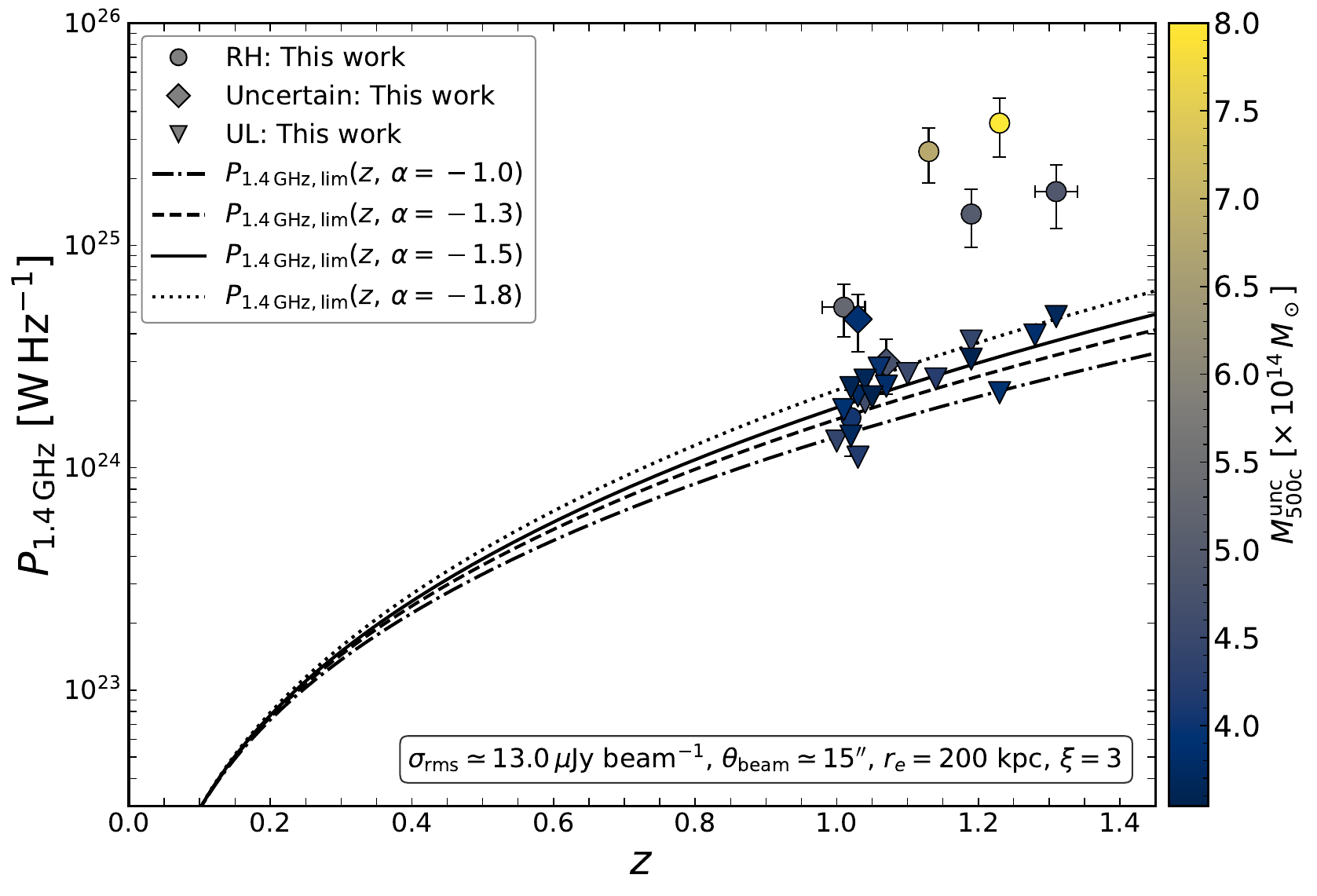}
     \caption{
    Limiting radio power at 1.4\,GHz as a function of redshift for a MeerKAT 1.28\,GHz observation. The black curves show the redshift evolution of the detection threshold $P_{1.4\,\mathrm{GHz,lim}}(z,\alpha)$ for a model radio halo with an exponential brightness profile and e-folding radius $r_{\mathrm{e}}=200$\,kpc, assuming spectral indices $\alpha=-1.8$ (dotted), $\alpha=-1.5$ (solid), $\alpha=-1.3$ (dashed), and $\alpha=-1.0$ (dash-dotted). Symbols show the MMDCS clusters, where circles mark confirmed radio halos, diamonds mark uncertain diffuse detections, and downward-pointing triangles are $3\sigma$ upper limits. Marker colour encodes $M_{\rm 500c}^{\mathrm{unc}}$.
    }
    \label{fig:P14_z_detection_limit}
\end{figure}

Figure~\ref{fig:P14_z_detection_limit} shows the resulting curves $P_{1.4\,\mathrm{GHz,lim}}(z,\alpha)$ over the range $0 < z < 1.4$, together with the measured powers and upper limits for the MMDCS clusters. Across $1 < z < 1.3$ the MeerKAT detection threshold lies at a few $\times 10^{23}$ to $\sim 10^{24}\,\mathrm{W\,Hz^{-1}}$, depending on the assumed spectral index, and the bulk of our upper limits lie on or just below these curves. This indicates that radio halos with intrinsic powers comparable to those of lower-redshift systems in moderately massive clusters would frequently fall below our detection limit once cosmological dimming and IC losses are taken into account.
\newline

In contrast, the detected halos in the MMDCS sample lie close to or above the predicted threshold. The fact that several of these halos sit well above the nominal detection limit, with powers comparable to, or exceeding, those of the most luminous intermediate-redshift systems, indicates that only clusters with unusually large turbulent energy densities and relatively strong magnetic fields can sustain such bright synchrotron emission at $z \sim 1$. In these environments, enhanced turbulent energy and amplified magnetic fields can partly counteract the stronger inverse-Compton cooling and the  $(1+z)^{-4}$ surface-brightness dimming that would otherwise suppress radio halos at high redshift \citep{2021NatAs...5..268D,2025A&A...695A.215D,2025A&A...698L..17S}. The lower detection fraction that we observe at $z \sim 1$ relative to intermediate-redshift samples is well accounted for by the combined effects of cosmological dimming, spectral steepening, and the finite sensitivity of our MeerKAT observations. Under these conditions only the most luminous or centrally concentrated halos remain detectable, while a substantial population of fainter diffuse sources is expected to be missed.
\newline

This analysis demonstrates that selection effects and surface-brightness dimming play a dominant role in shaping the observed radio-halo occurrence rate at $z>1$, and must be taken into account when detection fractions are compared across redshift. It also highlights the need for deeper, lower-frequency observations in order to uncover the full population of high-redshift cluster halos and relics whose peak synchrotron output is shifted below 1\,GHz.
\newline

Recent deep LOFAR observations have revealed a candidate radio mini-halo in the most distant cool-core cluster to date at $z=1.709$ \citep{2025ApJ...987L..40H}, whose physical conditions challenge conventional models of IC losses alone. The observed brightness and power of this mini-halo imply that, alongside strong magnetic fields and enhanced turbulence, clumping of the ICM and cosmic ray populations can boost synchrotron emission. \cite{2025ApJ...987L..40H} find that correlations among thermal gas, cosmic rays, and magnetic fields may partially compensate for cosmological dimming, allowing detectable diffuse emission even at extreme redshifts.

\subsection{Comparison with previous studies}

\begin{figure}
    \centering
    \includegraphics[width=\linewidth]{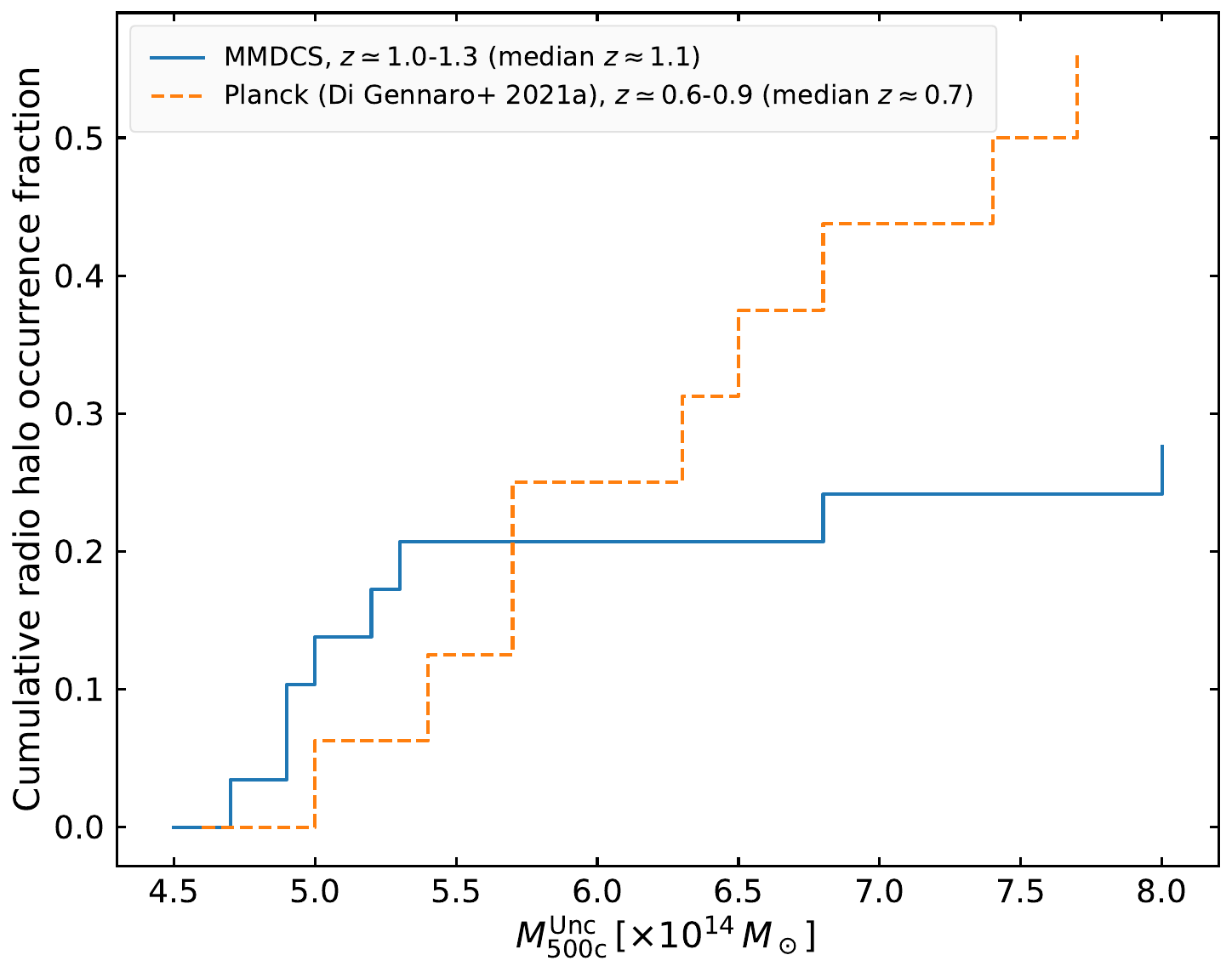}
    \caption{
    Cumulative radio halo occurrence fraction as a function of $M_{\rm 500c}^{\rm {Unc}}$
    for the mass-limited ($M_{\rm 500c}^{\rm {Unc}} \ge 4.5\times10^{14}\,M_\odot$) MMDCS
    sample at $z \simeq 1.0$--$1.3$ (solid blue line) and the
    \citet{2021NatAs...5..268D} \textit{Planck} sample at
    $z \simeq 0.6$--$0.9$ (dashed orange line). For each sample, the
    cumulative fraction is computed above the indicated mass threshold.
    At fixed mass, the \textit{Planck} clusters exhibit a higher halo fraction
    than the MMDCS clusters, suggesting a decline in the observable halo
    population towards $z > 1$.}
    \label{fig:halo_fraction_mass}
\end{figure}

For a quantitative comparison with previous studies, we restrict both our sample and literature samples to comparable mass ranges. Over the mass interval $M_{\rm 500c}^{\rm {Unc}} \geq 4.5\times10^{14}\,M_\odot$, our high-redshift subsample contains 29 clusters, of which 8 host radio halos, giving a halo fraction of $f_{\mathrm{RH}} = 0.28$. Recent MeerKAT and LOFAR studies report a decrease towards $z > 1$, dropping to $< 10$\% in the highest-redshift bins \citep{2025A&A...695A.215D,2024ApJ...966...38R,2021MNRAS.504.1749K}. In the \citet{2021NatAs...5..268D} \textit{Planck} sample at $0.6 < z < 0.9$, the same mass cut yields 16 clusters with 9 halos, corresponding to $f_{\mathrm{RH}} = 0.56$. Figure~\ref{fig:halo_fraction_mass} compares the cumulative halo occurrence fractions as a function of $M_{\rm 500c}^{\rm {Unc}}$ for these two mass-limited samples and shows that the halo fraction in MMDCS is systematically lower than in the intermediate-redshift \textit{Planck} sample across the overlapping mass range.
\newline

This difference is qualitatively consistent with the expected redshift dependence of halo occurrence, in which stronger IC losses and cosmological surface-brightness dimming suppress the synchrotron emission in high-redshift clusters. These results point to a lower observed incidence of radio halos beyond $z \sim 1$ even after accounting for mass, although many clusters in our sample could still host halos consistent with the \citet{2007MNRAS.378.1565C} scaling relation that remain undetected at the current MeerKAT $L$-band sensitivity. At the same time, our MeerKAT 1.28\,GHz observations are less sensitive to very steep-spectrum halos than LOFAR studies targeting higher-redshift clusters, which typically find halo fractions of order $\sim 10$ per cent in their highest-redshift bins \citep{2025A&A...695A.215D}. Taken together, these findings imply that the apparent drop in the observed radio halo population above $z \sim 1$ is mainly due to cosmological surface-brightness dimming, stronger IC losses, observing frequency, and survey sensitivity, and should therefore be regarded as a lower limit on the halo occurrence fraction rather than as unambiguous evidence for an intrinsic absence of halo emission in high-redshift clusters.

\subsection{Implications for cluster physics and future studies}

Our findings have significant implications for models of particle acceleration, magnetic field amplification, and the dynamical state of the ICM in the early universe. The clear division between radio halo detections and non-detections, as presented in Section \ref{sec:non-detections} and the left panel of Figure~\ref{fig:scaling_combined}, demonstrates that turbulence-driven re-acceleration and merging remain the principal mechanisms for halo formation. However, detection becomes less frequent and more dependent on cluster conditions at high-$z$. The lack of relics, the prevalence of compact AGN, varied radio source environments, and the influence of calibration artefacts illustrate the methodological challenges in searching for faint, extended radio emission with today’s instruments.
\newline

Taken together, these results indicate that the observability of turbulence-driven particle acceleration and large-scale magnetic field amplification in the ICM declines rapidly at $z > 1$, primarily because IC losses and cosmological surface-brightness dimming strongly suppress the detectable synchrotron signal at high-redshift. Observable radio halos at high-redshift are confined to the most massive and violently merging clusters, indicating that only clusters with exceptionally large turbulent energy densities and strong magnetic fields can produce visible synchrotron emission during this epoch. The observed decrease in halo and relic detections with redshift therefore offers initial, but still sensitivity-limited, constraints on the cosmological evolution of non-thermal plasma conditions in clusters, and provides useful guidance for models of particle acceleration and magnetic field growth.
\newline

In summary, the MMDCS provides rigorous constraints on diffuse radio emission in the most massive clusters at $z > 1$, confirming that most of these systems appear radio-quiet within current sensitivity limits. The key limiting factors such as surface-brightness dimming, IC losses, merger-driven shock strengths, the role of filaments, and observational depth present clear opportunities for future investigation. Continued improvements in calibration, imaging techniques, and telescope capabilities, together with complementary follow-up at other wavelengths, will be essential for uncovering the full physical picture of non-thermal processes in galaxy clusters throughout the history of the universe.

\section{Conclusions}
\label{sec:Conclusions}

We have presented results from the MeerKAT Massive Distant Cluster Survey (MMDCS), which targets the 30 most massive SZ-selected galaxy clusters at $z > 1$. Using deep MeerKAT $L$-band  observations, we conducted a systematic search for diffuse radio emission and derived stringent upper limits for non-detections, leveraging injection simulations that account for both sensitivity and artefacts.
\newline

Diffuse radio emission consistent with halos is detected in 8 clusters, corresponding to a $27\%$ detection rate. This includes 4 prominent and 4 faint radio halos, with the remainder of the sample yielding non-detections down to upper limits among the deepest yet reported for these redshifts. In the remaining 21 clusters ($70\%$) with reliable data, no diffuse emission is found, and $3\sigma$ upper limits on the 1.4 GHz radio power are established via the \texttt{MUVIT} injection method. No cluster-scale radio relics or mini-halos were detected in any system.
\newline

The detected radio halos scatter around the $P_{1.4\,\mathrm{GHz}}$–$M_{\rm 500c}^{\rm {Unc}}$ correlation measured for our sample, and a Bayesian regression using the detected and uncertain radio halos yields a slope and normalisation that, despite lying systematically above the low-redshift relations, remain statistically consistent with them within the large uncertainties. Most upper limits lie below this relation and populate the lower envelope of the $P_{1.4\,\mathrm{GHz}}$–$M_{\rm 500c}^{\rm {Unc}}$ plane, in line with the trend seen in previous samples that combine halos and non-detections. Several clusters, however, still have limits that allow halos with powers comparable to the correlation, so a fraction of the population may host undetected, moderately faint halos. These findings indicate that cosmological surface-brightness dimming, enhanced IC losses, and the finite depth of the MeerKAT data strongly influence the observed detection fraction at $z>1$, and that the underlying power–mass relation itself does not show clear evidence for evolution.
\newline

Our results provide key constraints for theoretical models of non-thermal processes in the ICM. They emphasize the importance of even deeper and lower-frequency radio surveys with next-generation facilities to fully explore the faint end of the radio halo (and relic) population at the highest redshifts. Integrating future radio datasets with multi-wavelength X-ray and optical/infrared studies will be essential for unrevealing the interplay between gas dynamics, particle acceleration, and magnetic field amplification during cluster formation.


\section*{Acknowledgements}
We acknowledge the use of the ilifu cloud computing facility \url{www.ilifu.ac.za}, a partnership between the University of Cape Town, the University of the Western Cape, Stellenbosch University, Sol Plaatje University and the Cape Peninsula University of Technology. The ilifu facility is supported by contributions from the Inter-University Institute for Data Intensive Astronomy (IDIA – a partnership between the University of Cape Town, the University of Pretoria and the University of the Western Cape), the Computational Biology division at UCT and the Data Intensive Research Initiative of South Africa (DIRISA). This work made use of the CARTA (Cube Analysis and Rendering Tool for Astronomy) software (DOI \url{10.5281/zenodo.3377984} - \url{https://cartavis.github.io}). The MeerKAT telescope is operated by the South African Radio Astronomy Observatory (SARAO), which is a facility of the National Research Foundation, an agency of the Department of Science and Innovation. The financial assistance of the SARAO
towards this research is hereby acknowledged. KK acknowledges funding support from SARAO under grant UID 97930. S.P.S.\ acknowledges financial support from the National Research Foundation (NRF; grant TTK240320210161), and M.H., and U.S.\ acknowledge support from the same funding agency.


\section*{Data Availability}

The raw MeerKAT datasets analysed in this paper can be accessed from the SARAO archive using proposal IDs SCI-20220822-MH-01 and SCI-20230907-MH-02 (\url{https://archive.sarao.ac.za}). The reduced MeerKAT images and derived source catalogues will be available at the CDS via anonymous ftp to \href{ftp://cdsarc.u-strasbg.fr}{cdsarc.u-strasbg.fr (130.79.128.5)} or via \url{https://cdsarc.cds.unistra.fr/viz-bin/cat/J/MNRAS/XXX/YYY} upon publication.



\bibliographystyle{mnras}
\bibliography{reference} 




\appendix
\section{Point-source subtraction diagnostics}
\label{app:subtraction_quality}

\begin{figure*}
    \centering
    \includegraphics[width=\textwidth]{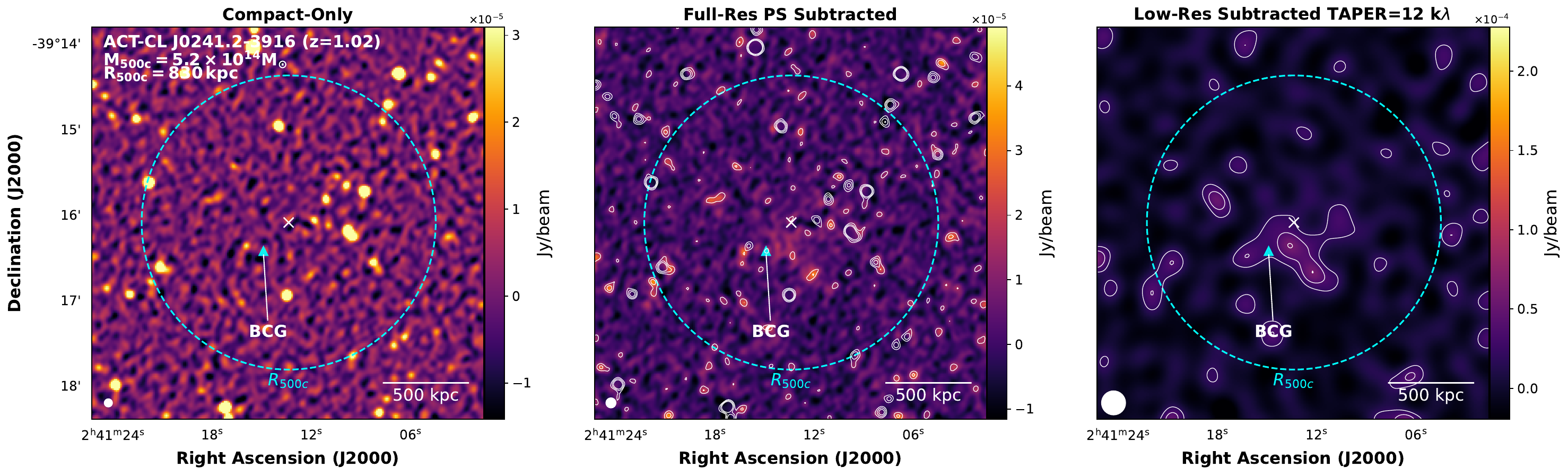}
    \caption{
    Compact and diffuse radio emission in ACT-CL~J0241.2$-$3916. The position of the BCG is indicated by a cyan triangle, while the white cross marks the location of the ACT SZ peak; the dashed cyan circle denotes $R_{\rm 500c}$.
     Left: high resolution  MeerKAT 1.28\,GHz image showing discrete radio sources.
     Middle: full-resolution image after point-source subtraction, with white contours from the compact-only map overlaid at levels of $\sigma_{\mathrm{HR}} \times [3,6,10]$ (where the $1\sigma$ noise level is $3.8\,\upmu\mathrm{Jy\,beam^{-1}}$), highlighting the locations of the subtracted sources. 
    Right: tapered, low-resolution point-source–subtracted image used for diffuse-emission detection, with its own contours overlaid at levels of $\sigma_{\mathrm{LR}} \times [3,6,10]$; the local $1\sigma$ rms is $5.5\,\upmu\mathrm{Jy\,beam^{-1}}$.}
    \label{fig:subtraction_0241}
\end{figure*}

\begin{figure*}
    \centering
    \includegraphics[width=\textwidth]{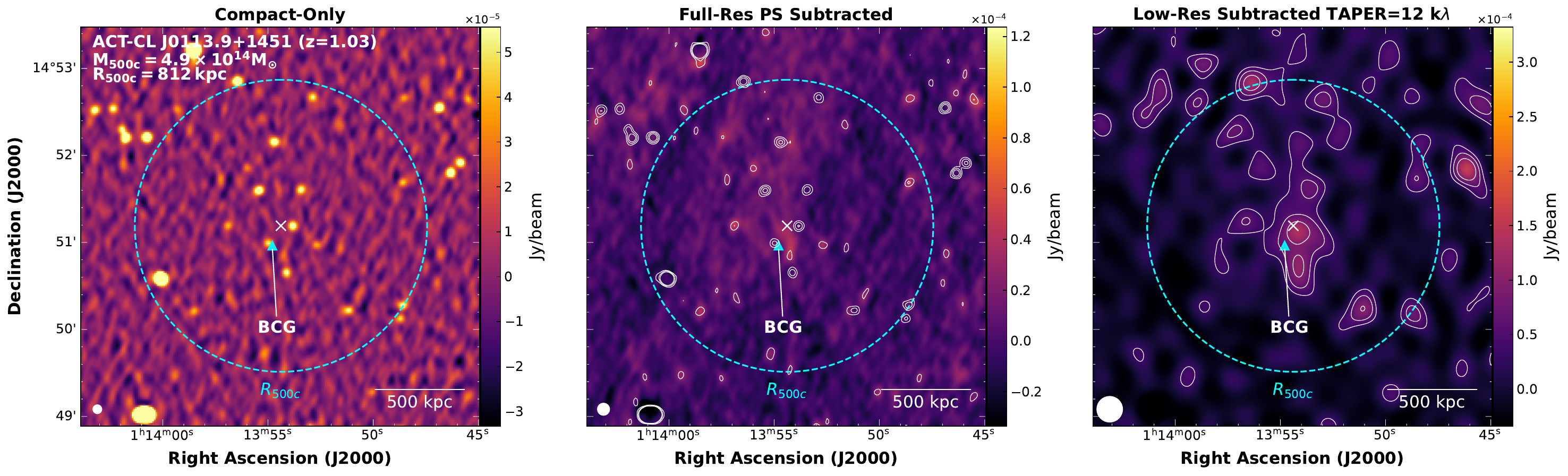}
    \caption{
     Compact and diffuse radio emission in ACT-CL~J0113.9+1451. The position of the BCG is indicated by a cyan triangle, while the white cross marks the location of the ACT SZ peak; the dashed cyan circle denotes $R_{\rm 500c}$.
    Left: HR MeerKAT 1.28\,GHz image showing discrete radio sources.
    Middle: full-resolution image after point-source subtraction, with white contours from the compact-only map overlaid at levels of $\sigma_{\mathrm{HR}} \times [3,6,10]$ (where the $1\sigma$ noise level is $7.0\,\upmu\mathrm{Jy\,beam^{-1}}$), highlighting the locations of the subtracted sources. 
    Right: tapered, low-resolution point-source–subtracted image used for diffuse-emission detection, with its own contours overlaid at levels of $\sigma_{\mathrm{LR}} \times [3,6,10]$; the local $1\sigma$ rms is $9.4\,\upmu\mathrm{Jy\,beam^{-1}}$.}
    \label{fig:subtraction_0113}
\end{figure*}

Here we illustrates the quality of the compact-source subtraction for the two clusters hosting the newly identified diffuse-emission candidates discussed in Section~\ref{sec:Clusters with detected diffuse emission}. Figures~\ref{fig:subtraction_0241} and \ref{fig:subtraction_0113} compare the high-resolution MeerKAT 1.28\, GHz images, the point-source--subtracted full-resolution images, and the tapered low-resolution images used in the diffuse-emission search. These panels demonstrate that the residual extended emission discussed in the main text is not simply driven by unsubtracted compact sources, although low-level residuals remain in the vicinity of some bright objects.

\section{SZ--radio overlays for the halo candidates}

\label{app:SZmaps}

\begin{figure*}
    \centering
    \includegraphics[width=0.48\textwidth]{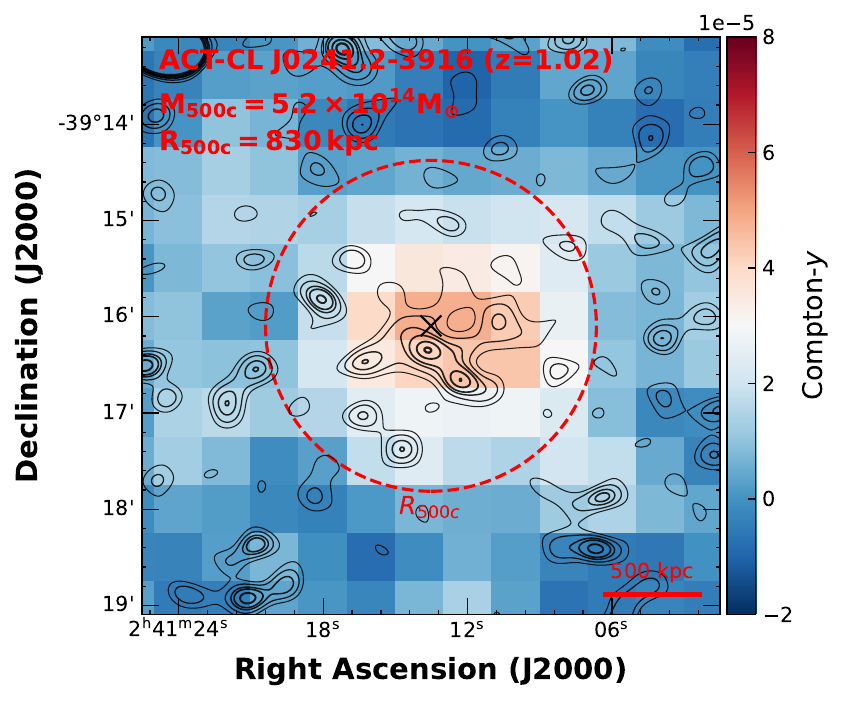}
    \includegraphics[width=0.48\textwidth]{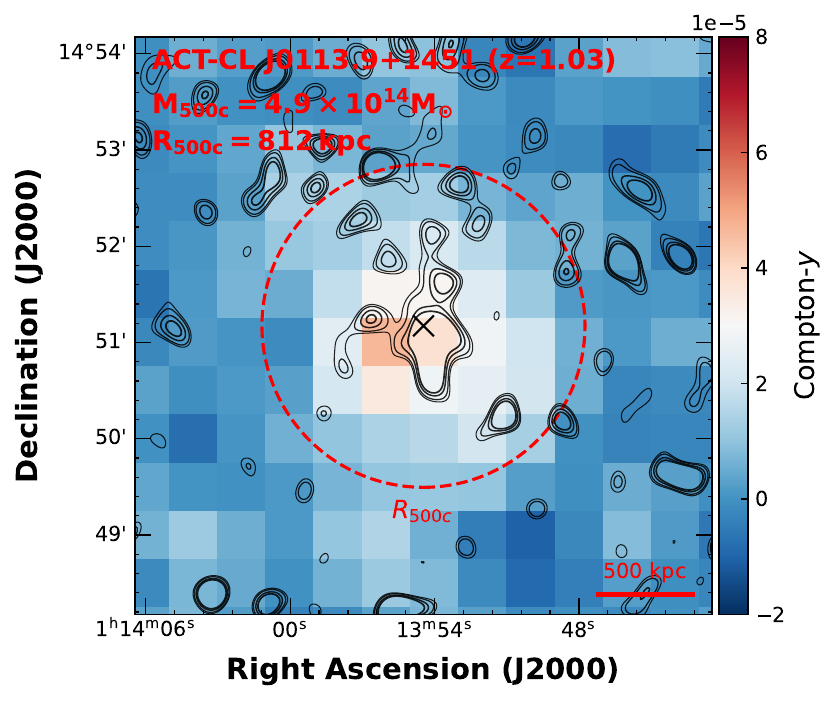}
    \caption{\textit{Planck}+ACT DR6 Compton-$y$ maps of ACT-CL~J0241.2$-$3916 (left) and
    ACT-CL~J0113.9+1451 (right), shown in colour, with MeerKAT
    1.28\,GHz low-resolution radio contours overlaid. The colour scale
    represents the dimensionless Compton-$y$ parameter, and the contours
    trace the diffuse radio emission. The black cross marks the ACT SZ
    peak, and the dashed red circle indicates $R_{\rm 500c}$. In both clusters
    the diffuse radio emission is broadly centred on, and comparable in
    extent to, the SZ signal within $R_{\rm 500c}$, consistent with a
    cluster-scale radio halo.}
    \label{fig:SZ_overlays}
\end{figure*}

This appendix presents zoomed-in ACT Compton-$y$ maps for the two clusters that host newly detected diffuse radio emission, in order to illustrate the correspondence between the thermal SZ signal and the MeerKAT diffuse radio emission. Figure~\ref{fig:SZ_overlays} shows the central regions of ACT-CL~J0241.2$-$3916 and ACT-CL~J0113.9+1451, with a fixed colour scale in units of $y\times10^{6}$ chosen to enhance the contrast of the SZ structure in the cluster cores and facilitate comparison with the radio morphology.

\section{Mock halo injection}

\begin{figure*}
\centering
\begin{tabular}{ccc}
    \includegraphics[width=0.355\textwidth]{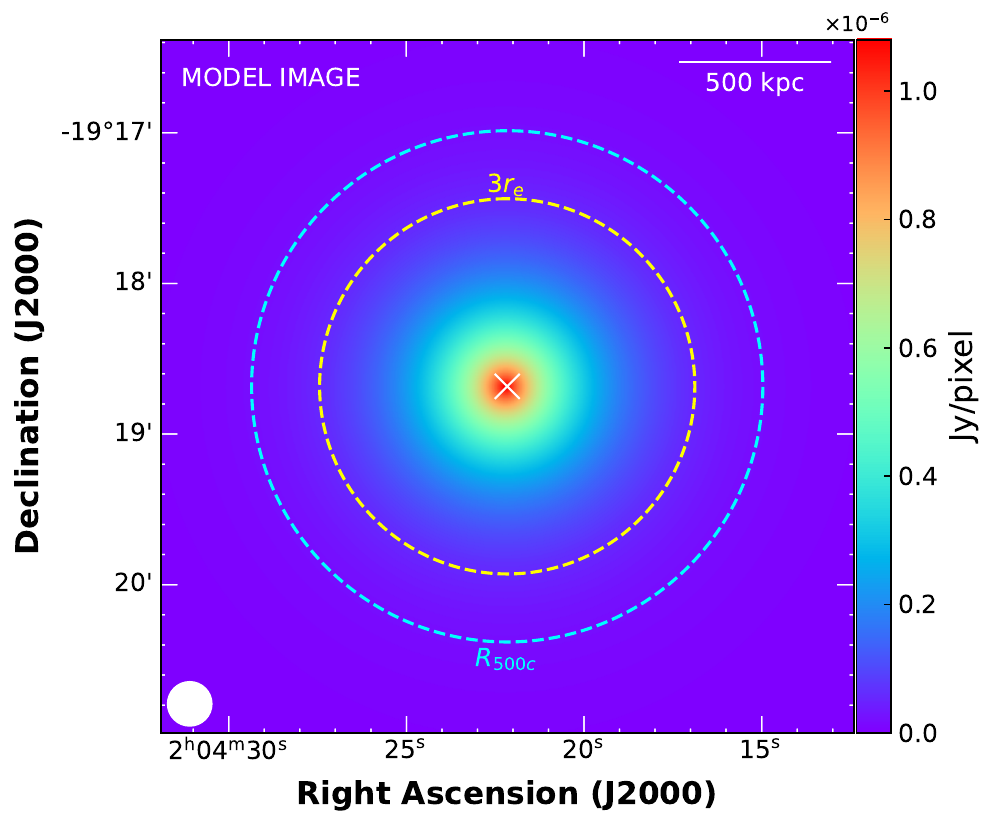} &
    \includegraphics[width=0.29\textwidth]{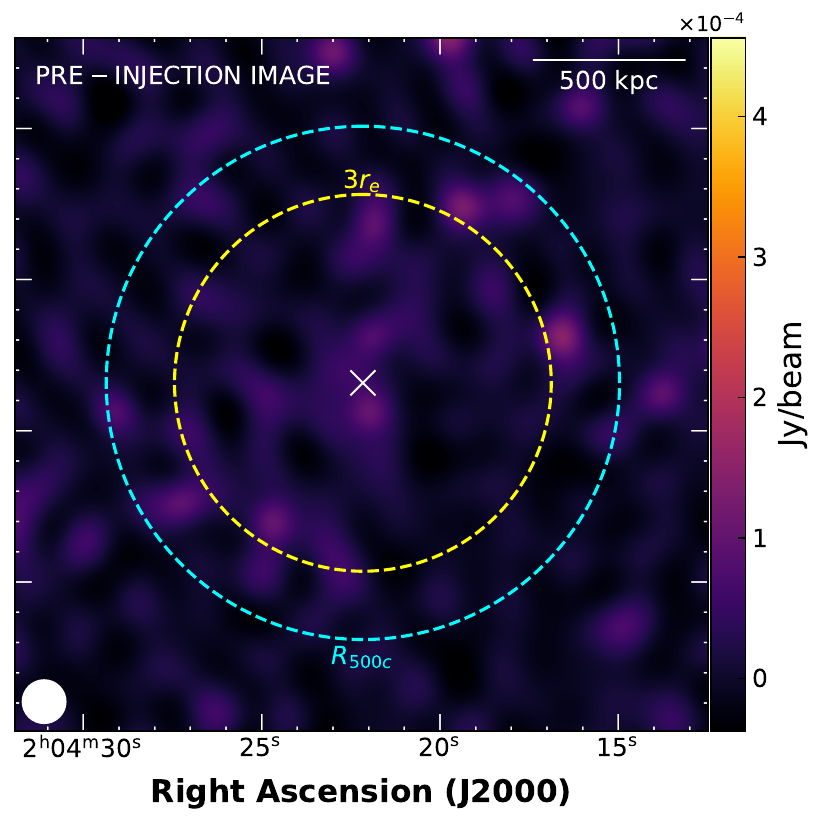} &
    \includegraphics[width=0.29\textwidth]{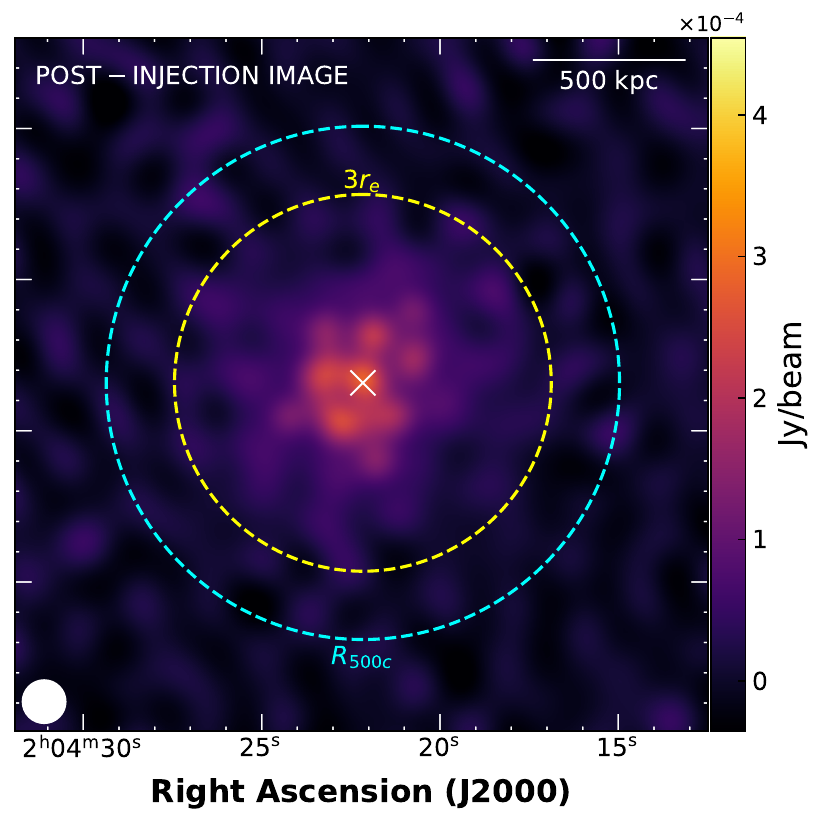}\\

\end{tabular}
\caption{Illustration of the mock halo injection. Left: model image at 1.28\,GHz of the injected halo, assuming an exponential surface-brightness profile. Middle: tapered, low-resolution image before injection. Right: corresponding image after halo injection. The yellow dashed circle, centred at (RA$_{\mathrm{inj}}$, Dec$_{\mathrm{inj}}$) and indicated by a white cross, has a radius $r = 3r_{e}$, and the cyan dashed circle indicates $R_{\rm 500c}$.}
\label{fig:injection_process}
\end{figure*}

\begin{figure*} 
\centering
\begin{tabular}{ccc}
    \includegraphics[width=0.29\textwidth]{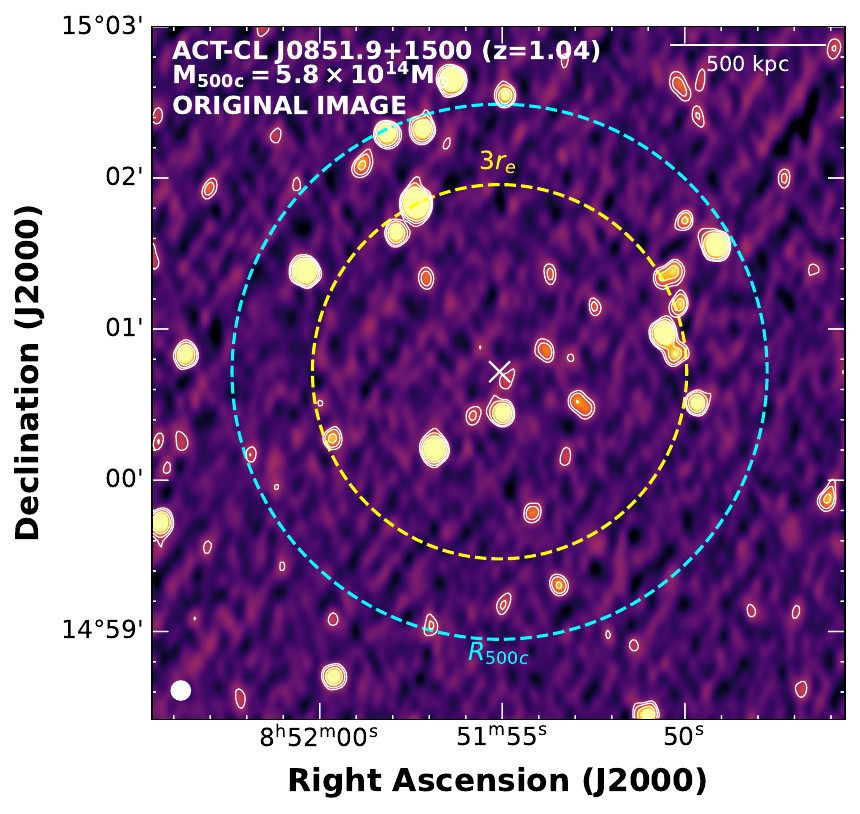} &
    \includegraphics[width=0.245\textwidth]{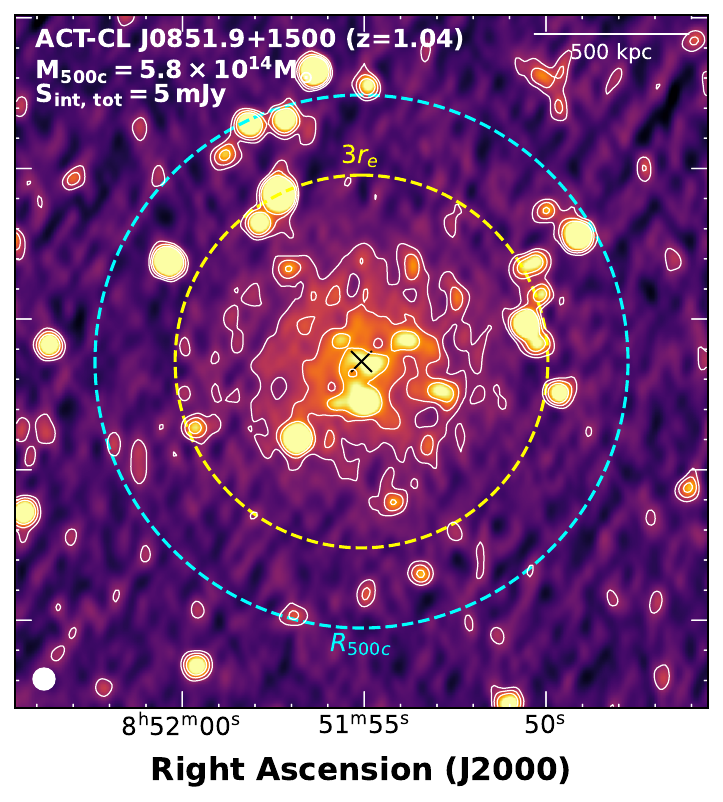} &
    \includegraphics[width=0.286\textwidth]{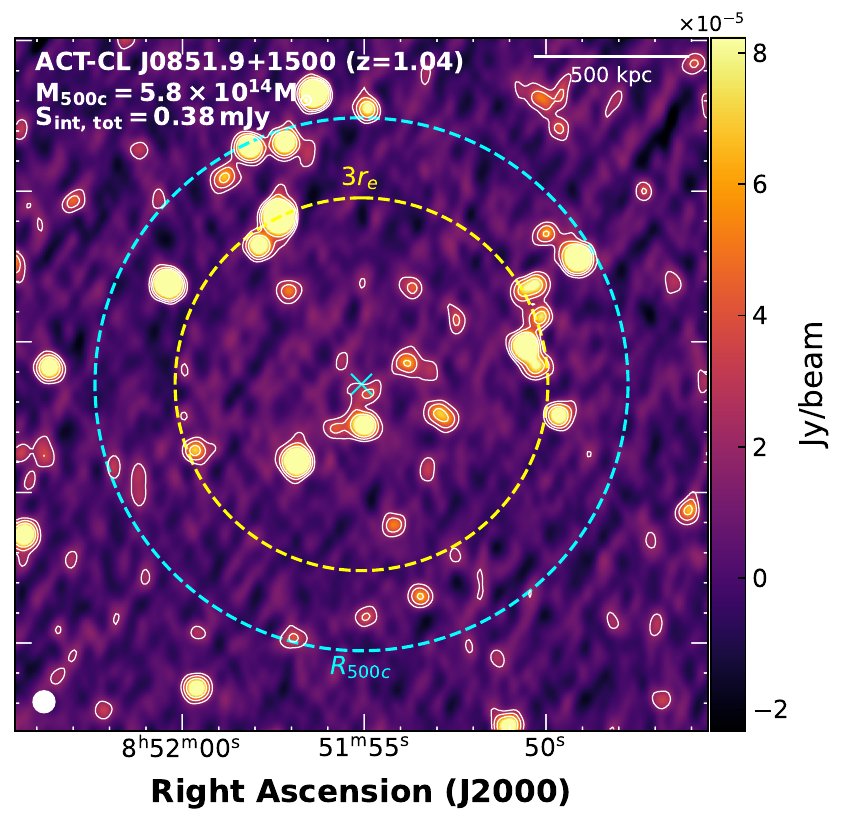}\\

    \includegraphics[width=0.29\textwidth]{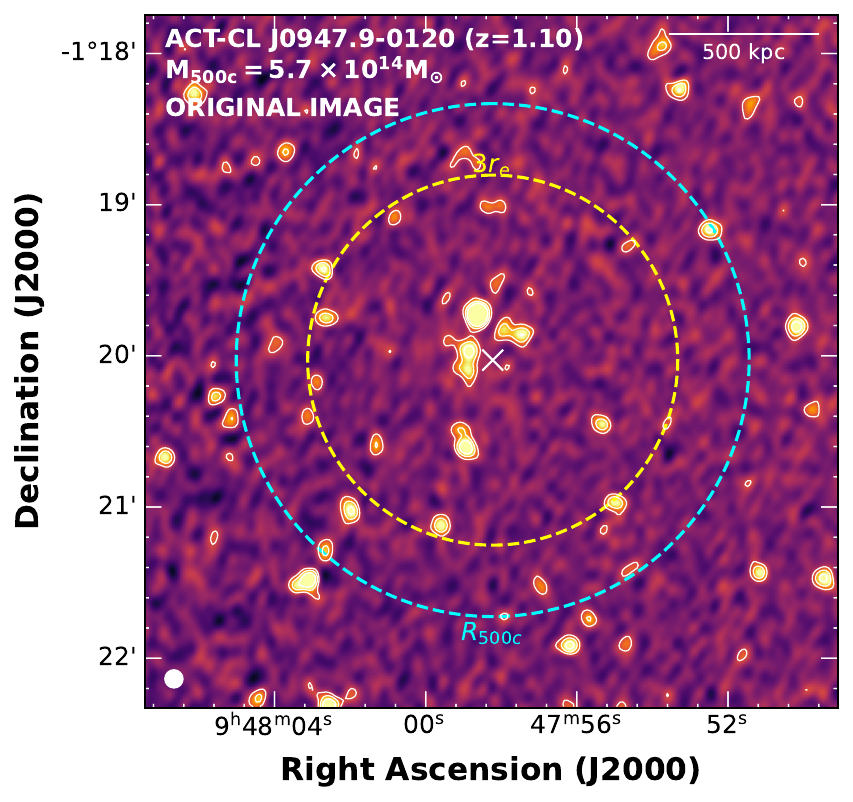} &
    \includegraphics[width=0.245\textwidth]{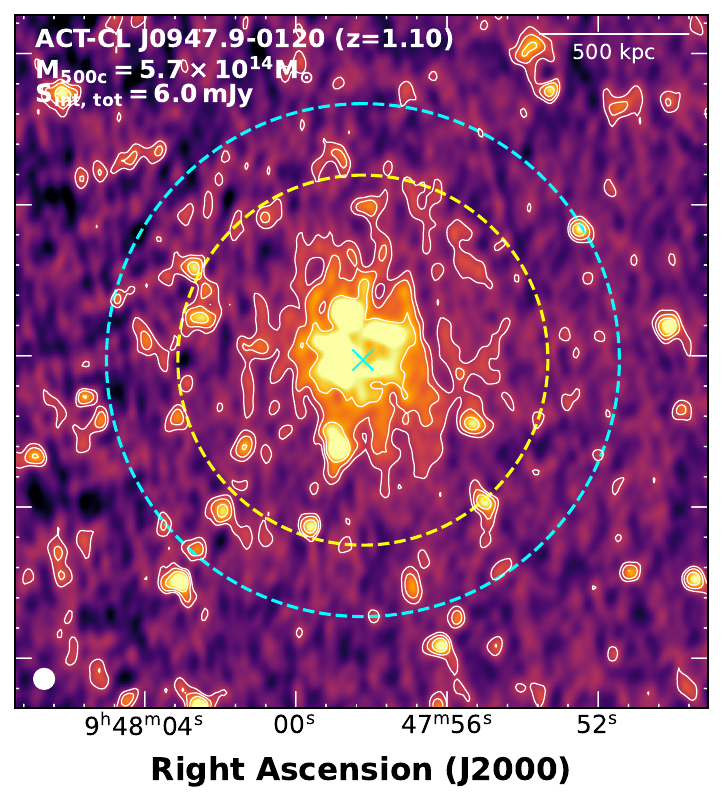} &
    \includegraphics[width=0.286\textwidth]{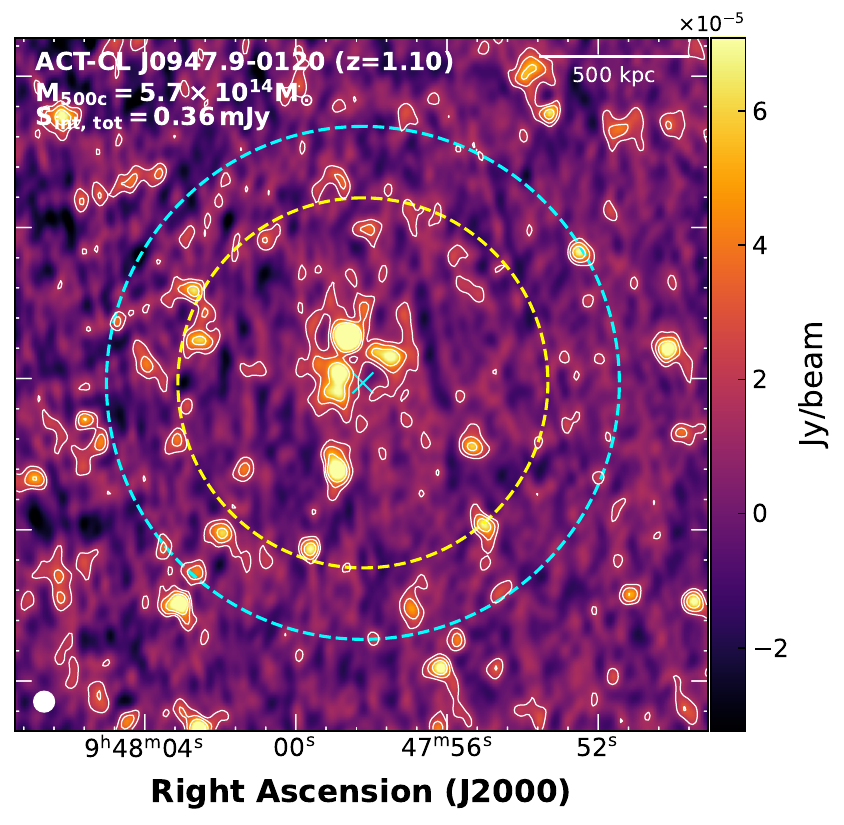}\\

    \includegraphics[width=0.293\textwidth]{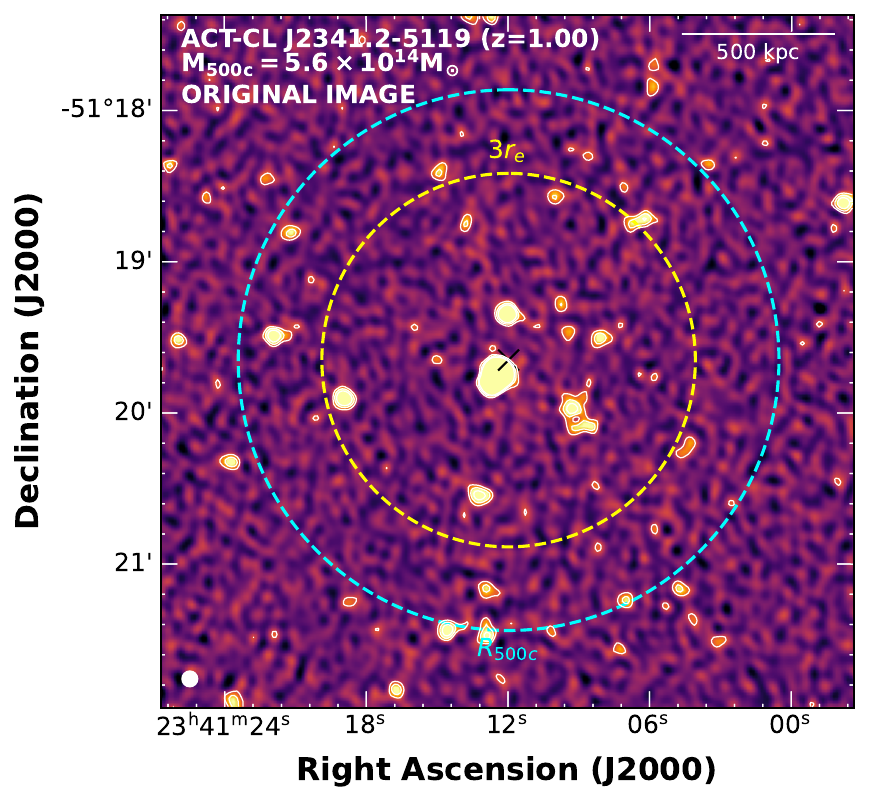} &
    \includegraphics[width=0.245\textwidth]{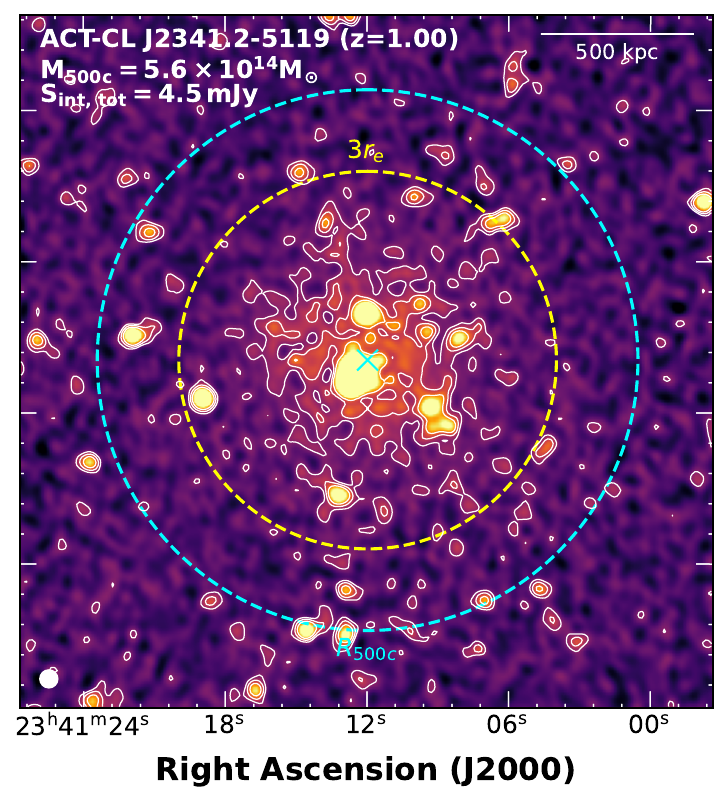} &
    \includegraphics[width=0.286\textwidth]{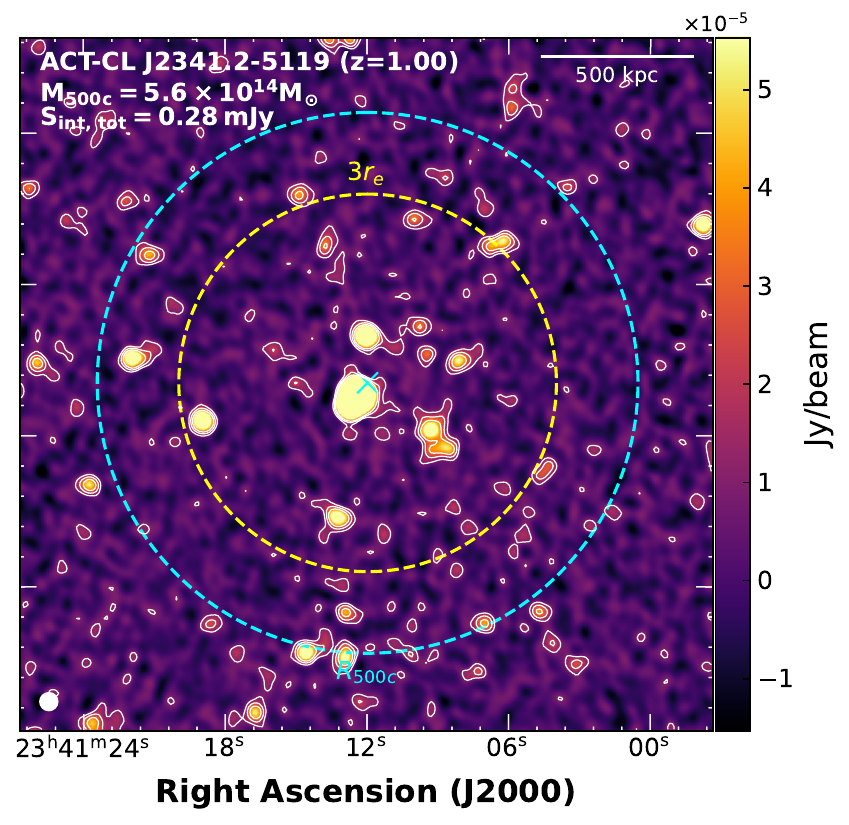} \\

    \includegraphics[width=0.30\textwidth]{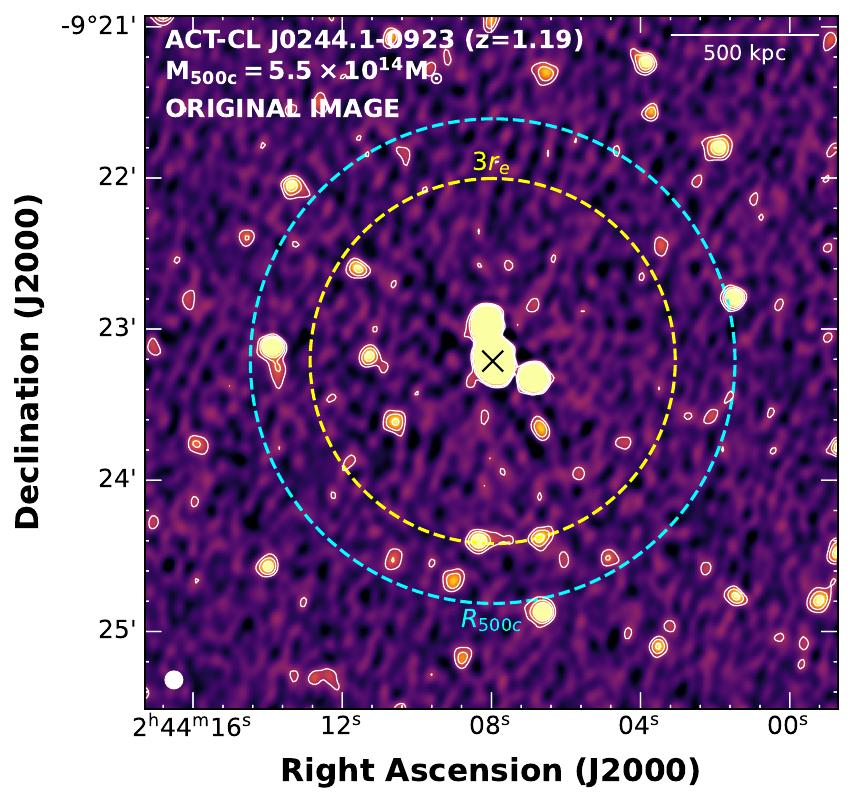} &
    \includegraphics[width=0.26\textwidth]{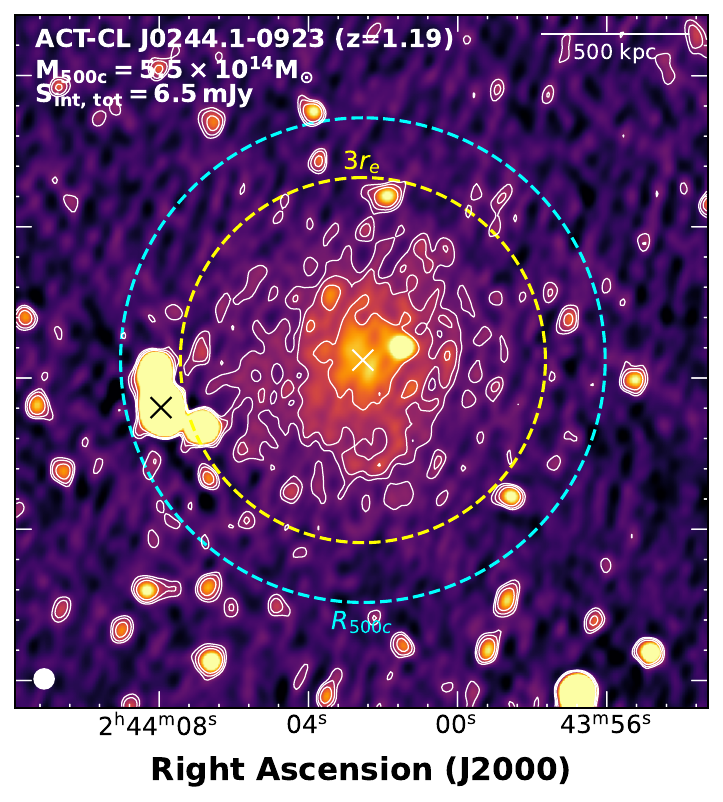} &
    \includegraphics[width=0.29\textwidth]{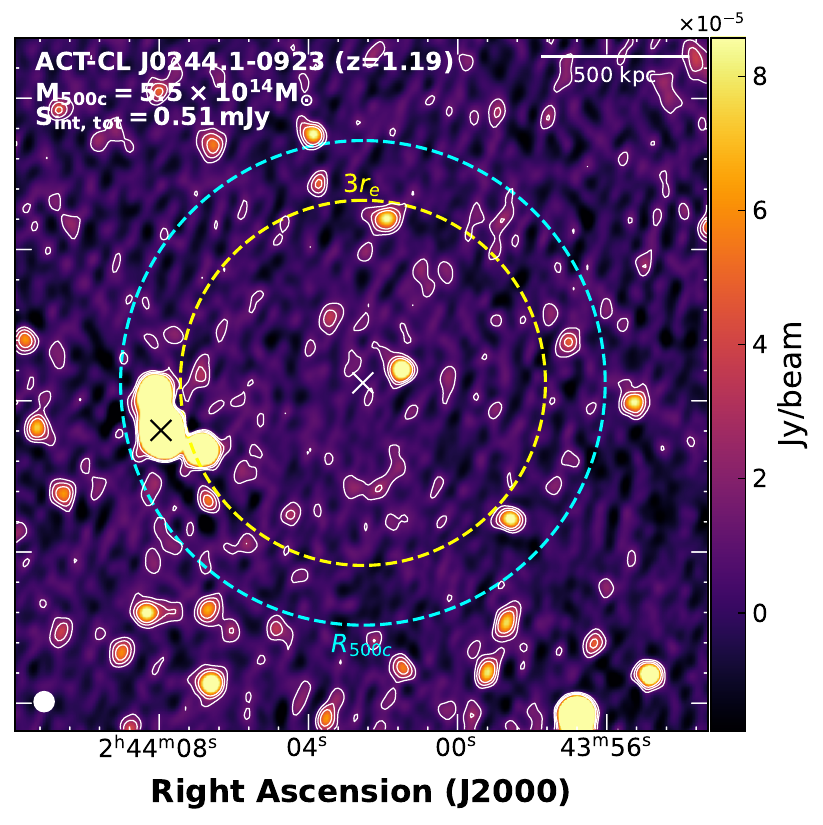} \\
\end{tabular}
\caption{Rows correspond to clusters ACT-CL J0851, J0947, J2341, and J0244. For each cluster, the left panel shows the original MeerKAT 1.28~GHz full-resolution image, with contours starting at $3\sigma$. The middle and right panels show the results of mock halo injections at different total flux densities. The yellow circle, centred at (RA$_{\mathrm{inj}}$, Dec$_{\mathrm{inj}}$), indicates a radius of $3r_e$, and the upper limits are obtained with $S_{\rm inj, tot} = 0.38$, $0.36$, $0.28$, and $0.51$ mJy, respectively.
}
\label{fig:mockhalo}
\end{figure*}

\begin{figure*} 
\centering
\begin{tabular}{ccc}

    \includegraphics[width=0.285\textwidth]{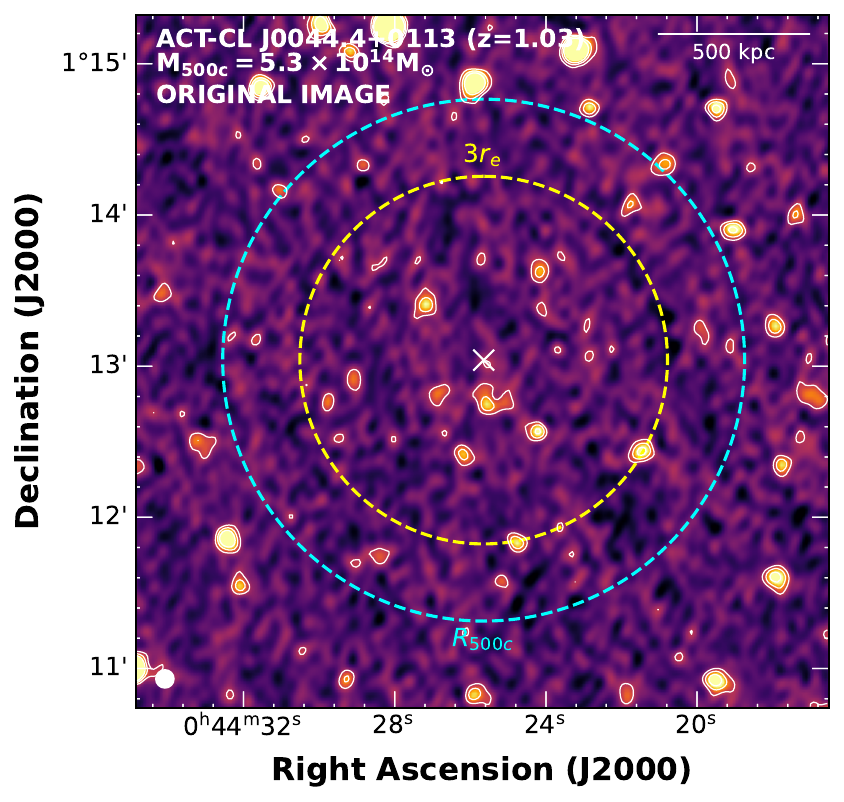} &
    \includegraphics[width=0.245\textwidth]{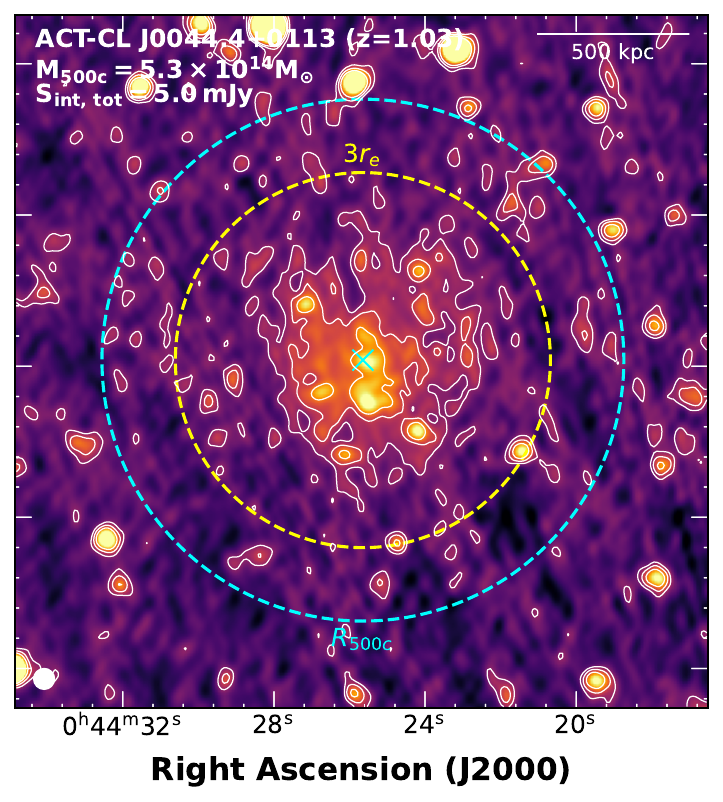} &
    \includegraphics[width=0.285\textwidth]{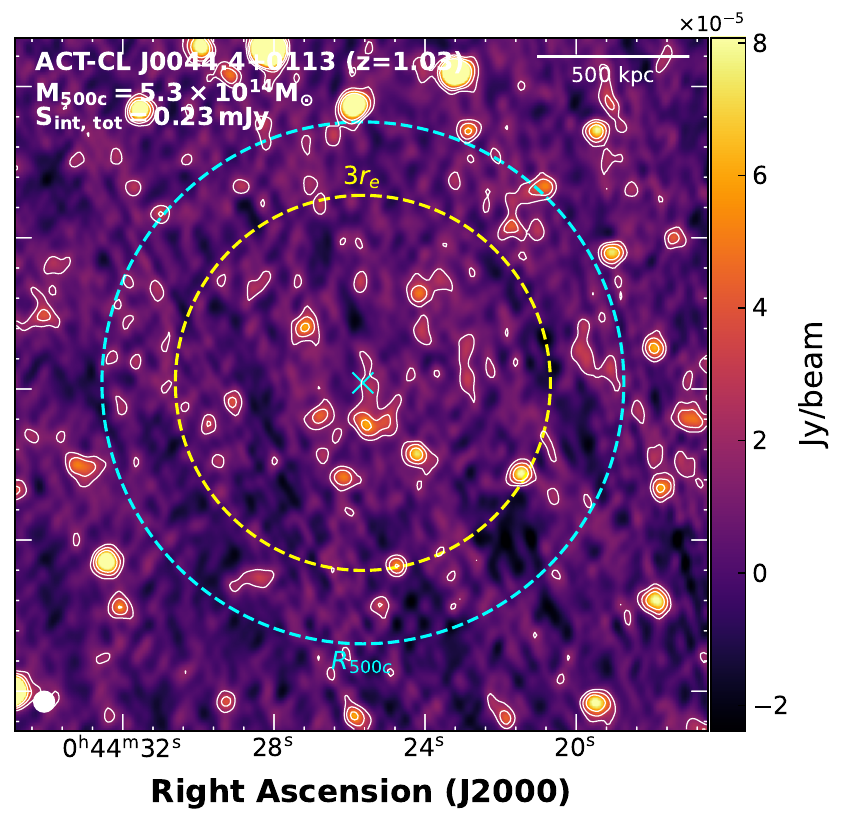}\\

    \includegraphics[width=0.29\textwidth]{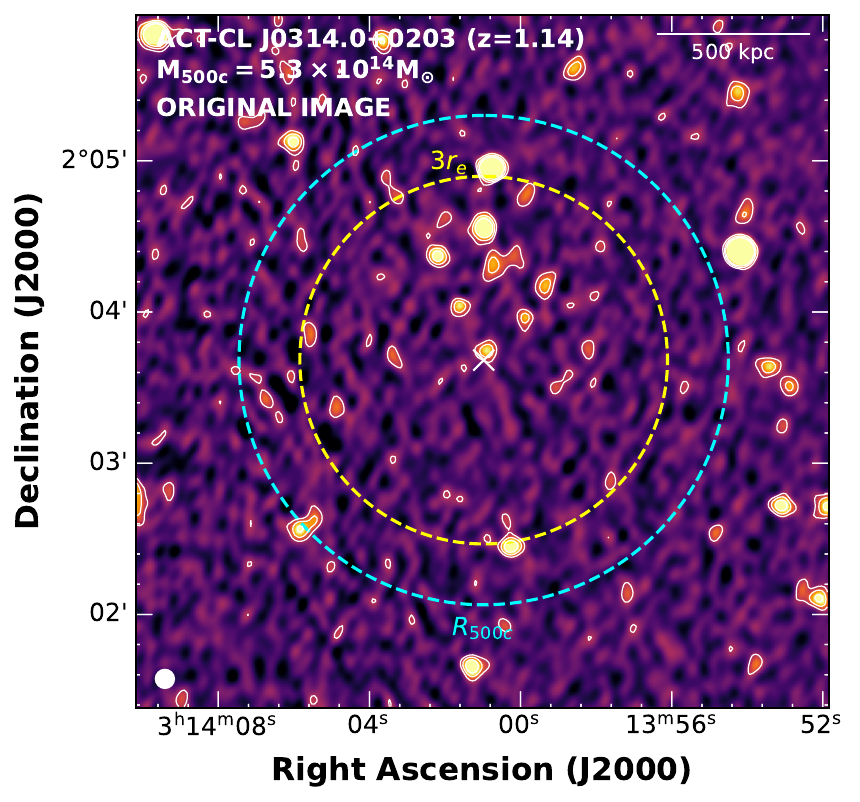} &
    \includegraphics[width=0.25\textwidth]{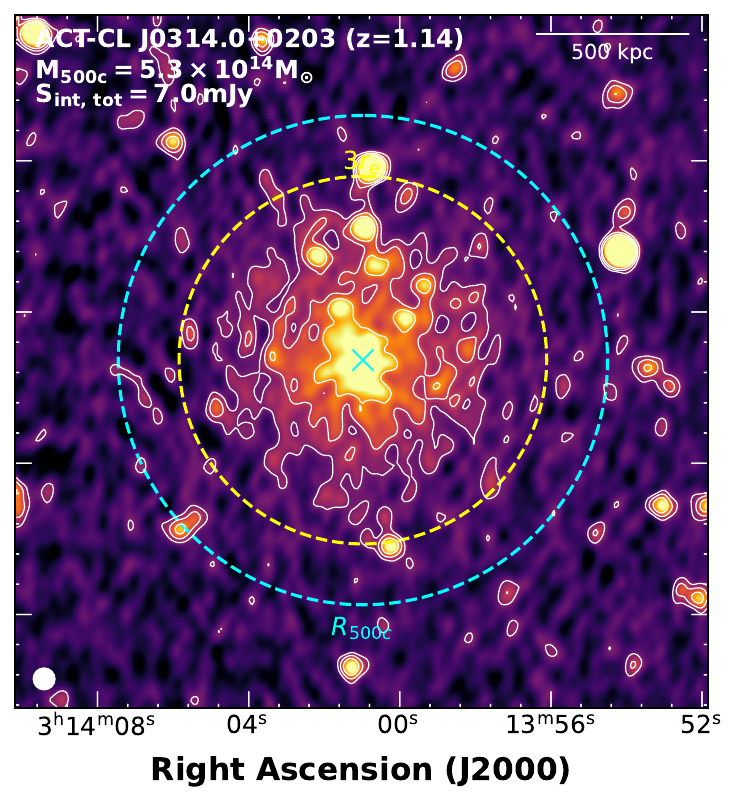} &
    \includegraphics[width=0.28\textwidth]{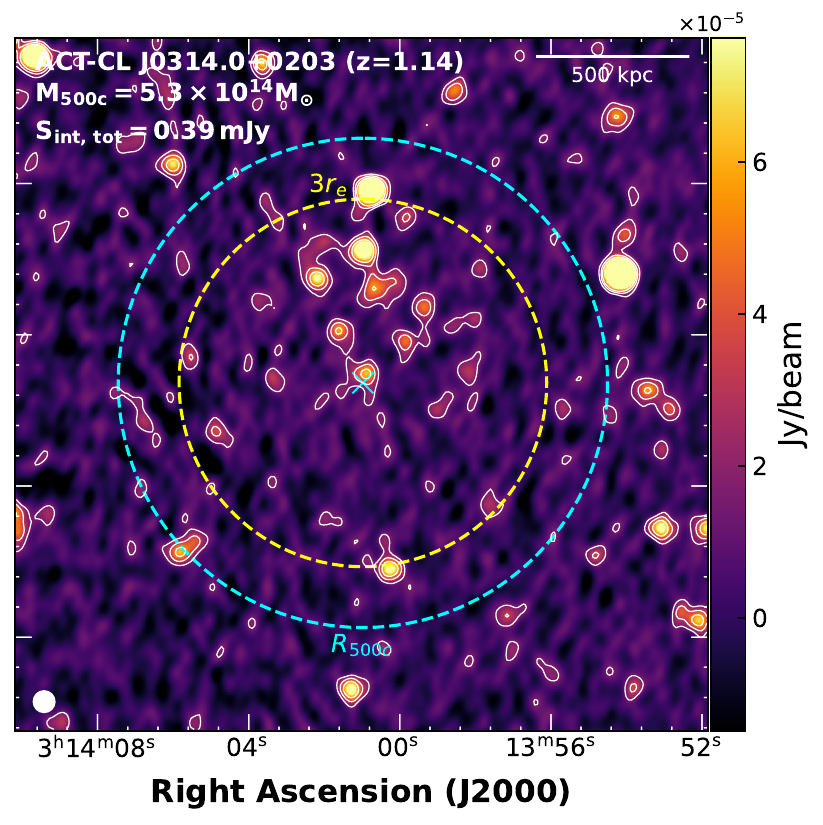}\\

    \includegraphics[width=0.29\textwidth]{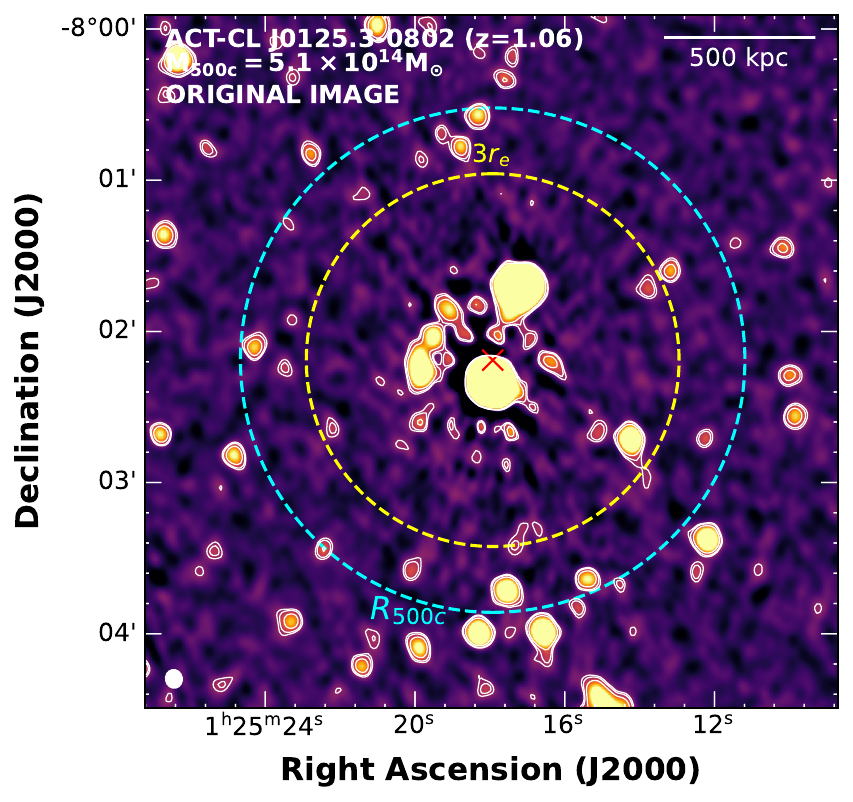} &
    \includegraphics[width=0.248\textwidth]{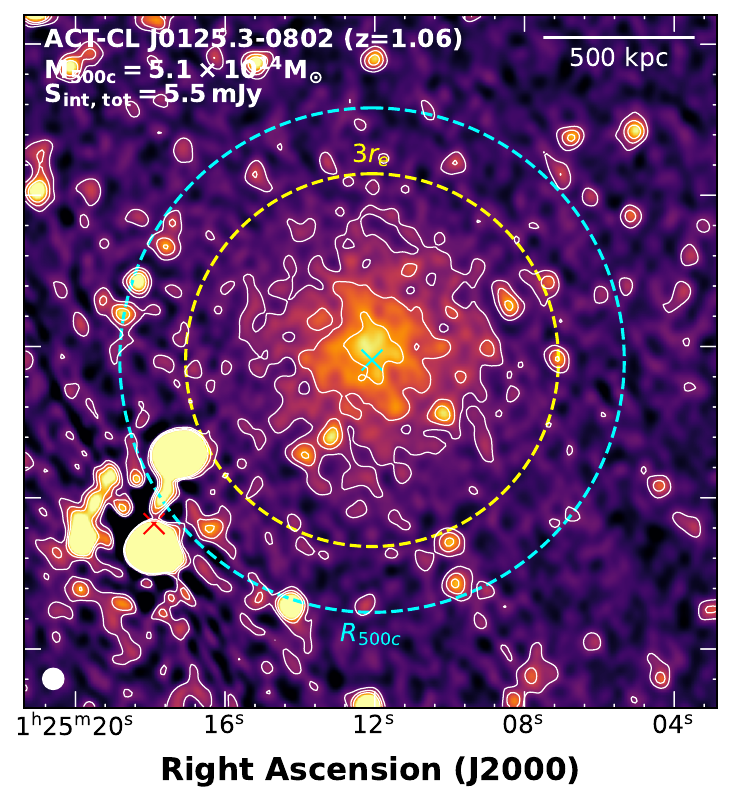} &
    \includegraphics[width=0.29\textwidth]{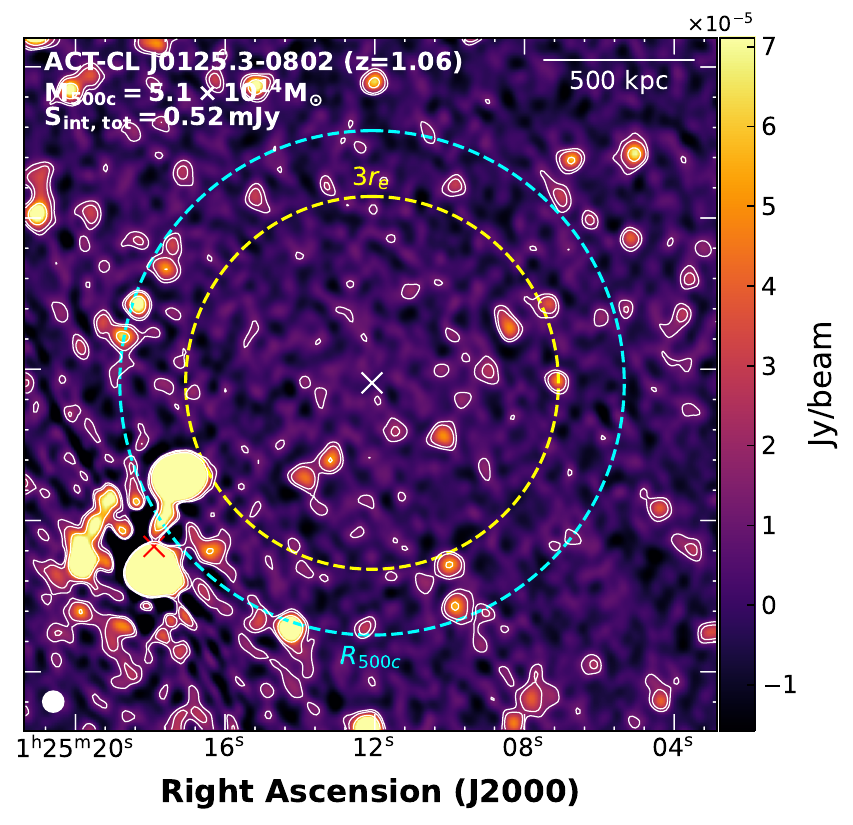} \\

    \includegraphics[width=0.287\textwidth]{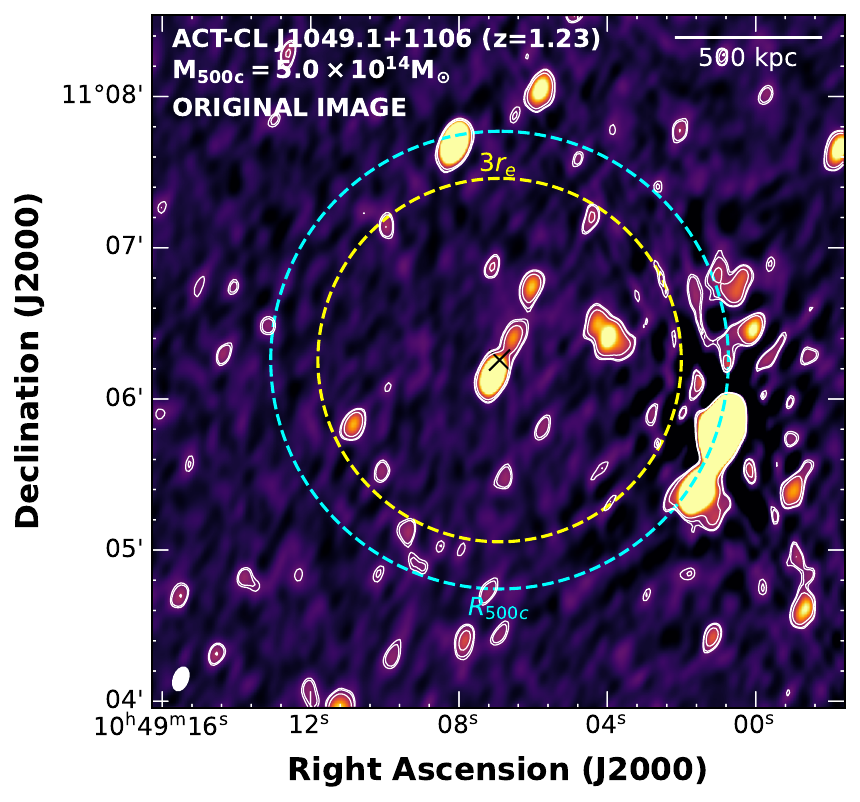} &
    \includegraphics[width=0.26\textwidth]{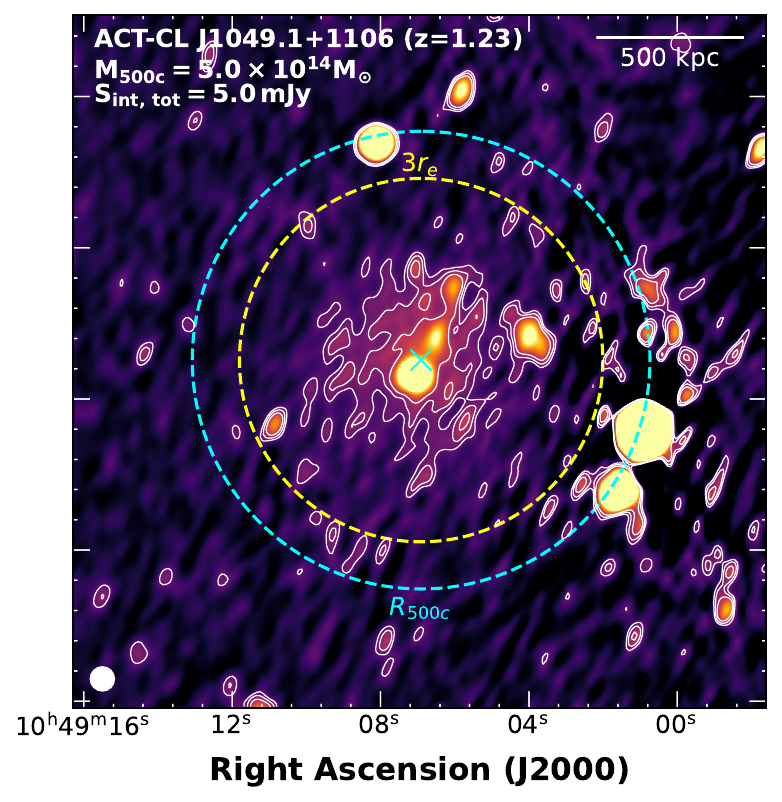} &
    \includegraphics[width=0.308\textwidth]{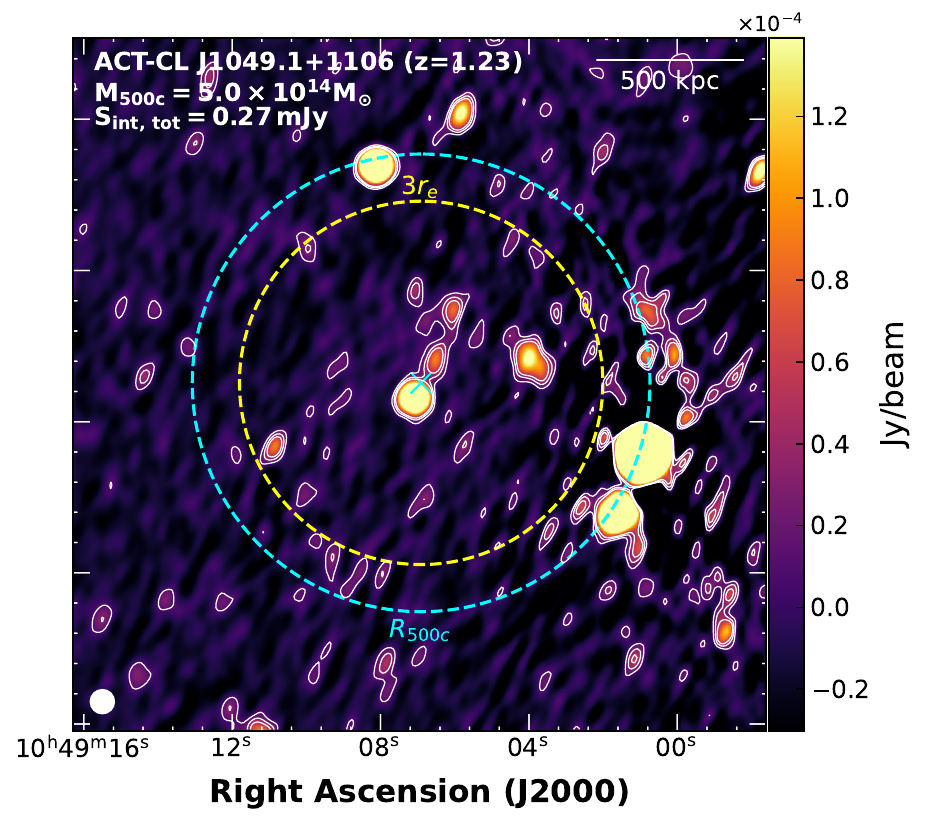} \\

\end{tabular}
\caption{Rows correspond to clusters ACT-CL J0044, J0314, J0125, and J1049. For each cluster, the left panel shows the original MeerKAT 1.28~GHz full-resolution image, with contours starting at $3\sigma$. The middle and right panels show the results of mock halo injections at different total flux densities. For ACT-CL J0244, the injection centre was shifted away from the cluster core because severe artefacts remained in the central region after point-source subtraction in the original image. The yellow circle, centred at (RA$_{\mathrm{inj}}$, Dec$_{\mathrm{inj}}$), indicates a radius of $3r_e$, and the upper limits are obtained with $S_{\rm inj, tot} = 0.23$, $0.39$, $0.52$, and $0.27$ mJy, respectively.}
\label{fig:mockhalo1}
\end{figure*}

\begin{figure*} 
\centering
\begin{tabular}{ccc}

    \includegraphics[width=0.285\textwidth]{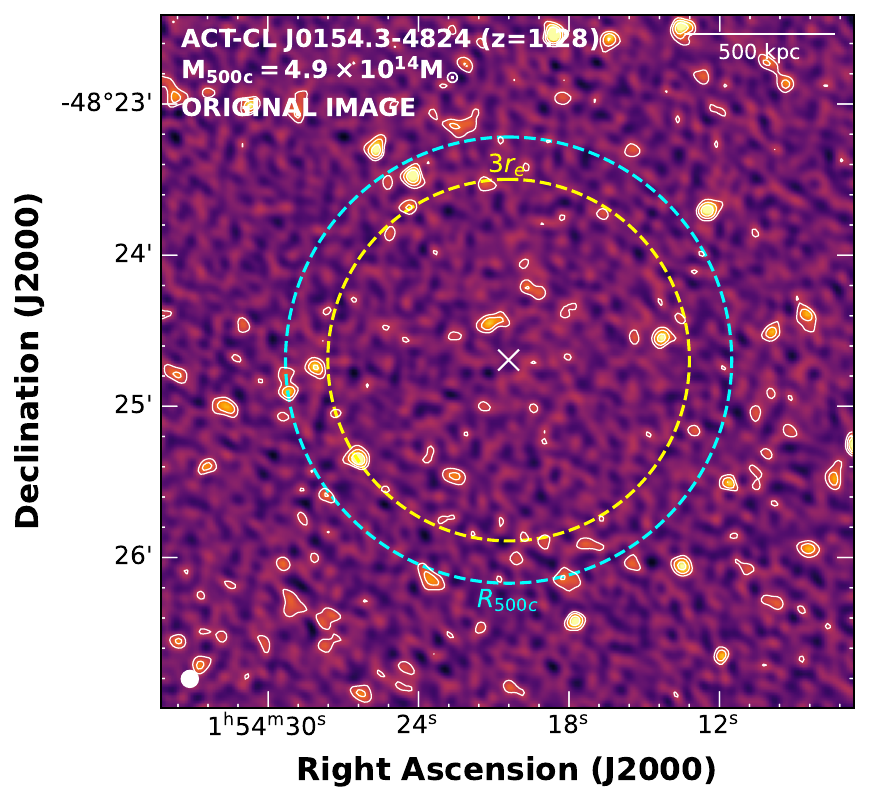} &
    \includegraphics[width=0.245\textwidth]{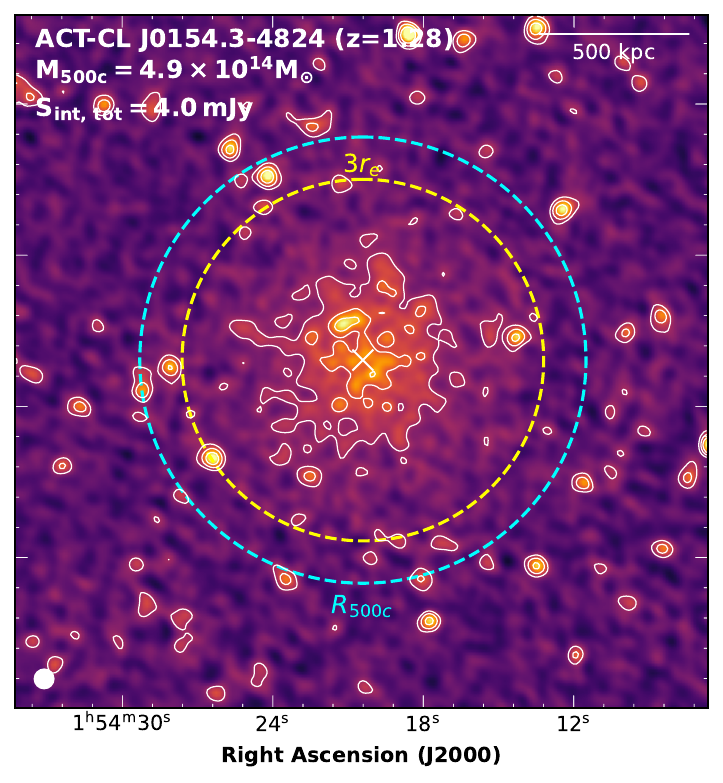} &
    \includegraphics[width=0.28\textwidth]{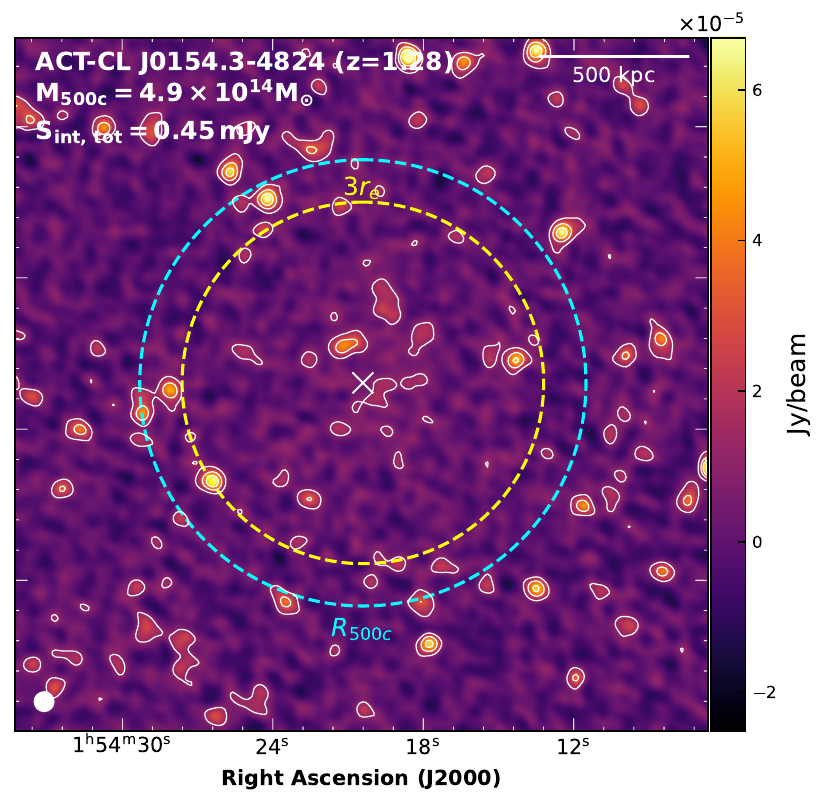}\\

    \includegraphics[width=0.288\textwidth]{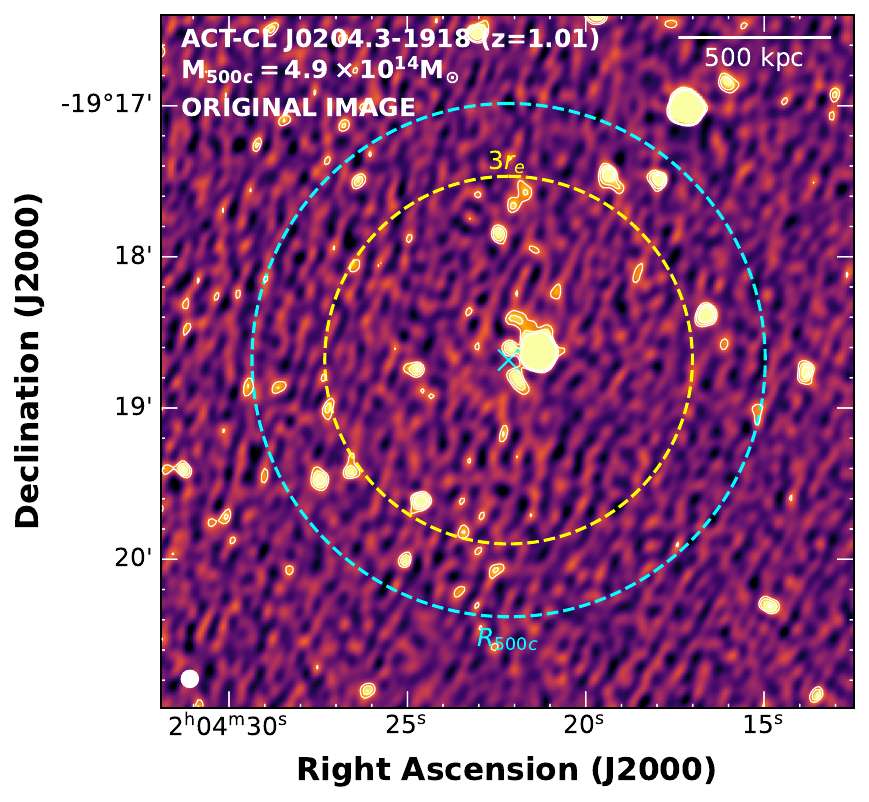} &
    \includegraphics[width=0.24\textwidth]{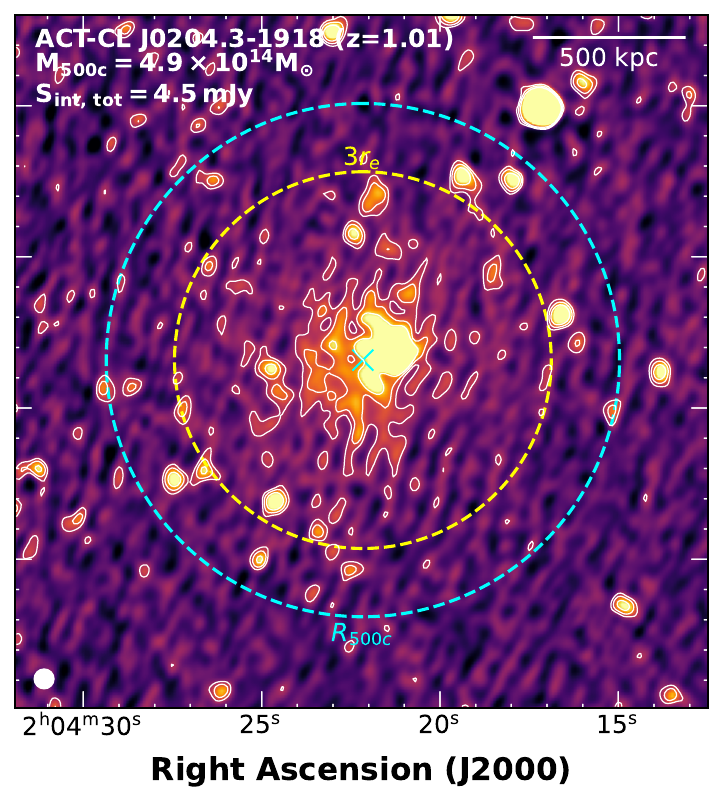} &
    \includegraphics[width=0.28\textwidth]{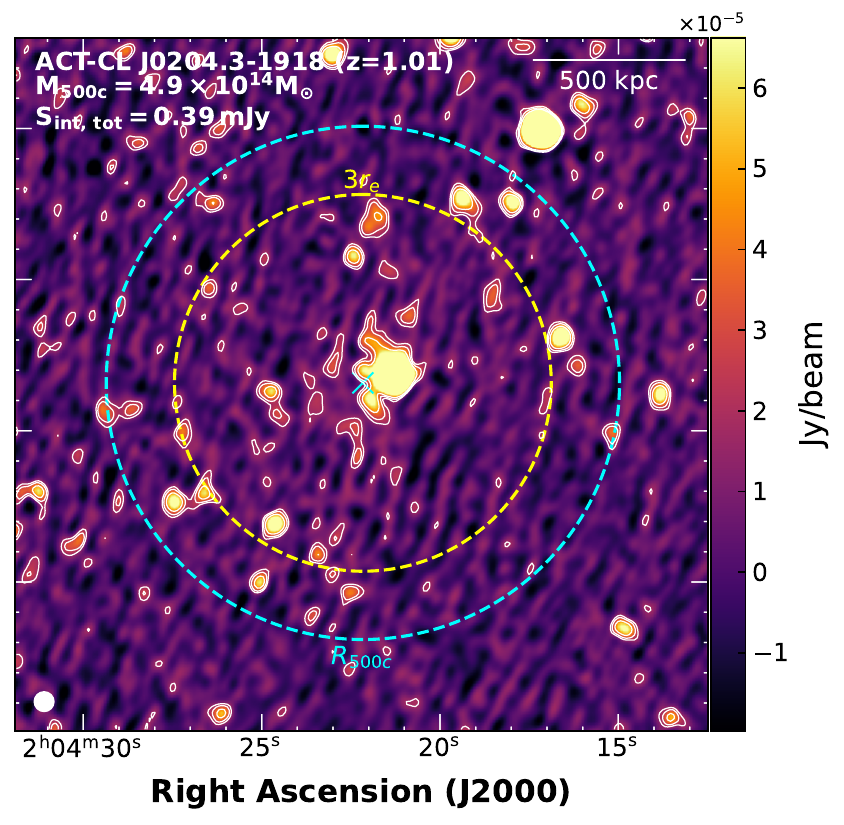}\\

    \includegraphics[width=0.292\textwidth]{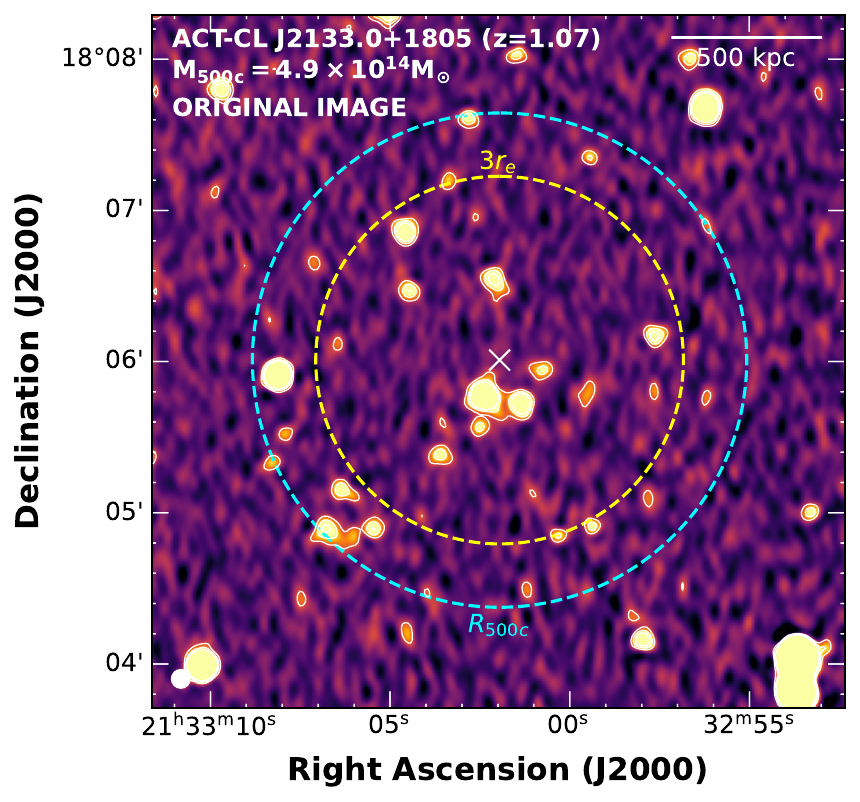} &
    \includegraphics[width=0.25\textwidth]{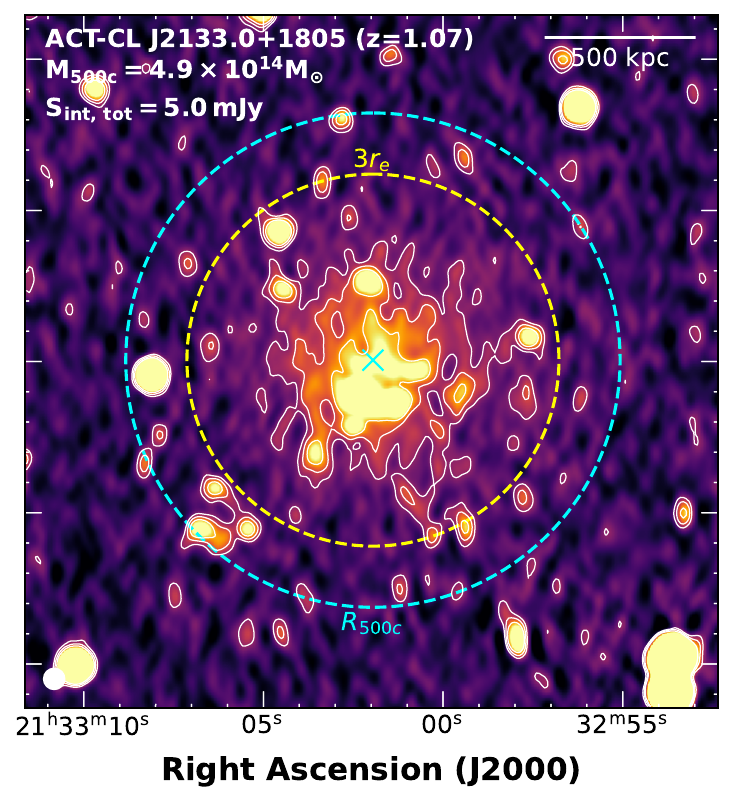} &
    \includegraphics[width=0.29\textwidth]{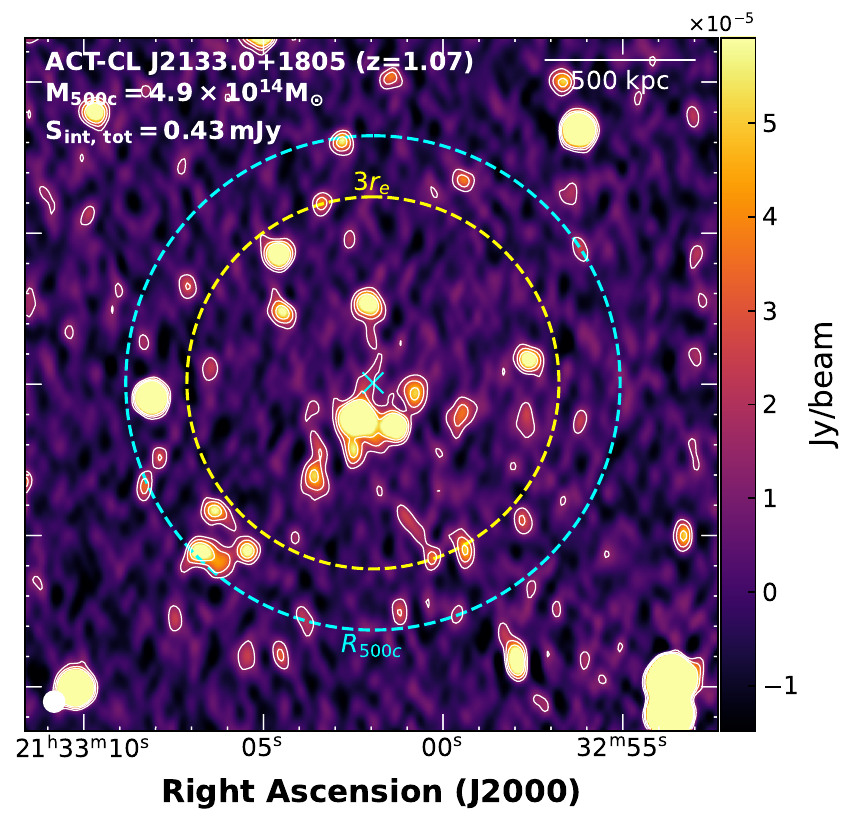} \\

    \includegraphics[width=0.297\textwidth]{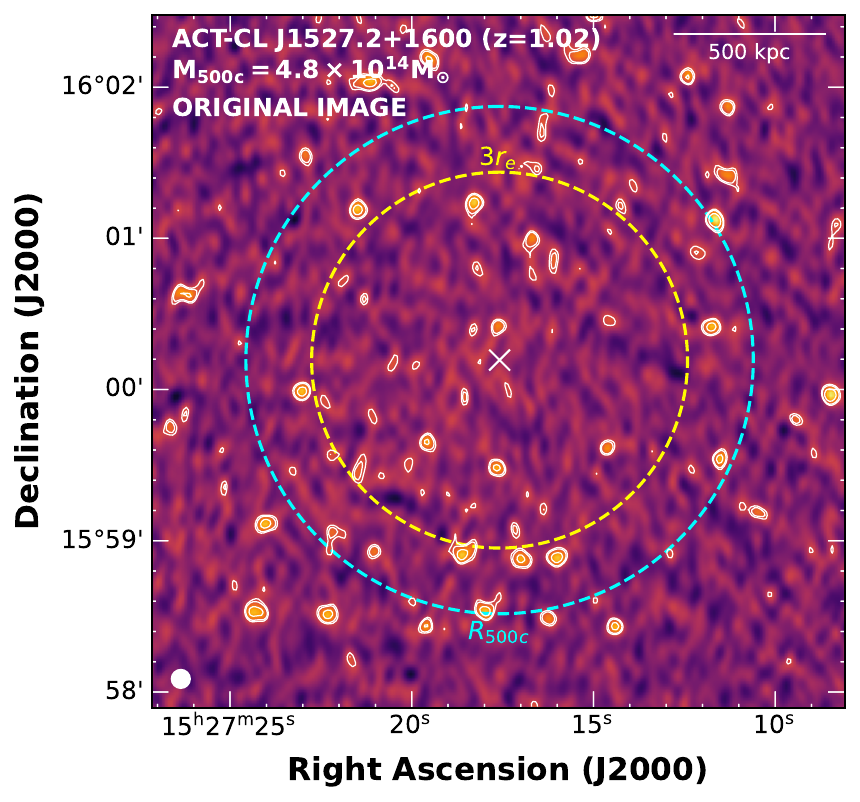} &
    \includegraphics[width=0.25\textwidth]{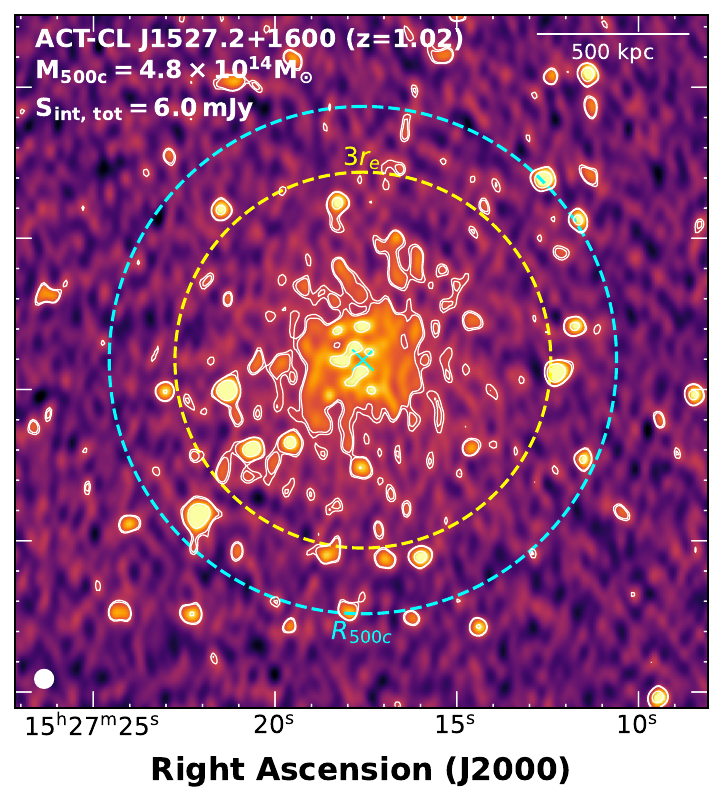} &
    \includegraphics[width=0.29\textwidth]{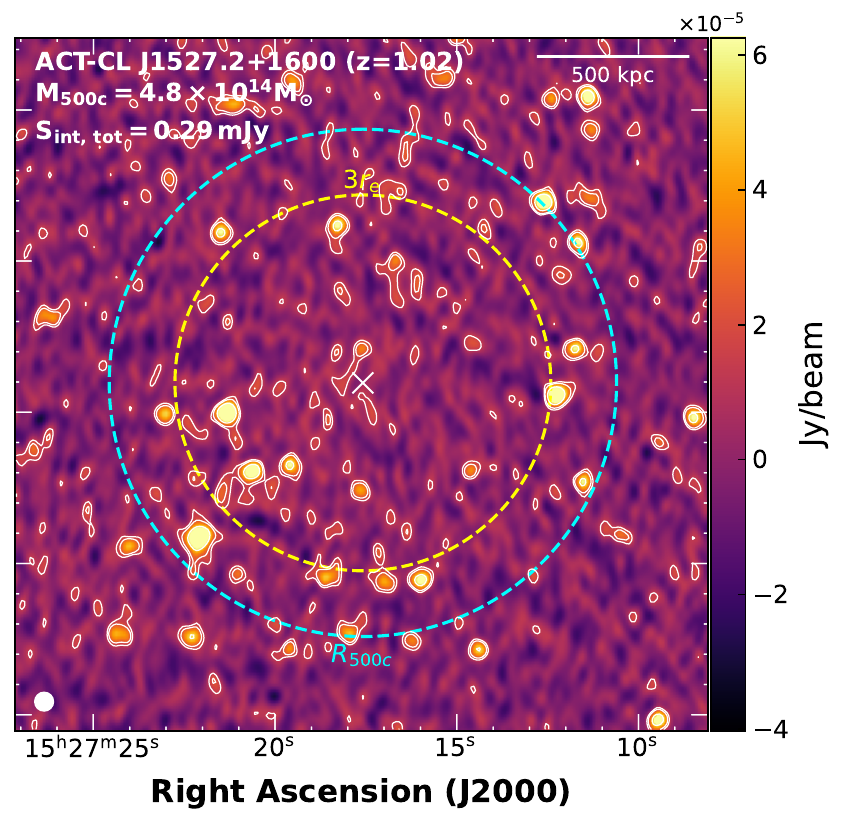} \\
\end{tabular}
\caption{Rows correspond to clusters ACT-CL J0154, J0204, J2133, and J1527. For each cluster, the left panel shows the original MeerKAT 1.28~GHz full-resolution image, with contours starting at $3\sigma$. The middle and right panels show the results of mock halo injections at different total flux densities. For ACT-CL J0125, the injection centre was shifted away from the cluster core because severe artefacts remained in the central region after point-source subtraction in the original image. The yellow circle, centred at (RA$_{\mathrm{inj}}$, Dec$_{\mathrm{inj}}$), indicates a radius of $3r_e$, and the upper limits are obtained with $S_{\rm inj, tot} = 0.45$, $0.39$, $0.43$, and $0.29$ mJy, respectively.}
\label{fig:mockhalo2}
\end{figure*}

\begin{figure*} 
\centering
\begin{tabular}{ccc}

    \includegraphics[width=0.296\textwidth]{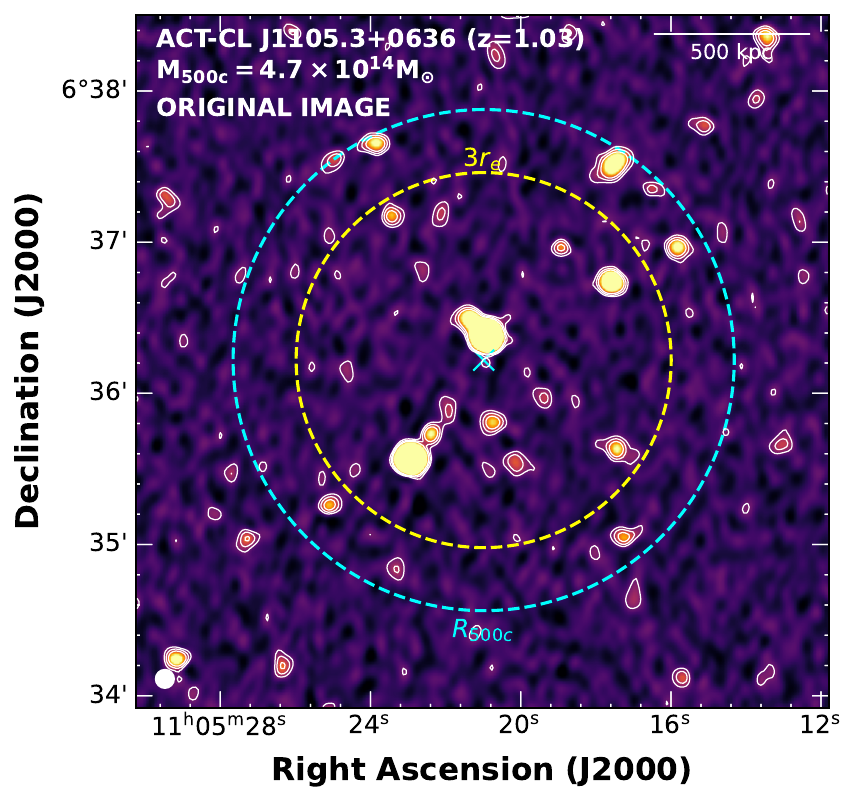} &
    \includegraphics[width=0.255\textwidth]{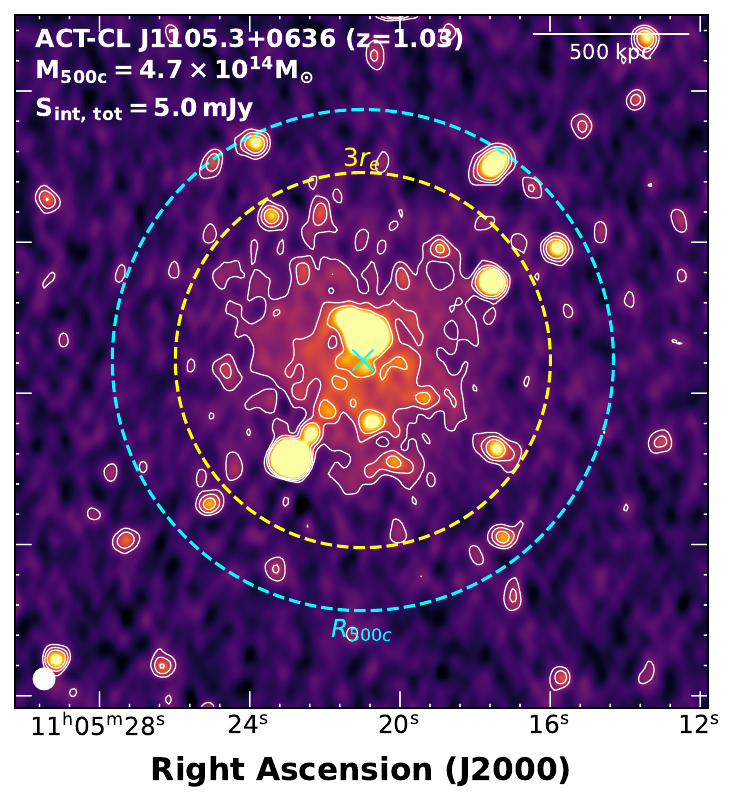} &
    \includegraphics[width=0.285\textwidth]{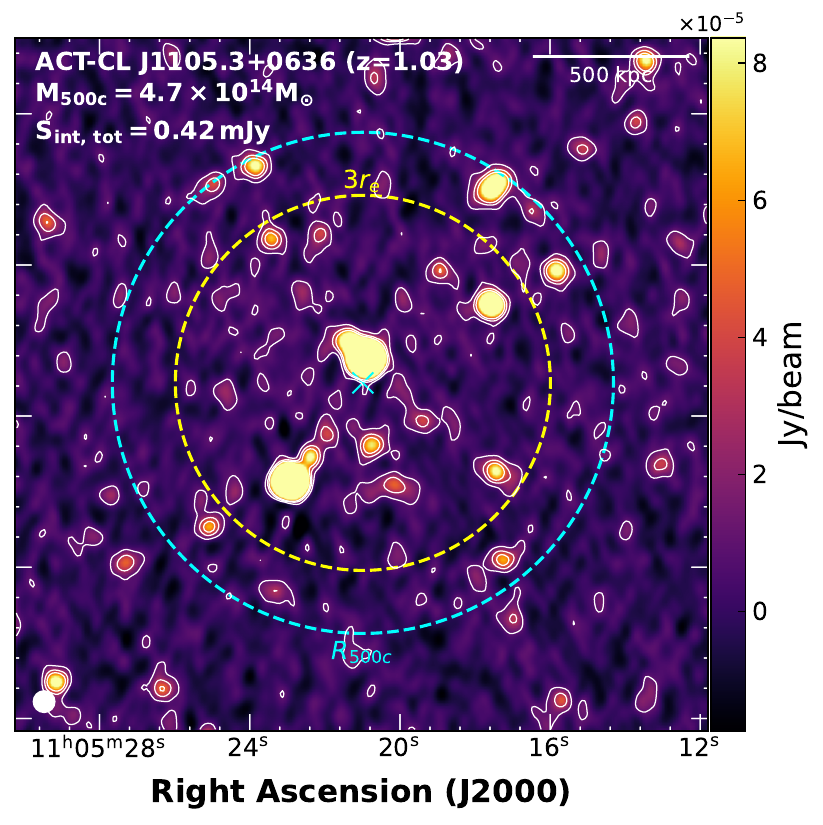}\\

    \includegraphics[width=0.295\textwidth]{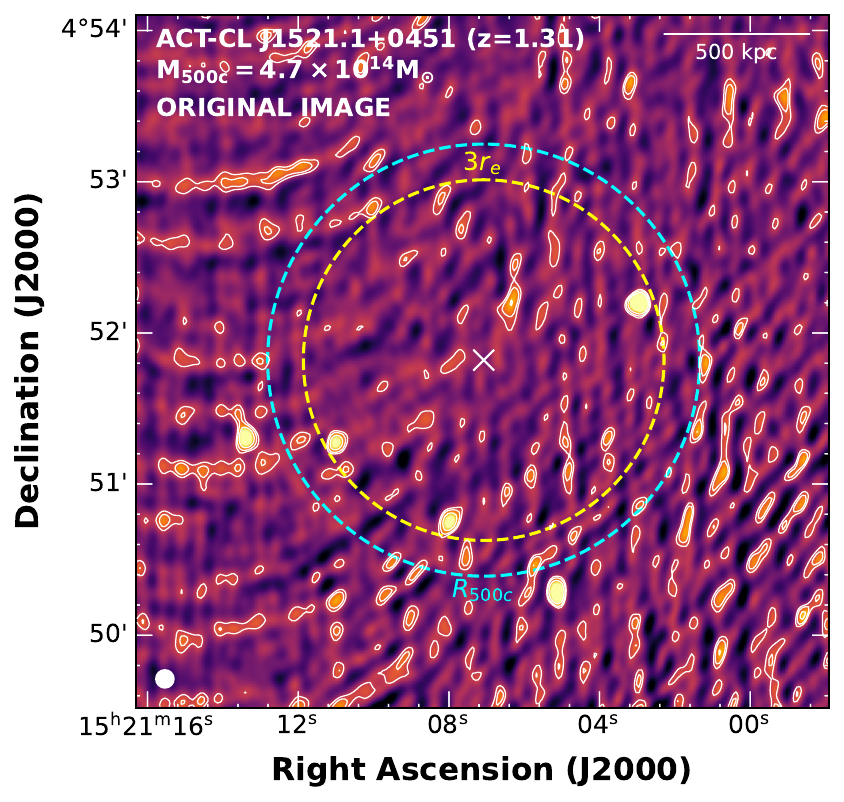} &
    \includegraphics[width=0.275\textwidth]{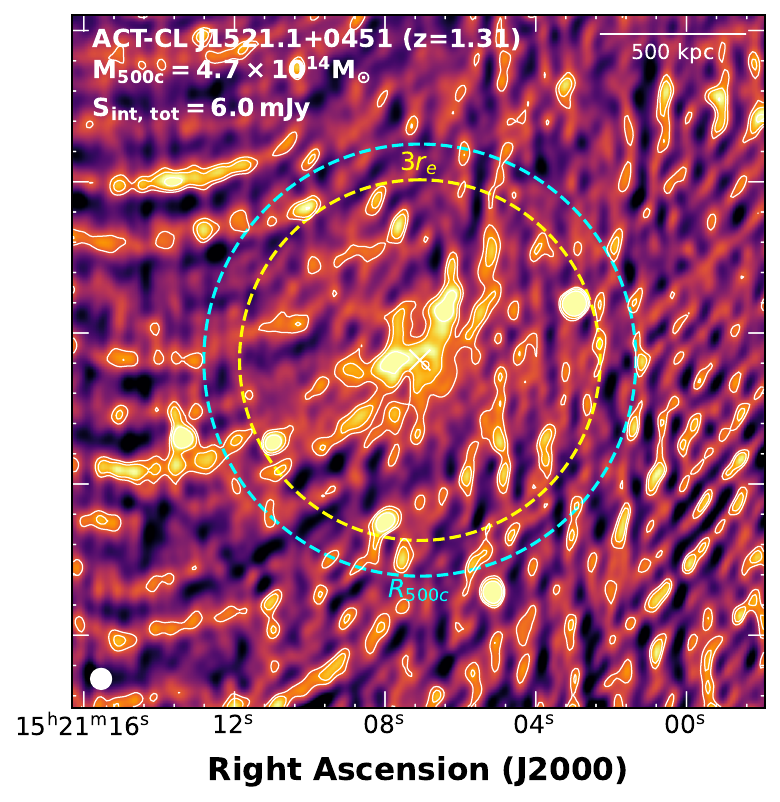} &
    \includegraphics[width=0.333\textwidth]{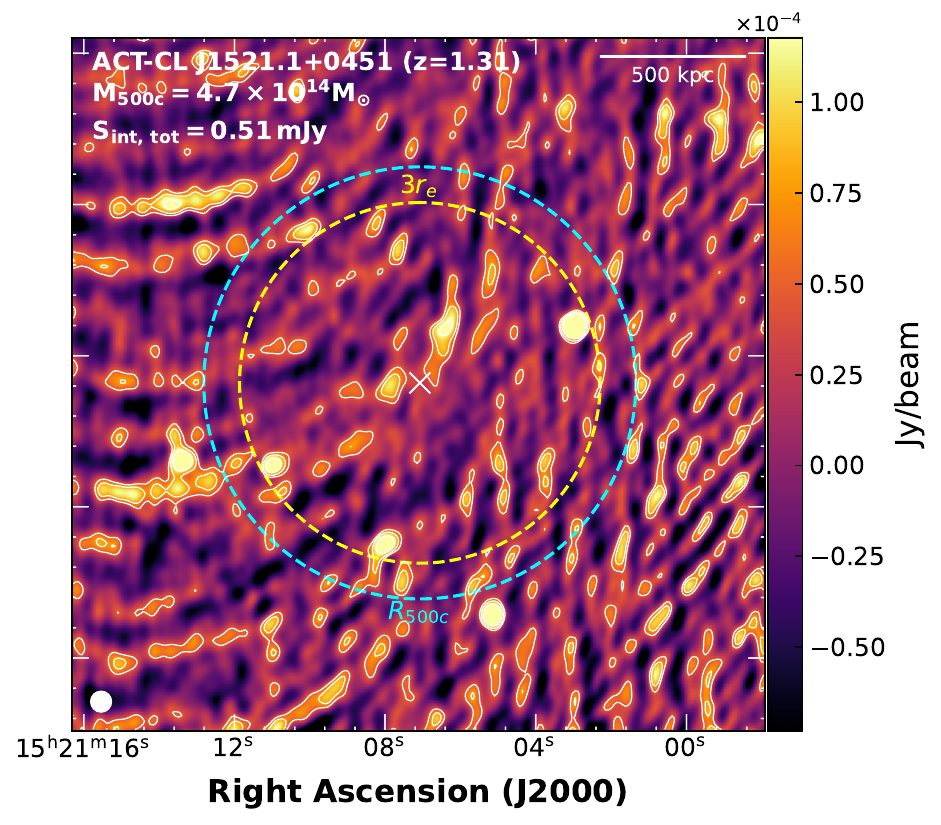}\\

    \includegraphics[width=0.295\textwidth]{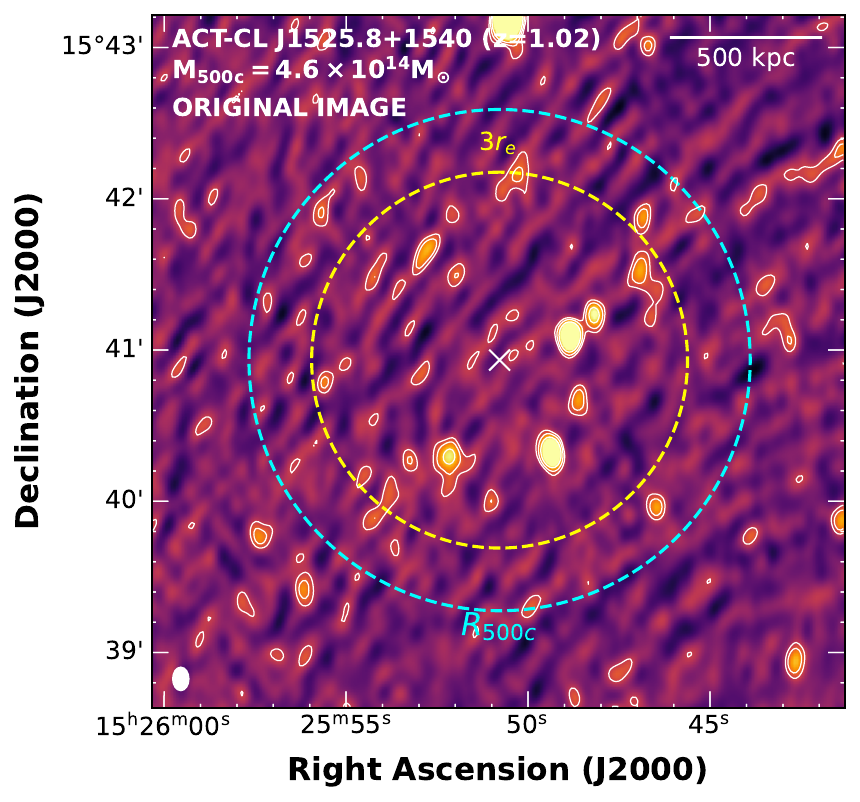} &
    \includegraphics[width=0.265\textwidth]{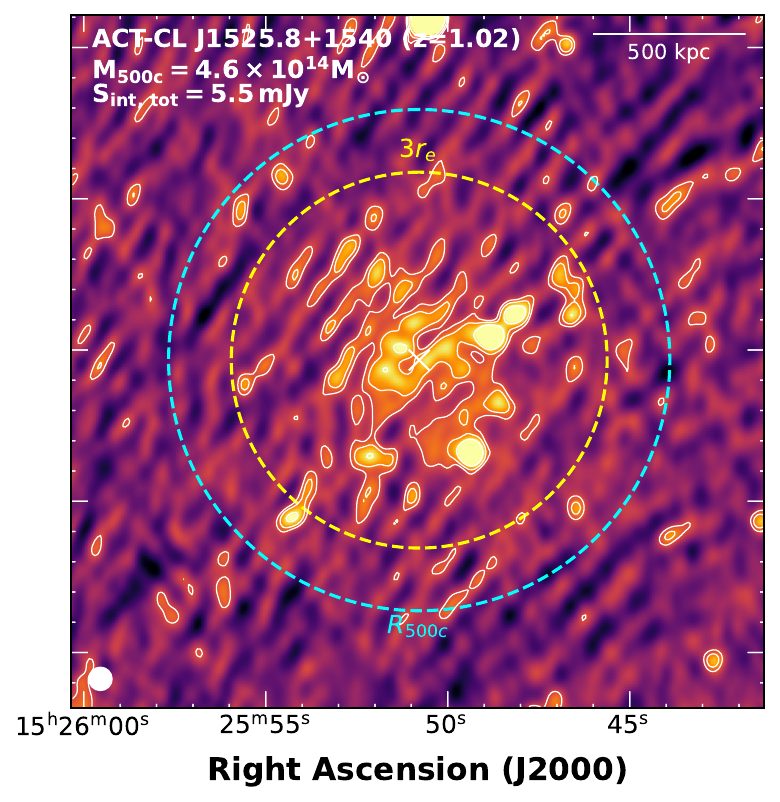} &
    \includegraphics[width=0.315\textwidth]{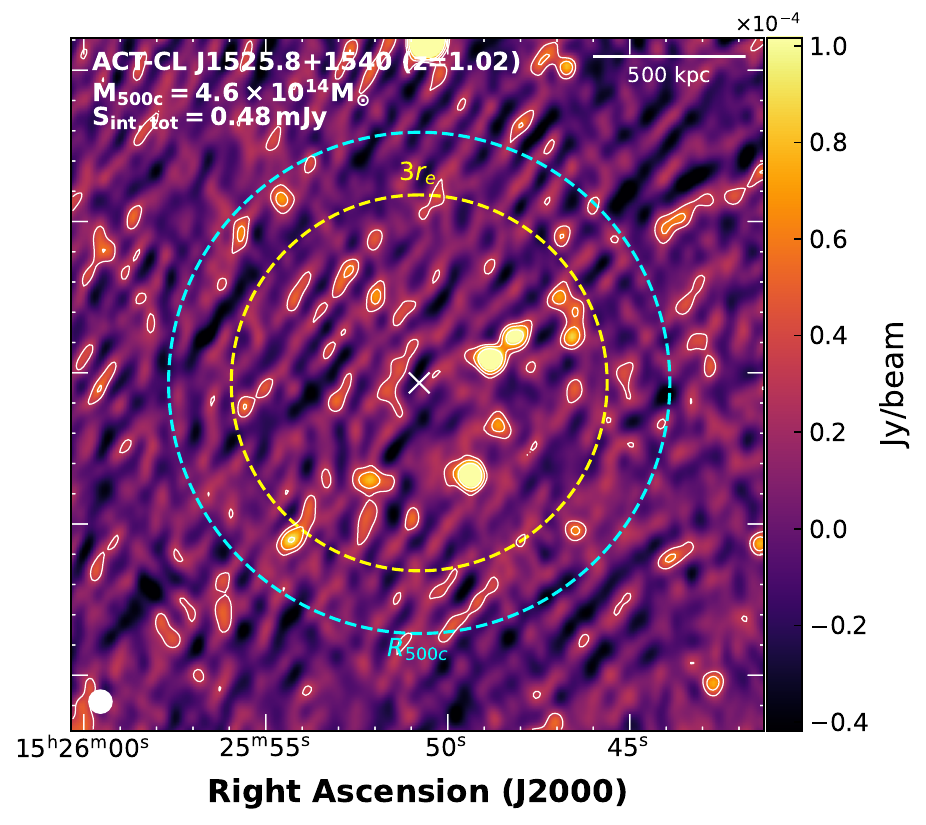} \\

    \includegraphics[width=0.298\textwidth]{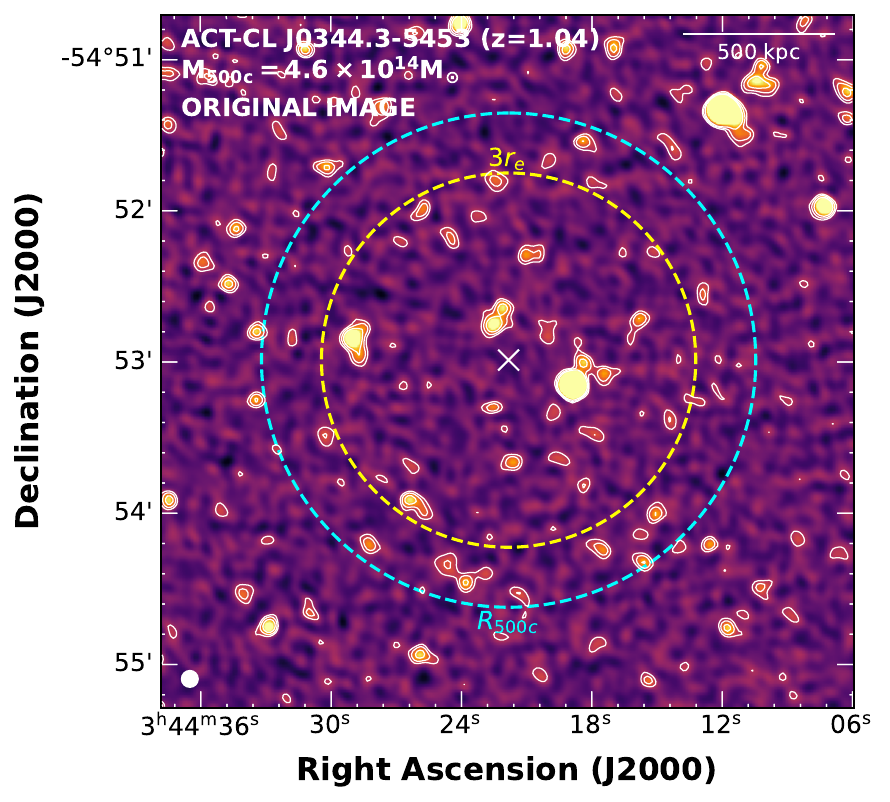} &
    \includegraphics[width=0.255\textwidth]{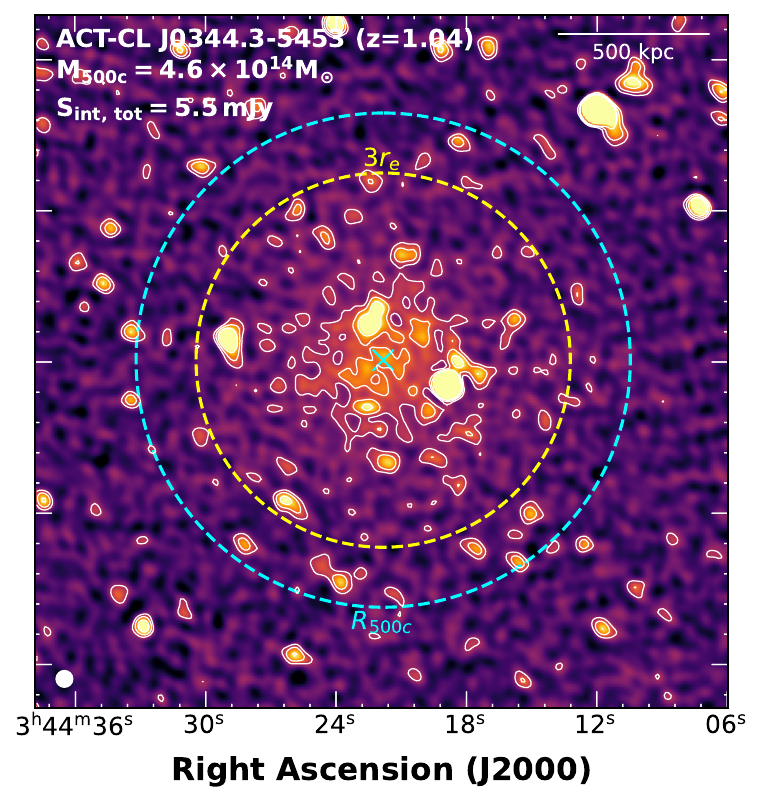} &
    \includegraphics[width=0.29\textwidth]{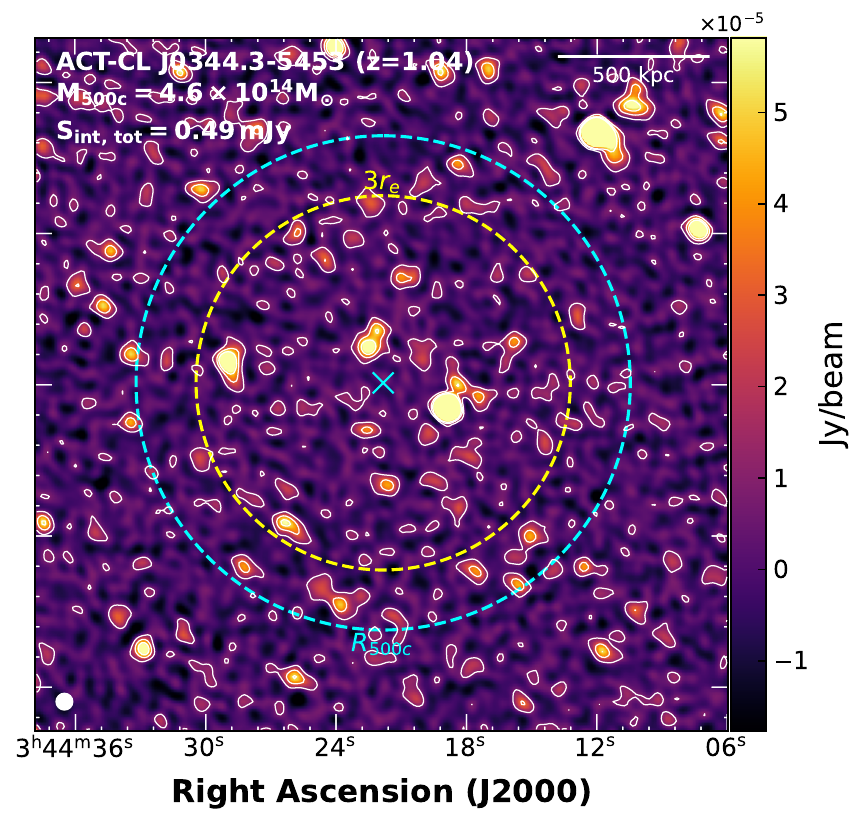} \\
\end{tabular}
\caption{Rows correspond to clusters ACT-CL J1105, J1521, J1525, and J0344. For each cluster, the left panel shows the original MeerKAT 1.28~GHz full-resolution image, with contours starting at $3\sigma$. The middle and right panels show the results of mock halo injections at different total flux densities. The yellow circle, centred at (RA$_{\mathrm{inj}}$, Dec$_{\mathrm{inj}}$), indicates a radius of $3r_e$, and the upper limits are obtained with $S_{\rm inj, tot} = 0.42$, $0.51$, $0.48$, and $0.49$ mJy, respectively.}
\label{fig:mockhalo3}
\end{figure*}

\begin{figure*}
\centering
\begin{tabular}{ccc}

    \includegraphics[width=0.292\textwidth]{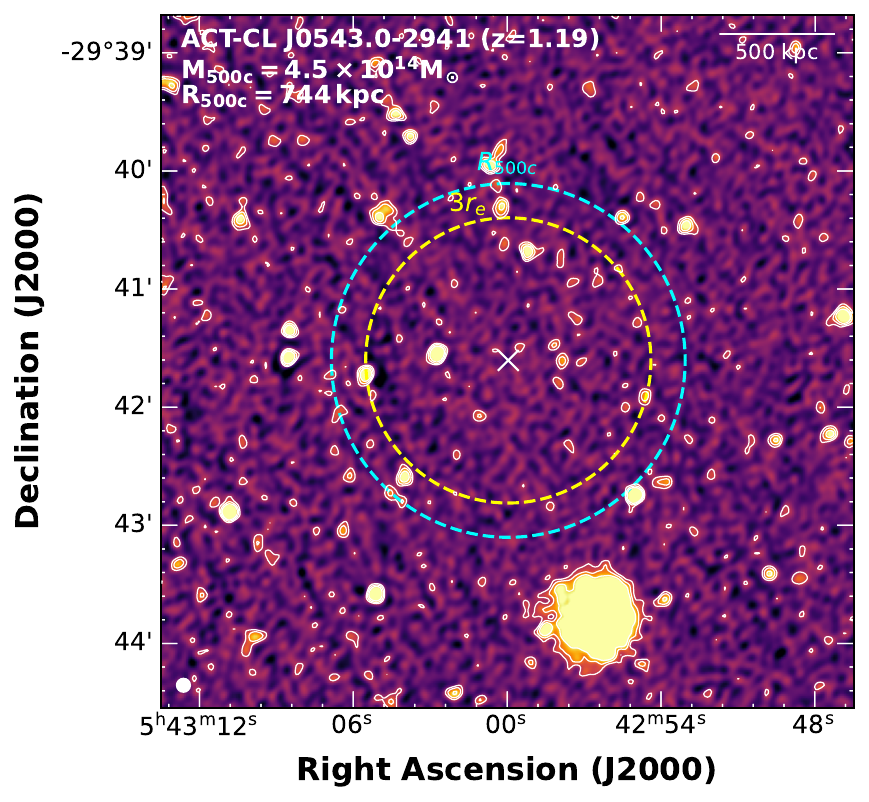} &
    \includegraphics[width=0.25\textwidth]{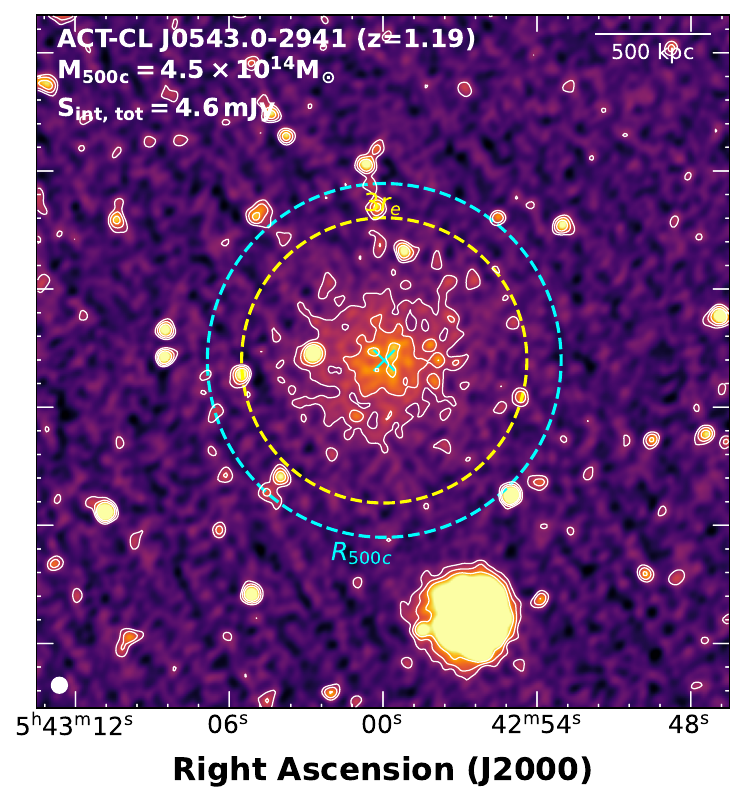} &
    \includegraphics[width=0.29\textwidth]{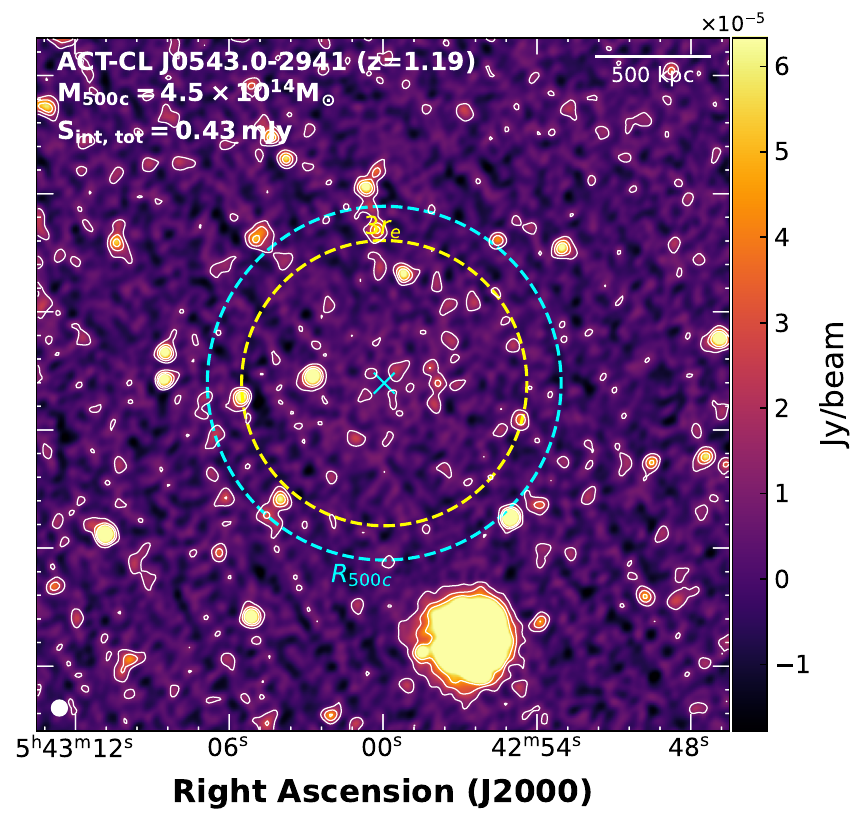}\\

    \includegraphics[width=0.275\textwidth]{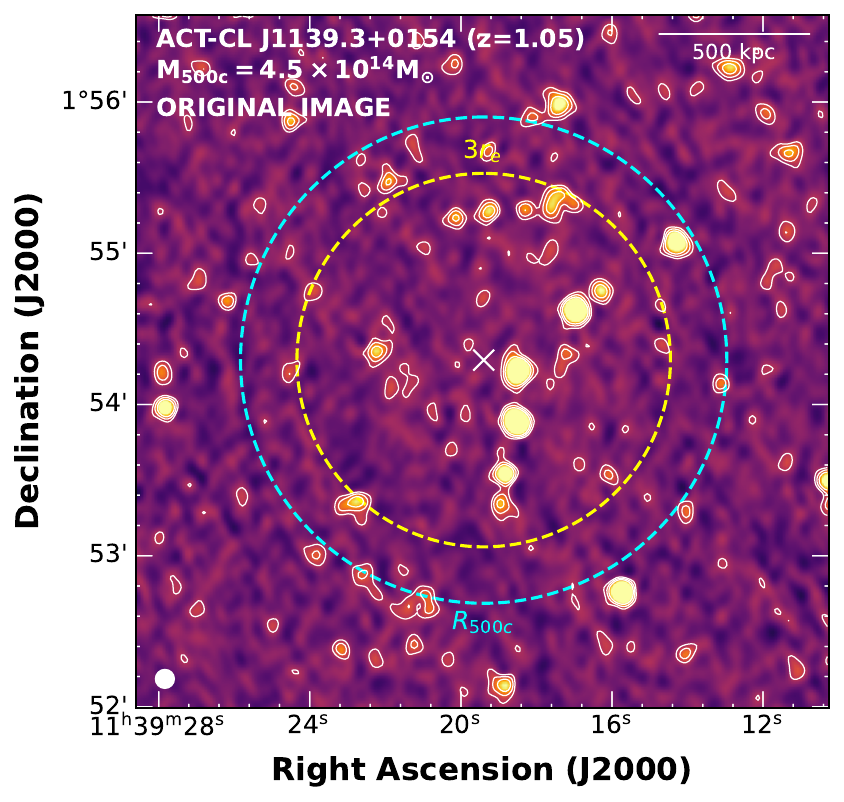} &
    \includegraphics[width=0.25\textwidth]{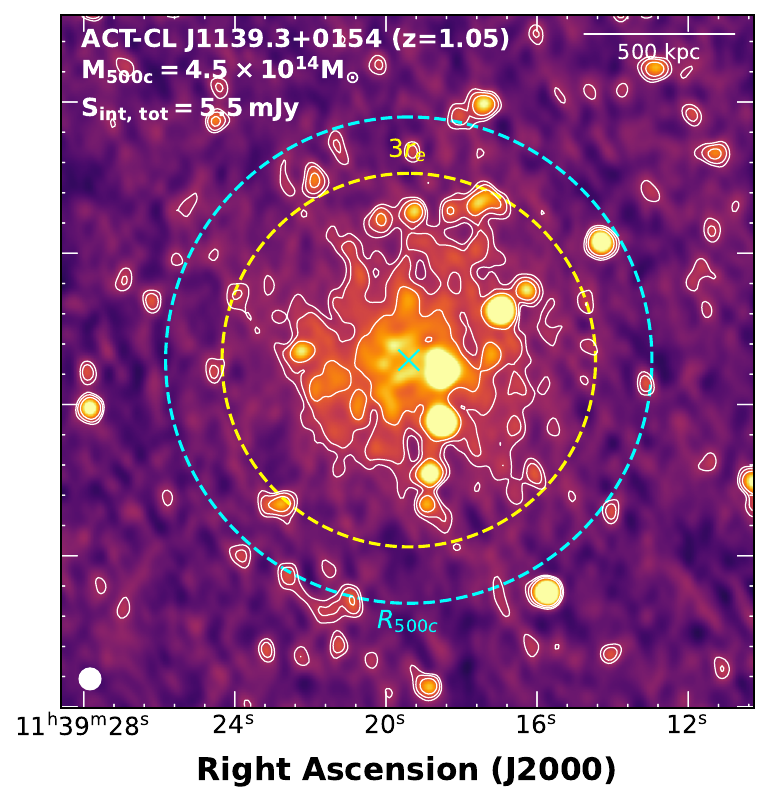} &
    \includegraphics[width=0.29\textwidth]{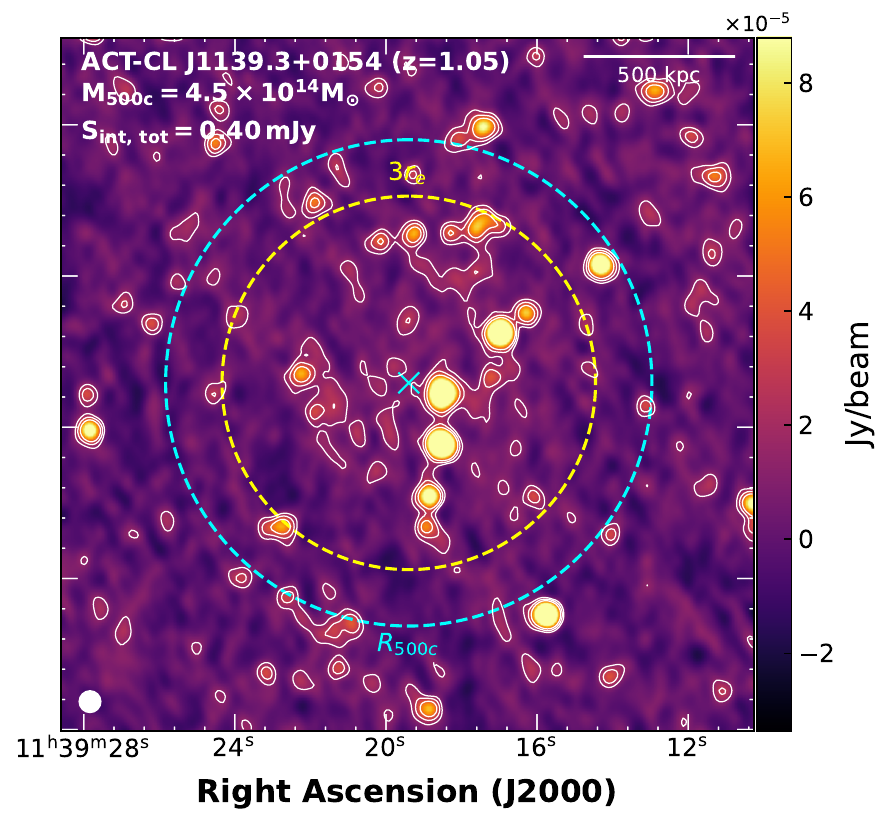} \\
\end{tabular}
\caption{Rows correspond to clusters ACT-CL J0543, and J1139. For each cluster, the left panel shows the original MeerKAT 1.28~GHz full-resolution image, with contours starting at $3\sigma$. The middle and right panels show the results of mock halo injections at different total flux densities. The yellow circle, centred at (RA$_{\mathrm{inj}}$, Dec${_\mathrm{inj}}$), indicates a radius of $3r_e$, and the upper limits are obtained with $S_{\rm inj, tot} = 0.43$, and $0.40$ mJy, respectively.
}
\label{fig:mockhalo4}
\end{figure*}

This appendix provides examples of the mock halo injection procedure used to derive upper limits for clusters without significant diffuse-emission detections. Figure~\ref{fig:injection_process} illustrates the injection workflow for a representative cluster, showing the model halo, the original tapered image, and the image after injection. Figures~\ref{fig:mockhalo}--\ref{fig:mockhalo4} present the corresponding injection results for the non-detection subsample used in the upper-limit analysis. These figures show how the visibility of the injected halo changes with total flux density and demonstrate the basis on which the adopted upper limits were determined.

\section{Radio source catalogues}
\label{app:catalogs}

Compact radio sources in each field were characterised using the
Python Blob Detection and Source Finder \citep[\texttt{PyBDSF}, v1.9.2;][]{2015ascl.soft02007M}\footnote[16]{\url{https://pybdsf.readthedocs.io/en/latest/}}
applied to the full-resolution MeerKAT 1.28\,GHz images. Source finding was performed on primary-beam
corrected maps, after masking bright artefacts where necessary.
We adopted a pixel detection threshold of $5\sigma$ and an island
threshold of $3\sigma$, where $\sigma$ is the local rms noise
estimated by \texttt{PyBDSF} in sliding boxes. Only components with
peak surface-brightness $\ge 5\sigma$ were retained in the final
catalogues.
\newline

For each detected component, \texttt{PyBDSF} provides positions,
integrated and peak flux densities, deconvolved sizes, local noise
estimates, and basic shape parameters from Gaussian fits. Sources
classified by \texttt{PyBDSF} as multiple Gaussians belonging to a
single physical object were grouped into islands and assigned a
common source identifier.

\bsp	
\label{lastpage}
\end{document}